\pdfoutput=1
\documentclass[11pt,a4paper]{article}
\usepackage{jheppub}
\usepackage[caption=false]{subfig}
\usepackage[export]{adjustbox} 
\usepackage{xcolor}

\newcommand{\beq}{\begin{eqnarray}}
\newcommand{\eeq}{\end{eqnarray}}
\newcommand{\non}{\nonumber\\}

\newcommand{\p}{\partial}
\def\Tr{\qopname\relax o{Tr}}
\def\tr{\qopname\relax o{tr}}

\newcommand{\bx}{\mathbf{x}}
\newcommand{\bxhat}{\hat{\mathbf{x}}}

\newcommand{\dvarphi}{\delta{\mkern-1mu}\varphi}
\newcommand{\dPhi}{\delta{\mkern-1mu}\Phi}
\newcommand{\df}{\delta{\mkern-2.5mu}f}

\newcommand{\dlambda}{\delta{\mkern-2.5mu}\lambda}
\newcommand{\dU}{\delta U}

\newcommand{\dtheta}{\delta{\mkern-1mu}\theta}

\newcommand{\Og}{\qopname\relax o{O}}
\newcommand{\SU}{\qopname\relax o{SU}}
\newcommand{\SL}{\qopname\relax o{SL}}
\newcommand{\Lag}{\mathcal{L}}
\newcommand{\calE}{\mathcal{E}}
\renewcommand{\d}{{\mathrm{d}}}
\renewcommand{\i}{\mathrm{i}}
\renewcommand{\div}{{\qopname\relax o{div}}}

\subheader{\rm\hfill IFUP-TH-2022}
\title{Near-BPS Skyrmions} 

\author{Sven Bjarke Gudnason$^1$,}
\affiliation{$^1$Institute of Contemporary Mathematics, School of
  Mathematics and Statistics, Henan University, Kaifeng, Henan 475004,
  P.~R.~China}
\emailAdd{gudnason(at)henu.edu.cn}
\author{Marco Barsanti$^{2}$ and}
\emailAdd{marco.barsanti(at)phd.unipi.it}
\author{Stefano Bolognesi$^{2}$}
\affiliation{$^2$Department of Physics ``E. Fermi'', University of
  Pisa and  INFN, Sezione di Pisa, 
  Largo Pontecorvo, 3, Ed. C, 56127 Pisa, Italy}
\emailAdd{stefanobolo(at)gmail.com}

\abstract{
  We consider the Skyrme model in the near-BPS
  limit. The BPS part is made of the sextic term plus a potential
  and the deformation is made of the standard massive Skyrme model
  controlled by a small parameter $\epsilon\ll1$. 
  In order to keep the perturbation under theoretical and
  computational control, we find a model for which BPS Skyrmions have compact support,
  henceforth denoted as compactons, and the spherically
  symmetric $B=1$ Skyrmion represents the most stable solution.
  We use the $\epsilon$-expansion scheme to systematically calculate
  the corrections to the energy and compare with the exact numerical
  computations in the $B=1$ sector.
  Finally, we use the $\epsilon$-expansion scheme to calculate the
  bound state of two $B=1$ Skyrmions and its
  binding energy, which corresponds, prior to quantization, to the deuteron in our model.
}

\begin{document}
\maketitle

\section{Introduction}

The Skyrme model \cite{Skyrme:1961vq,Skyrme:1962vh} is a
field-theoretic approach to nuclear physics based on the symmetries of
the strong interactions and the topology of chiral symmetry breaking,
providing the stability of the baryon as a topological soliton.
The topological soliton, also known as the Skyrmion, is exactly the
baryon of large-$N_c$ QCD \cite{Witten:1983tw,Witten:1983tx}.
A similar correspondence between the baryon and the instanton is
realized in holographic QCD models, such as the Witten-Sakai-Sugimoto
model \cite{Witten:1998zw,Sakai:2004cn}.
In most varieties of the Skyrme model, the binding energies come out
too large by roughly an order of magnitude with respect to the
phenomenological ones.
This not only has the effect that the ground state energies are
imprecise, but also leads to the illusion that the Skyrmions in each
topological sector are well separated in field space by an energy
barrier, which in turn validates the rigid-body quantization
\cite{Adkins:1983ya} and harmonic vibrational quantization
\cite{Halcrow:2015rvz,BjarkeGudnason:2018bju} approaches to the
quantum problem of nuclei -- although this turns out not even to be
true for the standard Skyrme model with a massive pion for large
baryon numbers \cite{Gudnason:2022jkn}.

The main approach to lowering binding energies in the Skyrme model is
to find a suitable BPS limit of the model, since in such a limit the
energy or mass is directly proportional to the topological degree or
baryon number, thus yielding vanishing binding energies at the
classical level.
The idea is then that a suitable small perturbation
around the BPS limit would be the right place to look for a
phenomenologically viable model.
There are several known BPS versions of the Skyrme model:
1) The mode expansion of 5D Yang-Mills theory in flat space as the Skyrme
model coupled to an infinite tower of vector mesons
-- this model is called the Sutcliffe model \cite{Sutcliffe:2010et}.
2) The replacement of the standard Skyrme model with a sextic derivative
term, which is the topological charge density squared, and a suitable
potential -- this model is often called the BPS-Skyrme model
\cite{Adam:2010fg,Adam:2010ds}.
3) The exclusion of the kinetic term and the altercation of the pion
mass term from the first to the fourth power -- this model is called
the lightly-bound Skyrme model \cite{Harland:2013rxa}.
4) The promotion of the coupling constants to being functions of the
isospin conserving part of the chiral Lagrangian field (i.e.~the sigma
field), which is inspired by the dielectric deformation of Maxwell
theory -- this model is called the dielectric Skyrme model
\cite{Adam:2020iye}.
1) The near-BPS limit of the Sutcliffe model is made by truncating the
infinite tower of vector mesons; this truncation breaks conformal
symmetry and introduces a scale in the model and numerical
computations suggest that two or three vector mesons are needed to reach
phenomenologically viable binding energies
\cite{Sutcliffe:2011ig,Naya:2018mpt}.
2) The near-BPS limit of the BPS-Skyrme model is taken by adding the
standard massive Skyrme model to the BPS-Skyrme sector with a suitably
small coefficient, which however complicates numerical computations
because very large field derivatives are naturally occurring in this
limit \cite{Gillard:2015eia}. A reason of interest in the BPS-Skyrme
model is due to the fact that solutions appear as liquid drops of
incompressible baryonic matter sharing, therefore, the same features
of real nuclei.
3) The near-BPS limit of the lightly-bound Skyrme model is taken by
adding the kinetic term and the pion mass term to the model; the
resulting Skyrmions become point-particle like \cite{Gillard:2015eia}
and are hence quite different from ordinary Skyrmions with larger
symmetries \cite{Manton:2004tk}.
4) The near-BPS limit of the dielectric Skyrme model is taken by
altering the form of the dielectric coupling constant (function)
\cite{Adam:2020iye}, but the Skyrmions again become point-particle
like in the limit where the binding energies become phenomenologically
viable \cite{Gudnason:2020ftf}.

In this paper, we will study the case 2), i.e.~that of the BPS-Skyrme
model and its near-BPS deformation. For the BPS part, we will take a
potential providing no contribution to the pion mass, whereas the
perturbation is chosen to be the standard massive Skyrme model --
i.e.~with the kinetic, the Skyrme and the pion mass terms -- all
multiplied by a small control parameter, $\epsilon$.
This work is a continuation of the $\epsilon$-expansion scheme that we
have developed for the case of the baby Skyrme model in the case of
compactons (Skyrmions with compact support in the BPS limit)
\cite{Gudnason:2020tps} and in the case of baby Skyrmions with
exponential (or rather Gaussian) tails \cite{Gudnason:2021gwc}.
In the baby Skyrme model case, we were able to check the precision and
validity of the $\epsilon$-expansion scheme as a perturbative approach
to near-BPS solitons by performing very large brute force numerical
computations -- only possible in the 2-dimensional case.
We also found in the previous work, that the precision of the
$\epsilon$-expansion scheme is better in the case of compactons as
compared to the solitons with tails \cite{Gudnason:2021gwc}.
Using this result as a guideline, in this paper we search for a viable
BPS model with compactons.
Due to the BPS solution for compactons having a discontinuous
derivative at the compacton boundary, as opposed to the true near-BPS
solution, we found in ref.~\cite{Gudnason:2020tps} that a certain cusp
condition must be imposed at the compacton boundary, making the total
field smooth there.
This becomes highly nontrivial if the compacton has a complicated
shape and we thus limit our search to stable $B=1$ compactons with
spherical symmetry.
These criteria limit our model to a rather specific choice,
with essentially only one parameter to dial -- namely $\epsilon$.
Finally, we compute the bound state of two spherically symmetric $B=1$
compactons by performing PDE solutions within the framework of the
semi-analytic $\epsilon$-expansion scheme and compute the binding
energy.

The analysis of the near-BPS Skyrme model has been performed in the
literature firstly in the series of works
\cite{Bonenfant:2010ab,Bonenfant:2012kt,Gillard:2015eia,Speight:2014fqa}. Various
difficulties emerged from these studies, so that only a partial
exploration of the model could be carried out. 
In refs.~\cite{Bonenfant:2010ab,Bonenfant:2012kt}, a first attempt of
an analytic approximation for the near-BPS model has been made using
an axially symmetric BPS solution.
Then, once inserted into the Lagrangian, a first approximation of the
near-BPS energy can be evaluated.
Starting from that result and after an appropriate quantization
procedure, the binding energies for the various nuclei have been
obtained, showing a reasonable agreement with experimental data
(mostly for large nuclei). Despite this result, the validity of the
entire analysis is questioned in
refs.~\cite{Gillard:2015eia,Speight:2014fqa}. 
In fact, as it was proven in the latter references, not all the BPS
solutions can be used as first approximation to the near-BPS
field. The proper BPS solution must, in fact, respect a mathematical
criterion called the restricted harmonic criterion
\cite{Speight:2014fqa}.
As that theorem is not respected by the choices made in
refs.~\cite{Bonenfant:2010ab,Bonenfant:2012kt}, the entire work must
be revisited. 
Generically, the moduli space -- present in the BPS limit -- is lifted
by a shallow effective potential and as long as the perturbation
parameter is sufficiently small, the near-BPS solutions reside close
to the BPS solutions in field space.
In the case of the BPS-Skyrme model, however, the moduli space is that
of volume preserving diffeomorphisms and is infinite dimensional,
drastically complicating the problem -- both mathematically and
numerically.
A rigorous mathematical formulation of the variational approach to the
problem with volume preserving diffeomorphisms has been studied in
ref.~\cite{Speight:2014fqa} and the case of adding the kinetic term to
the BPS-Skyrme model is dubbed the restricted harmonic problem.
Restricted refers to being in the infinite moduli space of volume
preserving diffeomorphisms and harmonic is the minimizer of the
kinetic term.
In the lack of a better term, we will denote the perturbation by
further terms than the kinetic term as generalized restricted harmonic 
(GRH).

Attacking the problem from a different angle, in
ref.~\cite{Gillard:2015eia} a full numeric attempt of solving the
near-BPS equations of motion has been performed.
In that work, exact near-BPS solutions have been found, pushing the
parameter $\epsilon$ to a small value, around $\epsilon\sim 0.2$.
On the contrary, for smaller values of $\epsilon$ ($\epsilon< 0.2$),
all results are so far numerically inaccessible.
In that range, indeed, the numerical solutions developed unwanted
spike-like singularities.
Such limitation was unfortunate for the study of the binding energy of
the system.
In fact, a proper estimate yields $\epsilon\sim 0.01$ for fitting the
physically small binding energies of nuclei.
Again, that analysis is carried out without a full understanding of
the near-BPS system -- we will comment more on this in the
conclusion.

Since the full numerical approach is extremely difficult, in this work
we propose a semi-analytical method for exploring the near-BPS Skyrme
model, building on the work of
refs.~\cite{Gudnason:2020tps,Gudnason:2021gwc}. 
The strategy is based on the expansion of the near-BPS field around a
BPS solution, however, with respect to
refs.~\cite{Bonenfant:2010ab,Bonenfant:2012kt}, two more steps are
considered. Firstly, we face the problem of the restricted harmonic
map.
Secondly, once that problem is resolved, we explore the system at the
next orders of the expansion.
As we will see, the results from the next-to-next-to-leading order
(N$^2$LO) are necessary for extracting the binding energy of a
multi-Skyrmion configuration.
In light of the previous works in the literature, in order to
implement our analysis, we have anticipated various technical and
methodological difficulties.
Moreover, without the possibility to perform numerical checks, it is
even more difficult to establish when, and under what circumstances, a
given approximation method could fail. To this end, instead of
considering immediately the complicated 3D analysis, we previously
performed our investigation in the 2D near-BPS baby Skyrme model
\cite{Gudnason:2020tps,Gudnason:2021gwc}. For the 2D case, we were
able to implement a new semi-analytical method for the near-BPS
analysis and, simultaneously, check it with full numerics. The
knowledge acquired from those studies serves as a guide in the 3D
case studied in this paper.

In light of our previous works, we have chosen the model in such a way
to possess the best features of both the previous 2D cases.
This means that, for what concerns the BPS sector, we choose a
compacton-type BPS model.
In this way, we can easily guess the restricted harmonic maps we need
for both the cases of a single and multi-Skyrmion configurations.
On the other hand, the BPS-deformation is taken to be the original
Skyrme model with the pion mass potential.
With this choice, a physical pion mass (not depending on $\epsilon$)
is included in the system.
In this work, we apply the successful techniques developed in
refs.~\cite{Gudnason:2020tps,Gudnason:2021gwc}.
In particular, applying the perturbative scheme to the case of a
single near-BPS Skyrmion, we find very good agreement with the exact
full-numerical solution.
This achievement confirms again the accuracy of our method.
Such a result, however, has been obtained only for the topological
sector $B=1$ due to the mathematical difficulties in finding the
restricted harmonic solution for $B>1$.

The BPS property of a theory is often, but not always, related to the
existence of a supersymmetric extension. This is, however, not the
case for the Skyrme model in 3+1 dimensions, because its target space
is given by the coset of the chiral symmetry breaking (for two
flavors of quark), i.e.~$\SU(2)\times\SU(2)/\SU(2)\simeq\SU(2)$, which
is topologically a 3-sphere and not a K\"ahler manifold
\cite{Zumino:1979et}.
The first attempt at finding a supersymmetric extension of the Skyrme
model indeed ended up with a target space that was effectively
compactified to $\mathbb{C}P^1\simeq S^2$ which is K\"ahler
\cite{Bergshoeff:1984wb}.
The supersymmetrized model differs from the Skyrme model in the
bosonic sector by containing extra terms in the Lagrangian, in
particular four time derivatives, which are absent in the Skyrme
model.
The Skyrme model with a 2-sphere for a target space (in 2 dimensions)
is known as the baby Skyrme model
\cite{Bogolubskaya:1989ha,Bogolyubskaya:1989fz,Piette:1994ug}
and was supersymmetrized with $\mathcal{N}=1$ supersymmetries (or two
supercharges) \cite{Adam:2011hj,Bolognesi:2014ova}.
The baby Skyrmions in the BPS limit are BPS states, but turn out to
preserve only a quarter of the supersymmetries (one supercharge)
\cite{Nitta:2014pwa,Nitta:2015uba}.
A successful construction of the supersymmetric
BPS-Skyrme model that eliminates the kinetic term by means of the
auxiliary field solution was found in ref.~\cite{Gudnason:2015ryh}; it
however has an enlarged target space of $\SL(2,\mathbb{C})$ instead of
$\SU(2)$.
This particular supersymmetric Skyrme model contains solitons, but
they are not preserving any supersymmetry \cite{Gudnason:2016iex}.

The paper is organized as follows.
In sec.~\ref{sec:model} we set up the model and notation, find the BPS
solutions and calculate the generic energy bound.
In sec.~\ref{sec:perturb}, we calculate the corrections to the energy
of the Skyrmions in the near-BPS limit within the $\epsilon$-expansion
scheme to leading order (LO), next-to-leading order (NLO) and
next-to-next-to-leading order (N$^2$LO). The latter two orders utilize a
linearized perturbation field.
We finally compute the explicit energy corrections to the $B=1$
spherically symmetric compacton.
In sec.~\ref{sec:binding}, we set up the calculation of the bound
state between two spherically symmetric $B=1$ compactons in the
attractive channel and perform the numerical calculations of the
perturbation fields, yielding the binding energies of the bound state.
In sec.~\ref{sec:physunits}, we convert the physical quantities to
physical units.
Finally, we conclude with a discussion in sec.~\ref{sec:conclusion}.

\section{The model}\label{sec:model}

The model is based on the BPS Skyrme
model \cite{Adam:2010fg,Adam:2010ds} with small non-BPS deformations
with a coefficient $\epsilon\ll1$.
The deformation-part of the Lagrangian is taken, generically, to be
the massive \cite{Battye:2004rw,Battye:2006tb} Skyrme
model \cite{Skyrme:1961vq,Skyrme:1962vh}.
We thus have
\begin{align}
  \Lag &= \Lag_{\rm BPS} + \epsilon\Lag_{\rm deform} + \Lag_\lambda\non
  &= \left(c_6\Lag_6 + \mu^2\Lag_0\right)
  + \epsilon\left(c_2\Lag_2 + c_4\Lag_4 - m_\pi^2V_{1,1}(U)\right)
  + \Lag_\lambda,
  \label{eq:L}
\end{align}
with the kinetic (Dirichlet) term, the Skyrme term, sextic term and
Lagrange multiplier term\footnote{The Lagrange multiplier term vanishes exactly since $U$ is an $\SU(2)$ field, but we include it here so that the vector formulation of the Skyrme model restricts the four-vector $\Phi$ to the 3-sphere, where $U=\Phi^0\mathbf{1}_2+\i\tau^a\Phi^a$, $a=1,2,3$.}
\begin{align}
  \Lag_2 &= \frac14\tr(L_\mu L^\mu), \\
  \Lag_4 &= \frac{1}{32}\tr([L_\mu,L_\nu][L^\mu,L^\nu]),\\
  \Lag_6 &= \frac{1}{144}\eta_{\mu\mu'}\epsilon^{\mu\nu\rho\sigma}\epsilon^{\mu'\nu'\rho'\sigma'}
  \tr(L_\nu L_\rho L_\sigma) \tr(L_{\nu'}L_{\rho'}L_{\sigma'}),\\
  \Lag_\lambda &= \frac{\lambda}{2}(\det U - 1),
\end{align}
where $L_\mu\equiv U^\dag\p_\mu U$ is the left-invariant chiral
current and $U$ is the Skyrme field, related to the pions as
\beq
U = \sigma\mathbf{1}_2 + \i\tau^a\pi^a, \qquad a=1,2,3,
\label{eq:U}
\eeq
and the potential $\Lag_0$, written in the form
\beq
-\Lag_0 = V_{s,p}(U) = \frac{1}{s p}\left(1 - \left(\frac{\tr U}{2}\right)^s\right)^p,
\eeq
that should not contribute to the pion mass,
whereas $V_{1,1}(U)=(1-\frac12\tr U)$ is the standard pion mass term.
The metric convention we use in this paper is of the mostly positive
signature, the spacetime indices run as $\mu,\nu,\rho,\sigma=0,1,2,3$,
$\eta_{\mu\nu}$ is the flat Minkowski metric and we take
$\epsilon^{0123}=1$.
The BPS sector consists of a sixth-order derivative term, which is the
topological current squared, as well as a potential term which we take
not to be the pion mass term.
The deformation sector, on the other hand, consists of the normal
Skyrme model with a pion mass term.

The potentials are consistent with the boundary condition
\beq
\lim_{|x|\to\infty} U = \mathbf{1}_2,
\eeq
which effectively point compactifies 3-space to a 3-sphere:
$\mathbb{R}^3\cup\{\infty\}\simeq S^3$.
The nonlinear sigma model constraint $\det U=1$ makes the target space
$\SU(2)$ which as a manifold is also a 3-sphere; this is imposed in
the model via the Lagrangian multiplier term $\Lag_\lambda$.
A static configuration, $U:S^3\to S^3$, is thus characterized by the
topological charge $B \in \pi_3(S^3)=\mathbb{Z} $, where $B$ is called
the baryon number and can be calculated as
\beq
B = -\frac{1}{24\pi^2}\int\epsilon_{ijk}\tr[L_iL_jL_k]\;\d^3x.
\eeq
Using the parametrization
\beq
U = \cos f\mathbf{1}_2 + \i\tau^a \hat{n}^a\sin f, \qquad
\hat{n} = \frac{1}{1+|u|^2}\left(u+\bar{u},-\i(u-\bar{u}),1-|u|^2\right),
\label{eq:Ansatz_fuub}
\eeq
with $f$ a real function and $u$ a complex function of spacetime, the
Lagrangian components read 
\begin{align}
  \Lag_2 &= -\frac12\p_\mu f\p^\mu f
  -\frac{2\sin^2f}{(1+|u|^2)^2}\p_\mu u\p^\mu\bar{u},\\
  \Lag_4 &= -\frac{2\sin^2f}{(1+|u|^2)^2}
  \left(\p_\mu f\p^\mu f \p_\nu u\p^\nu\bar{u} - \p_\mu f\p_\nu f \p^\mu u\p^\nu \bar{u}\right)\non
  &\phantom{=\ }
  -\frac{2\sin^4f}{(1+|u|^2)^4}
  \left((\p_\mu u\p^\mu\bar{u})^2 - \p_\mu u \p_\nu\bar{u}\p^\mu u\p^\nu\bar{u}\right),\\
  \Lag_6 &= -\frac{4\sin^4f}{(1+|u|^2)^4}
  \eta_{\mu\mu'}\epsilon^{\mu\nu\rho\sigma}\epsilon^{\mu'\nu'\rho'\sigma'}
  \p_\nu f\p_\rho u\p_\sigma\bar{u} \p_{\nu'} f\p_{\rho'}u\p_{\sigma'}\bar{u}.
\end{align}
The static energy reads
\begin{align}
  E &= E_{\rm BPS} + \epsilon E_{\rm deform}\non
  &= \left(c_6 E_6 - \mu^2\int_{\mathbb{R}^3}\Lag_0\;\d^3x\right)
+ \epsilon\left(c_2 E_2 + c_4 E_4 + m_\pi^2\int_{\mathbb{R}^3}V_{1,1}\;\d^3x\right),
\label{eq:Estatic}
\end{align}
with the components
\begin{align}
  E_6 &= \int_{\mathbb{R}^3}\frac{4\sin^4f}{(1+|u|^2)^4}\left(\i\epsilon_{i j k}\p_if\p_j u\p_k\bar{u}\right)^2\;\d^3x,\label{eq:E6}\\
  E_2 &= \int_{\mathbb{R}^3}\bigg[
  \frac12(\p_i f)^2
  +\frac{2\sin^2f}{(1+|u|^2)^2}|\p_i u|^2\bigg]\d^3x,\label{eq:E2}\\
  E_4 &= \int_{\mathbb{R}^3}\bigg[
    \frac{2\sin^2f}{(1+|u|^2)^2}\left((\p_if)^2|\p_ju|^2 - \p_if\p_jf\p_iu\p_j\bar{u}\right)\non&\phantom{=\int_{\mathbb{R}^3}\bigg[\ }
    +\frac{2\sin^4f}{(1+|u|^2)^4}\left(|\p_iu|^4 - (\p_iu\p_j\bar{u})^2\right)
  \bigg]\d^3x.\label{eq:E4}
\end{align}
In this paper, we consider BPS potential leading to compacton-type
solutions of the type 
\beq
-\Lag_0 = V_{s,p}(U) = \frac{1}{s p}\left(1 - \left(\frac{\tr U}{2}\right)^s\right)^p
= \frac{1}{s p}(1-\cos^s f)^p,
\label{eq:Vsp}
\eeq
with $(s,p)=(1,2)$, $(2,1)$ and $(2,2)$ (see the next
subsection).
Obviously, also the pion mass term, given by $(s,p)=(1,1)$, generates
a compacton-type soliton but such a potential is already included
in the BPS perturbation. 

The topological charge in the parametrization \eqref{eq:Ansatz_fuub}
reads 
\beq\label{topc}
B = -\frac{1}{2\pi^2}\int\frac{2\sin^2 f}{(1+|u|^2)^2}\i\epsilon_{i j k}
\p_i f \p_j u \p_k \bar{u}\; \d^3x.
\eeq

\subsection{BPS solution}
Taking the limit $\epsilon = 0$, we can write the static energy as
\begin{align}
  E &= \int_{\mathbb{R}^3}
  \left(\frac{4c_6\sin^4f}{(1+|u|^2)^4}\left(\i\epsilon_{i j k}\p_if\p_j u\p_k\bar{u}\right)^2
  +\mu^2V_{s,p}\right)\d^3x\non
  &= \int_{\mathbb{R}^3}
  \left(\sqrt{c_6}\frac{2\sin^2f}{(1+|u|^2)^2}\i\epsilon_{i j k}\p_if\p_j u\p_k\bar{u}
  +\mu\sqrt{V_{s,p}}\right)^2\d^3x\non
  &\phantom{=\ }
  -4\mu\sqrt{c_6}\int_{\mathbb{R}^3}\frac{\sin^2f}{(1+|u|^2)^2}\i\epsilon_{i j k}\p_if\p_j u\p_k\bar{u}\sqrt{V_{s,p}}
  \;\d^3x,\label{eq:Bogo}
\end{align}
where we have performed a Bogomol'nyi trick in the second equality.
The BPS equation is
\beq
\sqrt{c_6}\frac{2\sin^2f}{(1+|u|^2)^2}\i\epsilon_{i j k}\p_if\p_j u\p_k\bar{u}
= -\mu\sqrt{V_{s,p}},
\label{eq:BPS}
\eeq
and the Bogomol'nyi mass is given by the last line of
eq.~\eqref{eq:Bogo}.

Using the axially symmetric Ansatz for $u$:
\beq
u = \tan\left(\frac{\theta}{2}\right)e^{\i N\phi},
\label{eq:axial_Ansatz}
\eeq
the BPS equation reads
\beq
\sin^2(f) f_r =
-\frac{\mu\sqrt{V}r^2}{\sqrt{c_6}N},
\label{eq:BPS_axial_f}
\eeq
and for the potential $(s,p)=(1,p)$, we have
\beq
\cos^2\left(\frac{f}{2}\right)\sin^{2-p}\left(\frac{f}{2}\right) f_r
=-\frac{2^{\frac{p}{2}-2}\mu r^2}{N\sqrt{p c_6}}.
\eeq
Integrating with respect to $r$ yields
\begin{align}\label{bge}
&\frac{2^{2-p}\tan\big(\frac{f}{4}\big)^{1-p}}{(p-4)(p-2)}
\bigg[
2{}_2F_1\left(\frac{1-p}{2},2-p;\frac{3-p}{2};-\tan^2\left(\frac{f}{4}\right)\right)\non
&+\left(p-4-2(p-2)\cos\left(\frac{f}{2}\right)+(p-2)\cos f\right)
\sec^{2p-4}\left(\frac{f}{4}\right)
\bigg]
=-\frac{2^{\frac{p}{2}-2}\mu r^3}{3N\sqrt{p c_6}} + \kappa,
\end{align}
where ${}_2F_1$ is the standard hypergeometric function and $\kappa$
is an integration constant. 
If the limit $f\to 0$ of the left-hand side of the above equation
diverges, then the soliton has a tail that tends to infinity.
If not, the soliton is of compacton type.
Taylor expanding the left-hand side yields
\beq
-\frac{2^{p-2}f^{3-p}}{p-3} + \mathcal{O}(f^{5-p}),
\eeq
which reveals that the soliton is indeed a compacton for $p<3$.
The constant $\kappa$ of the eq.~\eqref{bge} must be chosen so as make
$f$ respect the boundary conditions
\beq
f(0)=\pi, \qquad f(R)=0,
\label{eq:BCf}
\eeq
that guarantee that the soliton bears a nontrivial topological charge
\eqref{topc}.

For $p=1$, which corresponds to the standard pion mass term, the
left-hand side of (\ref{bge}) is invertible
\beq
\frac{2}{3}\cos^3\left(\frac{f}{2}\right)
=\frac{\mu r^3}{6N\sqrt{2c_6}} - \kappa,
\eeq
yielding the explicit compacton solution
\beq
f = 2\arccos\left(\frac{r}{R}\right),
\label{eq:BPSsol1}
\eeq
where we have defined the compacton radius
\beq
R\equiv\sqrt[3]{\frac{4N\sqrt{2c_6}}{\mu}},
\eeq
and set $\kappa:=0$.

For other values of $p$ (with $s=1$), the potential does not give an
invertible function that enables us to write explicit solutions for
$f$. 
It will be useful, however, to consider the $p=2$ solution, for which the BPS
solution reduces to
\beq
f + \sin f = \pi\left(1 - \frac{r^3}{R^3}\right),
\label{eq:BPSsol2}
\eeq
with the compacton radius defined by
\beq
R\equiv\sqrt[3]{\frac{3\pi\sqrt{2c_6}N}{\mu}},
\label{eq:Rsol2}
\eeq
and we have set $\kappa:=\frac{\pi}{2}$. 
The solution is implicit but still simple.

Considering instead the potential $(s,p)=(2,1)$ in the axially
symmetric Ansatz \eqref{eq:axial_Ansatz}, the BPS equation reads
\beq
\sin(f)f_r =-\frac{\mu r^2}{\sqrt{2c_6}N}.
\eeq
Integrating with respect to $r$ yields
\beq
f = \arccos\left(\frac{2r^3}{R^3}-1\right),
\label{eq:BPSsol02}
\eeq
where the compacton radius now is
\beq
R = \sqrt[3]{\frac{2N\sqrt{2c_6}}{\mu}},
\eeq
where we have set $\kappa:=1$.

Finally, let us consider the potential \eqref{eq:Vsp} with
$(s,p)=(2,2)$ in the axially symmetric Ansatz \eqref{eq:axial_Ansatz},
for which the BPS equation reduces to
\beq
f_r = -\frac{\mu r^2}{2\sqrt{c_6}N}.
\eeq
Integrating with respect to $r$ gives
\beq
f = \pi\left(1 - \frac{r^3}{R^3}\right),
\label{eq:BPSsol22}
\eeq
where the compacton radius is
\beq
R = \sqrt[3]{\frac{6\pi\sqrt{c_6}N}{\mu}},
\eeq
and we have set $\kappa:=\pi$.

\subsection{BPS energy}

The Bogomol'nyi mass is given by the total derivative
\beq
M_{\rm BPS}^{(s,p)} = 
-4\mu\sqrt{c_6}\int_{\mathbb{R}^3}\frac{\sin^2f}{(1+|u|^2)^2}\i\epsilon_{i j k}\p_if\p_j u\p_k\bar{u}\sqrt{V_{s,p}}
  \;\d^3x,\label{eq:BPSmass}
\eeq
which is the lower bound for the static energy
\beq
E \geq M_{\rm BPS}^{(s,p)}.
\eeq
Considering the potential \eqref{eq:Vsp} with $s=1$ and using the
axially symmetric Ansatz \eqref{eq:axial_Ansatz}, the Bogomol'nyi mass
reads
\begin{align}
  M_{\rm BPS}^{(1,p)} &= -2^{\frac{p}{2}+3}\mu N\sqrt{\frac{c_6}{p}}
    4\pi\int_0^R\cos^2\left(\frac{f}{2}\right)\sin^{2+p}\left(\frac{f}{2}\right)
    f_r\;\d r \non
    &= 2^{\frac{p}{2}+4}\mu N\sqrt{\frac{c_6}{p}} \pi^{\frac{3}{2}}\frac{\Gamma\left(\tfrac{3+p}{2}\right)}{\Gamma\left(3+\tfrac{p}{2}\right)},
\end{align}
where we again have used the boundary conditions
\beq
f(0)=\pi, \qquad f(R)=0.
\label{eq:fBC}
\eeq
Notice that the Bogomol'nyi mass is proportional to the degree $N$ as
it must.
For $p=1,2$ we have
\beq
M_{\rm BPS}^{(1,1)} = \frac{128\pi}{15}\sqrt{2c_6} \mu N, \qquad
M_{\rm BPS}^{(1,2)} = 2\pi^2\sqrt{2c_6} \mu N.
\label{eq:MBPS12}
\eeq
Considering instead the potential \eqref{eq:Vsp} with $s=2$, $p=1$ and using the 
axially symmetric Ansatz \eqref{eq:axial_Ansatz}, the Bogomol'nyi mass
reads
\begin{align}
  M_{\rm BPS}^{(2,1)} &= -\mu N\sqrt{2c_6}
    4\pi\int_0^R\sin^3(f) f_r\;\d r \non
    &= \frac{16\pi}{3}\sqrt{2c_6} \mu N,
    \label{eq:MBPS02}
\end{align}
where we have used the boundary conditions \eqref{eq:fBC}.

Finally, let us consider the potential \eqref{eq:Vsp} with
$(s,p)=(2,2)$ with the axially symmetric Ansatz
\eqref{eq:axial_Ansatz}, for which the Bogomol'nyi mass reads
\begin{align}
  M_{\rm BPS}^{(2,2)} &= -\mu\sqrt{c_6}\mu
  4\pi\int_0^R\sin^4(f) f_r\;\d r\non
  &= \frac32\pi^2\sqrt{c_6} \mu N,
  \label{eq:MBPS22}
\end{align}
where we again have used the boundary conditions \eqref{eq:fBC}.

\subsubsection{Energy bound}

In this section, we will review the lower bound on the energy of the
generalized Skyrme model of ref.~\cite{Adam:2013tga} (see also
ref.~\cite{Harland:2013rxa}).
The general idea of the calculation utilizes the fact that the
generalized Skyrme model is the sum of different subsectors, each of
which has a known energy bound.
Then to find the total energy bound, an optimization between the
different bounds should be carried out. 

Given the Lagrangian \eqref{eq:L}, which corresponds to the static
energy \eqref{eq:Estatic}, we can write 
\beq
E=& \mu^2 E_0 + \epsilon c_2 E_2 + \epsilon c_4 E_4 + c_6 E_6,
\label{eq:Energy}
\eeq
where $E_2$, $E_4$, and $E_6$ are defined in eqs.~\eqref{eq:E2},
\eqref{eq:E4} and \eqref{eq:E6}, respectively and where we have defined
\beq
E_0=\int_{\mathbb{R}^3}\widetilde{V}(U)\;\d^3x
=\int_{\mathbb{R}^3}\left[
V_{s,p}(U)
+\frac{\epsilon m^2_{\pi}}{\mu^2}V_{1,1}(U)
\right]\;\d^3x.
\label{eq:E0def}
\eeq
Then, given the energy bounds:
\beq
\beta E_0 + E_6 &\geq& 4\pi^2\beta^{\frac12}\,\langle\widetilde{V}^{\frac12}\rangle\, |B|,\\
\beta E_0 + E_4 &\geq& 4\pi^2(2\beta)^{\frac14}\,\langle\widetilde{V}^{\frac14}\rangle \, |B|,\\
\beta E_2 + E_4 &\geq& 6\pi^2\beta^{\frac12}\, |B|,\\
\beta E_2 + E_6 &\geq& 8\pi^2\big(\tfrac{\beta}{2}\big)^{\frac34}\, |B|\,\,,
\eeq
in which we have defined $\langle\cdots\rangle$ as the target-space
average of a generic quantity $X$ as
\beq
\langle X\rangle \equiv -\frac{1}{24\pi^2B}\int_{\mathbb{R}^3}X\epsilon_{i j k}\tr[L_i L_j L_k]\;\d^3x,
\eeq
it is possible to rewrite the energy \eqref{eq:Energy} as a sum of the
above subsectors introducing four parameters $\alpha_{2i}$ with 
$0\leq\alpha_{2i}\leq 1$ for $i=0,\ldots,3$, that determine how each term
is split between the given bounds.
The general bound for the total energy as a function of $\{\alpha_i\}$ is thus:
\begin{align}
  E=&\ \big(\alpha_0 \mu^2 E_0 + \alpha_6 c_6 E_6\big)
    +\big( (1-\alpha_0) \mu^2 E_0+\alpha_4 \epsilon c_4 E_4\big)\non
    &+\big( \alpha_2 \epsilon c_2 E_2 + (1-\alpha_4) \epsilon c_4 E_4\big)
    +\big( (1-\alpha_2) \epsilon c_2 E_2 + (1-\alpha_6)c_6 E_6\big)\non
    \geq&\;2\pi^2\Big[\,
      2\mu(\alpha_0\alpha_6 c_6)^{\frac12}\langle\widetilde{V}^{\frac12}\rangle
      +2\sqrt{\mu}(\alpha_4\epsilon c_4)^{\frac34}\big(2(1-\alpha_0)\big)^{\frac14}\langle\widetilde{V}^{\frac14}\rangle\non
      &\ \
      +3\epsilon\big((1-\alpha_4)c_4\alpha_2c_2\big)^{\frac12}
      +4\big((1-\alpha_6)c_6\big)^{\frac{1}{4}}\big(\tfrac12(1-\alpha_2)\epsilon c_2\big)^{\frac{3}{4}}\,\Big]|B|.
\label{eq:bound}
\end{align}
Once we have chosen the potential $\widetilde{V}(U)$ and the
parameters $c_0$, $c_2$, $c_4$ and $c_6$, the strongest energy bound
for the system is the maximum of the functional \eqref{eq:bound},
which is a maximization problem in four variables ($\alpha_i$).
It is difficult to write down an analytic solution to the solution of
the maximization problem, but it is rather easy to find numerically.

As a consistency check, we can set $\mu=c_6=\alpha_4=0$,
$\alpha_2=\epsilon=1$ and $c_2=c_4=2$,
for which we obtain $E\geq12\pi^2|B|$, which is the standard
topological energy bound in Skyrme units
\cite{Skyrme:1961vq,Manton:2004tk}.
Notice that when $m_\pi>0$, the integral $\langle\widetilde{V}\rangle$
depends on $\epsilon$, see eq.~\eqref{eq:E0def}.

\section{Perturbation in \texorpdfstring{$\epsilon$}{epsilon}}\label{sec:perturb}

In this section we consider the $\epsilon$-expansion scheme around the
BPS solution. This technique has been developed and tested in the
previous works on the baby-Skyrme model
\cite{Gudnason:2020tps,Gudnason:2021gwc}, where comparison was made
with full brute-force numerical computations, hence establishing the
accurateness of the framework. Here we adapt it to a more
complex model; the idea remains the same but various modifications
have to be made.  The solution at zeroth order is just the BPS
solution. The leading-order correction to the mass is given by the
perturbation minimized and evaluated on the BPS solution. This can be
done if a certain finiteness condition applies and for the present
paper we restrict to this case. The minimization problem of the
perturbation is the so-called generalized restricted harmonic
problem. We briefly review the conditions and the solutions that are
known so far, essentially the $B=1$ Skyrmion and non-overlapping
multi-compactons solutions. We then discuss an approximate criterion,
using as a test the axially symmetric solutions, to test if other
preferred restricted harmonic solutions could exist. Once we have
selected the candidate model for which the $B =1+1+\dots$ is the most
probable restricted harmonic solution, we consider the expansion to
higher order in $\epsilon$ and the modifications to the leading-order
solution.

\subsection{Zeroth order}

We will now consider perturbing the BPS sector (i.e.~the model
\eqref{eq:L} with $\epsilon=0$) with a small perturbation,
$0<\epsilon\ll1$.
The deformation adds interactions among the Skyrmions leading to
bound states with low binding energy (as long as $\epsilon\ll1$).
In this so-called near-BPS limit, the field $U$ can be written as
\begin{equation}\label{expu}
  U(x)=U_0(x)+\dU(\epsilon,x)\qquad \text{with}\qquad
  \dU(0,x)=0,
\end{equation}
where $U_0(x)$ is a BPS solution and $\dU$ is a perturbation that
depends on $\epsilon$, but not necessarily in an analytic way.
Expanding the static energy in $\epsilon$, the zeroth order is simply
given by the BPS mass
\begin{equation}
    E^{(0)}=M_{\rm BPS}^{(s,p)},
\end{equation}
where $M_{\rm BPS}^{(s,p)}$ is given by eq.~\eqref{eq:BPSmass}.

\subsection{Leading-order correction}

The leading-order (LO) correction to the energy, is linear in
$\epsilon$ and is given by the perturbation part of the Lagrangian,
$\Lag_{\rm deform}$, evaluated on the background BPS solution
\begin{equation}\label{eq:LO}
  \epsilon M_{\rm LO}=-\epsilon \int_{\mathbb{R}^3}\Lag_{\rm deform}(U^{*}_0)\;\d^3x,
\end{equation}
where $U^{*}_0$ is the BPS configuration that minimizes the above
integral.
This solution is required to be \emph{generalized-restricted harmonic}
(GRH), using the definition of
refs.~\cite{Speight:2014fqa,Gudnason:2021gwc}.

Before analyzing this point, it is necessary to check the finiteness
of the LO energy contribution. For instance, the compacton solution
\eqref{eq:BPSsol1} gives a divergent contribution to the LO energy,
since the integral \eqref{eq:LO} diverges if $c_2>0$.
In this paper, in order to consistently implement the perturbative
method, we consider only the cases (the potentials) that lead to
finite contributions at every order in the expansion.
In the following, we provide a criterion for the LO finiteness, valid
for BPS solutions of compacton-type.

\subsubsection{Finiteness of LO energy}

Using as a test the axially symmetric Ansatz \eqref{eq:axial_Ansatz}, the LO
energy of a compact solution with radius $R$ reads
\begin{align}
  \epsilon M_{\rm LO}(N) &=
  -\epsilon c_2\int_{\mathbb{R}^3}\Lag_2\;\d^3x
  -\epsilon c_4\int_{\mathbb{R}^3}\Lag_4\;\d^3x
   +\epsilon m_\pi^2 \int_{\mathbb{R}^3} V_{1,1}\;\d^3x\non
  &= 2\pi\epsilon c_2\int_0^R\left(r^2f_r^2 + (1+N^2)\sin^2f\right)\;\d r\non
  &\phantom{=\ }
  +2\pi\epsilon c_4\int_0^R\left(
  (1+N^2)\sin^2(f)f_r^2 + \frac{N^2}{r^2}\sin^4f
  \right)\;\d r,\non
   &\phantom{=\ }
  +4\pi\epsilon m_\pi^2\int_0^R r^2(1-\cos f)\;\d r.
  \label{eq:E_LO_gen}
\end{align}
For several classes of BPS solutions, we find that the divergence is
due to the term $r^2f_r^2$ that tends to infinity at the border of the
compacton (in the limit $r\to R$). We therefore reduce the finite LO
energy condition to requesting that 
\begin{equation}\label{eq:crt}
  \int_{R-\delta}^R r^2 f_r^2\;\d r<\infty,\qquad\text{with}\qquad \delta\ll R.
\end{equation}
Note that in the Lagrangian $\Lag_4$ the quantity $f_r^2$ is
multiplied by $\sin^2f$, which alleviates the divergence of the
integral since $f\to 0$ for $r\to R$ (see eq.~\eqref{eq:BCf}).
Using the BPS equation \eqref{eq:BPS_axial_f} and the boundary condition
\eqref{eq:BCf}, we can manipulate the condition \eqref{eq:crt} as 
\begin{equation}\label{eq:manip}
  \int_{R-\delta}^R r^2 f_r^2\;\d r
  =\int_{\delta'}^0 r^2(f) \frac{\d f}{\d r}\;\d f
  =\frac{\mu}{|N|c_6} \int_{0}^{\delta'} r^4(f)\sqrt{\frac{V_{s,p}}{\sin^4f}}\;\d f
  <\infty,
\end{equation}
with $0<f(R-\delta)=\delta'\ll 1$.
Now, it is useful to expand the function $r(f)$ around $f=0$ as
\begin{equation}
  r(f)\simeq R
  +\left.\frac{\d r(f)}{\d f}\right\rvert_{0}f
  +\frac12\left. \frac{\d^2 r(f)}{\d f^2}\right\rvert_{0}f^2
  +\mathcal{O}(f^3).
\end{equation}
Every positive power of $f$ in this expansion improves the convergence
of the integral and thus if we have
\begin{equation}
  \frac{\mu R^4}{|N|c_6}\int_{0}^{\delta'}\sqrt{\frac{V_{s,p}}{\sin^4 f}}\d f<\infty
\end{equation}
then the condition \eqref{eq:manip} consequently holds true.
Considering a class of potentials $V_{s,p}$ of the type
\eqref{eq:Vsp}, for which $V_{s,p}\simeq f^{2p}$ near $f=0$, then we
have
\begin{equation}
  \frac{\mu R^4}{|N|c_6}\int_{0}^{\delta'}\sqrt{\frac{V_{s,p}}{\sin^4 f}}\;\d f
  \sim\frac{\mu R^4 s^{p/2}}{2^{p/2}|N|c_6\sqrt{sp} } \int_{0}^{\delta'} f^{p-2}\;\d f
  =\frac{\mu R^4s^{p/2}}{2^{p/2}|N|c_6\sqrt{sp} }\left[\frac{f^{p-1}}{p-1}\right]_0^{\delta'}<\infty.
\end{equation}
In order for this condition to hold true, we deduce that $p$ must be
greater than $1$, i.e. $p>1$.

Once this criterion is defined, it is an easy check to verify that
only the combinations $(s,p)=(1,2)$ and $(s,p)=(2,2)$ of the potential
\eqref{eq:Vsp} lead to a finite LO energy.
Hence, we will discard the choice $(s,p)=(2,1)$ that, analogously to
the pion-mass potential $(s,p)=(1,1)$, generates a divergent LO energy
for $c_2>0$.

\subsubsection{Generalized-restricted harmonic}

The generalized-restricted harmonic (GRH) solution $U^{\star}_0$ in
eq.~\eqref{eq:LO} represents the BPS configuration that extremizes the
LO energy within the whole (BPS) moduli space. In particular, to
implement the perturbative expansion of the field, we need $U^{\star}_0$
to be a minimum (at least locally) of the LO energy.
In ref.~\cite{Speight:2014fqa}, a criterion for the choice of these GRH
solutions is discussed for a perturbation of the type
$\Lag_2$ and $\Lag_2+\Lag_4$.
The validity of that criterion is not spoiled by the presence of the
pion-mass potential $V_{1,1}$, due to the volume-preserving
diffeomorphism invariance of the potential energy. In this section, we
analyze the GRH problem following different steps. Firstly, we use the
theorem developed in ref.~\cite{Speight:2014fqa} to identify the BPS
configuration that extremizes (and minimizes) the perturbation energy
due to $\Lag_2$. Then, using again the results of
ref.~\cite{Speight:2014fqa}, we check if such a configuration is a minimum
even for the combination $\Lag_2+\Lag_4$.
In the end, we comment on the trivial role of the potential $V_{1,1}$
in this context.

We briefly review the criterion of ref.~\cite{Speight:2014fqa}.
Given a smooth map $\phi$ from the manifold $\mathcal{M}$ to the
manifold $\mathcal{N}$, the Dirichlet energy is generally defined as
\begin{equation}\label{e2}
E_2 = \frac12\int_{\cal M}h_{ab}\,g^{ij}\,\p_i\phi^a\p_j\phi^b\sqrt{\det g_{kl}}\;\d^dx,
\end{equation}
where
$g=g_{ij}\,\d x^i\otimes\d x^j$ and $h=h_{ab}\,\d\phi^a\otimes\d\phi^b$
are the metrics of the manifold $\mathcal{M}$ and
$\mathcal{N}$, respectively, and $d$ is the number of (spatial)
dimensions (ignoring time here).
Using the map $\phi$, the pull-back $\phi^*h$ of the metric $h$ to
$\mathcal{M}$ is defined as 
\begin{equation}
\phi^*h = h_{ab}\,\frac{\p\phi^a}{\p x^i}\frac{\p\phi^b}{\p x^j}\,\d x^i\otimes\d x^j.
\end{equation}
Among all the maps $\phi$ with finite Dirichlet energy, connected by
volume-preserving diffeomorphisms, a map $\tilde{\phi}$ is restricted
harmonic if and only if the one-form 
\beq
\div\,\tilde{\phi}^*h   \quad {\rm on} \ 
\mathcal{M} \  {\rm is \  exact}.
\eeq
The divergence $\div$ of a symmetric $(0,2)$ tensor
$\eta=\eta_{ij}\,\d x^i\otimes\d x^j$ on $\mathcal{M}$ acts as
\begin{equation}
  \div\,\eta=D^i\eta_{ij} \d x^j
  =g^{ik}\big(\p_k\eta_{ij}-\Gamma^l_{ki}\eta_{lj}-\Gamma^l_{kj}\eta_{il}\big) \,\d x^j,
\end{equation}
where the connection is
$\Gamma^i_{jk}=\frac{1}{2}g^{il}\left(\p_k g_{lj}+ \p_j g_{lk} - \p_l g_{jk} \right)$.

In order to more easily use this criterion, we rewrite the Dirichlet
energy for the Skyrmions in the form \eqref{e2}.
To this end, the field $U\in\SU(2)$ can be decomposed in terms of four
scalar fields $\sigma$, $\pi^1$, $\pi^2$ and $\pi^3$ as given in
eq.~\eqref{eq:U}.
This relation allows us to define an $\Og(4)$ vector field $\Phi^a$ as
\begin{equation}
  \Phi^a=(\sigma,\pi^1,\pi^2,\pi^3) \qquad \text{with}\qquad
  \Phi^a\Phi^a=1,\qquad a=0,1,2,3.
  \label{eq:Phi}
\end{equation}
With this notation, the Dirichlet energy $E_2$ reads
\begin{equation}
  E_2=-\int_{\cal M}\Lag_2\;\d^3x
  =-\frac14\int_{\cal M}\Tr[L_i L_i]\;\d^3x
  =\frac12\int_{\cal M}h_{ab}\,g^{ij}\, \p_i\Phi^a\p_j\Phi^a\sqrt{\det g_{kl}}\;\d^3x,
\end{equation}where $h_{ab}=\delta_{ab}$ and $g^{ij}=\delta^{ij}$.

We now apply this theorem to the case of spherically symmetric
Skyrmions. In the following, we use spherical coordinates
$(r,\theta,\varphi)$ on $\mathcal{M}=\mathbb{R}^3$ and the vector 
notation $\Phi^a$ for the Skyrme field with the constraint
$\Phi^a\Phi^a=1$.
Then, we rewrite the spherically symmetric Ansatz
\eqref{eq:axial_Ansatz} for a generic $B=N$ compacton in the form
\begin{equation}
  \label{N1s}
  \Phi^a=
  \begin{pmatrix}
    \cos f(r)\\
    \sin f(r)\sin (\theta)\cos (N\varphi)\\
    \sin f(r)\sin (\theta)\sin (N\varphi) \\
    \sin f(r)\cos (\theta)
  \end{pmatrix},
\end{equation}
where the function $f$ depends only on the radial coordinate, $r$.
The metric $h$ is the standard Euclidean metric and the pull-back
$\Phi^*h$ of $h$ can be written as 
\begin{equation}\label{eq:pul1} 
\Phi^*h=\frac{\p\Phi^a}{\p \tilde{x}^i}\frac{\p\Phi^a}{\p\tilde{x}^j}\,\d\tilde{x}^i\otimes\d\tilde{x}^j
=f'{}^2\d r^2+\sin^2(f) \big(\d{\theta}^2+N^2\sin^2({\theta})\,\d\varphi^2\big),
\end{equation}
with $\d\tilde{x}^i=(\d r,\d{\theta},\d\varphi)$.

Taking the divergence of the tensor \eqref{eq:pul1}, we obtain the one-form
\begin{equation}\label{eq.1fom}
\begin{aligned}
  \div\,\Phi^*h&=\,D^i\,(\p_i\Phi^a\p_j\Phi^a)\,\d \tilde{x}^j\\
  &=\left(2f'f''+\frac{2}{r}f'^2-(1+N^2)\frac{\sin^2 f}{r^3}\right)\d r
  +(1-N^2)\frac{\sin^2 f}{r^2}\cot{\theta}\,\d\theta. 
\end{aligned}
\end{equation}
According to Poincar\'e's lemma, if the one-form \eqref{eq.1fom} is
closed then it is exact. 
Therefore, the solution \eqref{N1s} is restricted harmonic if
$\d(\div\,\Phi^*h)=0$, where $\d$ is the exterior derivative.
Explicitly,
\begin{equation}
\begin{aligned}
  \d(\div\,\Phi^*h)&=\frac12(\p_i\omega_j-\p_j\omega_i)\,\d x^i\wedge\d x^j\\
  &=(1-N^2)\frac{\d}{\d r}\left(\frac{\sin^2f}{r^2}\right)\cot{\theta}\,\d r\wedge\d\theta.
  \label{exp1}
\end{aligned}
\end{equation}
The ratio $\frac{\sin^2f}{r^2}$ cannot be a constant since that would be
incompatible with the boundary conditions \eqref{eq:BCf}. The only
possibility for eq.~\eqref{exp1} to vanish is therefore $N=\pm1$. We
conclude that a spherically symmetric compacton with arbitrary
orientation and topological charge $N=\pm1$ is a restricted harmonic
map. The same proof can be trivially extended to the case of a
composition of $B=1+1+1+\cdots$ spherically symmetric compactons placed
in $\mathbb{R}^3$ without overlapping one another.

Despite several attempts, we have not been able to analytically find any
restricted-harmonic maps different from the spherical $B=N=1$ BPS
configuration. 
Therefore, in the aim of correctly implementing the
perturbative method, in this paper, we will use only that background
solution and a multiple non-overlapping composition of it.

The use of the spherical $N=1$ compacton has an important convenience
due to a relevant result obtained in ref.~\cite{Speight:2014fqa}.
In particular, it has been proved that every hedgehog field is both
$\Lag_2$-restricted harmonic and restricted $\Lag_4$-critical, and
thus restricted $(\Lag_2+\Lag_4)$-critical.
This result allows us to say that the spherically symmetric $N=1$ BPS
solution is a stationary point of the LO energy $E_2+E_4$.
Moreover, again in ref.~\cite{Speight:2014fqa}, such a configuration
has been verified to be \textit{stable} restricted
$(\Lag_2+\Lag_4)$-critical, as we need for our purpose.
Once the spherical compacton $B=N=1$ is identified as a GRH solution,
we can easily verify that the presence of the pion-mass potential energy 
\begin{equation}\label{e033}
E_{\pi}=m_{\pi}^2\int\big( 1-\cos{f}\big)\sqrt{\det g_{ij}}\;\d^3x,
\end{equation}
 does not influence that result.
The LO energy
$E_{\pi}$ is diff-invariant and thus it does not
play any role in the choice of the GRH map.

To summarize, both the single spherical $B=1$ BPS compacton and the
composition of non-overlapping $B=1+1+1+\cdots$ spherical compactons
correctly respect the generalized restricted harmonicity criterion and
represent local minima of the LO energy.
However, being able to consider only this possibility, we will have no
indication about the stability or the meta-stability of such a
solution within each topological sector.
Due to this fact, in the aim of building stable nuclei, we want to
focus our analysis only on those near-BPS systems that lead to
energetically preferred configurations made by $B=1+1+1+\cdots$
Skyrmions.
That information can be extracted from the evaluation of $N_{\star}$ that
is, in the same way of refs.~\cite{Gudnason:2020tps,Gudnason:2021gwc},
the charge of the spherical configuration that minimizes the energy
per nucleon ($E/N$).
The value of $N_{\star}$ is specific for every type of near-BPS system and
thus will help us to choose a suitable potential.

In the next section, we will prove that, if $N_{\star}>1$, we certainly
know that a GRH map of charge $N>1$ minimizes the energy per nucleon
better than the spherical $N=1$ solution.
Therefore, considering such near-BPS model, a nucleus made of
$B=1+1+1+\cdots$ Skyrmions can be at best meta-stable.
On the contrary, if $N_{\star}\sim 1$ it is possible to have a stable
$B=1+1+1+\cdots$ nucleus.

\subsubsection{Explicit LO corrections}

The leading-order-in-$\epsilon$ correction to the energy comes from
plugging the BPS solution into the energy functional
\begin{align}
  \epsilon M_{\rm LO}^{(s,p)}(N) &=
  -\epsilon c_2\int_{\mathbb{R}^3}\Lag_2\;\d^3x
  -\epsilon c_4\int_{\mathbb{R}^3}\Lag_4\;\d^3x
  +\epsilon m_\pi^2 \int_{\mathbb{R}^3} V_{1,1}\;\d^3x\non
  &= 2\pi\epsilon c_2\int_0^R\left(r^2f_r^2 + (1+N^2)\sin^2f\right)\;\d r\non
  &\phantom{=\ }
  +2\pi\epsilon c_4\int_0^R\left(
  (1+N^2)\sin^2(f)f_r^2 + \frac{N^2}{r^2}\sin^4f
  \right)\;\d r\non
  &\phantom{=\ }
  +4\pi\epsilon m_\pi^2\int_0^R r^2(1-\cos f)\;\d r,
\end{align}
where we have used the axially symmetric Ansatz
\eqref{eq:axial_Ansatz}.

Besides the LO energy, analogously to the method of
refs.~\cite{Gudnason:2020tps,Gudnason:2021gwc}, we calculate the value
of $N_{\star}$ in this section, i.e.~the charge of the configuration that
minimizes the energy per nucleon ($E/N$).
To find such configuration, given the energy of a $B=N$ spherically
symmetric BPS solution $U_{0}^{\rm sph}$ 
\begin{equation}\label{nste}
    E(\epsilon, N)=E_{\rm BPS}(N)+\epsilon M_{\rm LO}(N, U_{0}^{\rm sph}),
\end{equation}
we must solve
\begin{equation}
  \frac{\d}{\d N}\left(\frac{E(\epsilon, N)}{N} \right)
  = \frac{\d}{\d N}\left(\frac{M_{\rm LO}(N, U_{0}^{\rm sph})}{N} \right)=0,
\end{equation}
and find $N_{\star}$ by solving for $N$.
Note that, in this calculation, the dependence of $\epsilon$ vanishes.

Before dealing with the explicit calculation of $N_{\star}$, we must point
out an important difference about the meaning of $N_{\star}$ between the
2D cases in refs.~\cite{Gudnason:2020tps,Gudnason:2021gwc} and here.
In refs.~\cite{Gudnason:2020tps,Gudnason:2021gwc}, once the near-BPS
baby Skyrme model is chosen, the value of $N_{\star}$ identifies which
$Q=N$ solution represents the most stable candidate to be the building
block of a nucleus (at least at the leading-order approximation).
In the baby Skyrme model case, all the axially symmetric solutions of
any topological charge are restricted-harmonic and thus, in the
calculation of $N_{\star}$, the LO energies of the different topological
sectors are correctly compared.
Here, on the contrary, the situation is different.
In fact, as shown in the previous section, only the $N=1$ spherically
symmetric compacton is (generalized) restricted harmonic.
Thus, for any $N>1$, the expression \eqref{nste} evaluated on a
spherical BPS compacton does not represent the correct LO energy of a
$B=N$ near-BPS Skyrmion.

From the above considerations, the calculation of $N_{\star}$ seems
meaningless in the 3D case.
The reason for carrying out this calculation is that, with such a
result, we can indirectly prove if an unknown GRH configuration (of
charge $N>1$) minimizes $E/N$ better than $N=1$.
We will verify this statement in the following.

Let us consider to have found a spherical BPS solution $U_{0}^{\rm sph}$
of charge $B=\widetilde{N}$, whose value of the ratio
$E/\widetilde{N}$ is smaller than the one calculated for the GRH solution
$U_{0}^{\rm sph}$ of charge $B=1$, i.e.,
\begin{equation}\label{enst}
  \frac{M_{\rm LO}(B=\widetilde{N}, U_{0}^{\rm sph})}{\tilde{N}}
  <\frac{M_{\rm LO}(B=1, U_{0}^{\rm sph})}{1}.
\end{equation}Then, we have
\begin{equation}\label{crNstar}
  \frac{M_{\rm LO}(B=\widetilde{N}, U_0^{\star})}{\widetilde{N}}
  <\frac{M_{\rm LO}(B=\widetilde{N}, U_0^{\rm sph})}{\widetilde{N}}
  <\frac{M_{\rm LO}(B=1, U_0^{\rm sph})}{1},
\end{equation}
where $U_0^{\star}$ is the unknown GRH solution of topological charge $B=\widetilde{N}$.
In eq.~\eqref{crNstar}, we used the fact that a GRH map minimizes the
LO energy better than any other BPS maps.

Finding a result of the type \eqref{enst} (that is equivalent of
finding $N_{\star}>1$), means that surely a GRH solution of charge $B>1$,
more energetically favored than the spherical $B=N=1$, exists.
As a consequence, in that case a near-BPS solution made of
$B=1+1+1+\cdots$ Skyrmions would be at best meta-stable. 

On the other hand, if we obtain $N_{\star}\sim 1$, we cannot
definitively prove that the configuration $B=1+1+1+\cdots$ is the one energetically
favored, but surely we avoid the previous counter argument.
Therefore, in the following we will select the proper constraints to
have $N_\star\sim 1$.

We will now calculate the LO energy and the value of $N_{\star}$
for the different near-BPS systems built with the BPS potential
$V_{s,p}$ and $(s,p)=(1,1),(1,2),(2,1),(2,2)$.

For this calculation, it is convenient to have an explicit BPS
solution, so we will first consider the case of the potential
\eqref{eq:Vsp} with $s=1$ and $p=1$, for which we have the Bogomol'nyi
mass \eqref{eq:MBPS12} and BPS solution \eqref{eq:BPSsol1}.
This potential is the pion mass and hence is not a potential that we
eventually would want to use, since we want the pion mass to be in
the deformation sector.
For this exercise, we set $m_\pi:=0$, since $V_{1,1}$ is included
instead in the BPS sector.
In particular, we get
\beq
f_r^2 = \frac{4}{R^2-r^2}, \qquad
\sin^2f = 4\left(1-\frac{r^2}{R^2}\right)\frac{r^2}{R^2},
\eeq
which means that the leading-order energy does not converge if
$c_2>0$ is turned on (due to the singularity in the integral
over $(rf_r)^2$.
Setting $c_2:=0$, we obtain
\beq
\epsilon M_{\rm LO}^{(1,1)} =
2\pi\epsilon c_4\left[(1+N^2)\frac{16}{3R} + N^2\frac{128}{105R}\right],
\eeq
where
\beq
R = \sqrt[3]{N}\tilde{R}, \qquad
\tilde{R} = \sqrt[3]{\frac{4\sqrt{2c_6}}{\mu}}.
\eeq
The leading-order mass per $N$ has a minimum at
\beq
N_\star^{(1,1)} = \sqrt{\frac{70}{43}} \simeq 1.276,
\eeq
and indeed the $N=1$ leading-order energy correction per $N$ is
smaller than that of the $N=2$.

Considering instead the potential \eqref{eq:Vsp} with $s=2$ and $p=1$,
for which we have the Bogomol'nyi mass \eqref{eq:MBPS02} and BPS solution
\eqref{eq:BPSsol02}, we have
\beq
f_r^2 = \frac{9r}{R^3-r^3}, \qquad
\sin^2f = 4\left(1-\frac{r^3}{R^3}\right)\frac{r^3}{R^3},
\eeq
which again means that the leading-order energy does not converge if 
$c_2>0$ is turned on (due to the singularity in the integral
over $(rf_r)^2$).
We have again set $m_\pi:=0$ since the BPS solution is massive (as $p=1$).
Setting $c_2:=0$, we obtain
\beq
\epsilon M_{\rm LO}^{(2,1)} =
2\pi\epsilon c_4\left[(1+N^2)\frac{36}{5R} + N^2\frac{36}{55R}\right],
\eeq
where
\beq
R = \sqrt[3]{N}\tilde{R}, \qquad
\tilde{R} = \sqrt[3]{\frac{2\sqrt{2c_6}}{\mu}}.
\eeq
The leading-order mass per $N$ has a minimum at
\beq
N_\star^{(2,1)} = \sqrt{\frac{11}{6}} \simeq 1.354,
\eeq
and indeed the $N=1$ leading-order energy correction per $N$ is
smaller than that of the $N=2$.

Although the two solutions we have considered now, conveniently have
explicit BPS solutions in terms of $f$, they both yield infinite
leading-order corrections to the kinetic term
(i.e.~$-\int\Lag_2\,\d^3x$).
Moreover, they have a contribution to the pion mass from the
BPS sector, which we want to avoid as we want the pion mass to scale
with $\epsilon$ in the near-BPS limit.
We will therefore consider the case of the potential \eqref{eq:Vsp}
with $(s,p)=(1,2)$, for which the BPS solution is given by
eq.~\eqref{eq:BPSsol2} and the Bogomol'nyi mass by \eqref{eq:MBPS12}.
Since the BPS solution \eqref{eq:BPSsol2} is not explicit, we have to
rewrite the integrals for the leading-order correction to the energy
as
\begin{align}
  \epsilon M_{\rm LO}^{(1,2)}(N) &=
  2\pi\epsilon c_2\int_\pi^0 \left(r^2\frac{\p f}{\p r}
  +(1+N^2)\sin^2f\frac{\p r}{\p f}\right)\d f\non
  &\phantom{=\ }
  +2\pi\epsilon c_4\int_\pi^0\left((1+N^2)\sin^2f\frac{\p f}{\p r}
  +\frac{N^2}{r^2}\sin^4(f)\frac{\p r}{\p f}\right)\d f\non
  &\phantom{=\ }
  +\frac{4\pi\epsilon m_\pi^2}{3}\int_\pi^0(1-\cos f)\frac{\p r^3}{\p f}\d f.
\end{align}
Using now that
\beq
r^3 = N\tilde{R}^3\left(1 - \frac{f+\sin f}{\pi}\right),
\eeq
we have
\begin{align}
  \frac{\p r}{\p f} &= -\frac{\tilde{R}N^{1/3}}{3\pi}
  \frac{1+\cos f}{\left(1 - \frac{f+\sin f}{\pi}\right)^{\frac23}},
  \non
  r^2\frac{\p f}{\p r} &= - 3\pi N^{1/3}\tilde{R}
  \frac{\left(1 - \frac{f+\sin f}{\pi}\right)^{\frac43}}{1+\cos f},
  \non
  \frac{\p r^3}{\p f} &= -\frac{N\tilde{R}^3}{\pi}(1+\cos f),
\end{align}
and can write the leading-order correction to the energy divided by
$2\pi\epsilon$ as 
\begin{align}
  \frac{M_{\rm LO}^{(1,2)}(N)}{2\pi} &=
  3\pi c_2\tilde{R} N^{1/3} a_1
  +\frac{c_2\tilde{R}(1+N^2)N^{1/3}}{3\pi} a_2
  +\frac{3\pi c_4(1+N^2)}{\tilde{R} N^{1/3}} a_3
  +\frac{c_4N^{5/3}}{3\pi\tilde{R}} a_4\non
  &\phantom{=\ }
  +\frac{m_\pi^2 N\tilde{R}^3}{3},
\end{align}
where we have defined the integrals
\begin{align}
  a_1&\equiv \int_0^\pi\frac{\left(1 - \frac{f+\sin f}{\pi}\right)^{\frac43}}{1+\cos f}\;\d f
  \simeq 0.4699,\non
  a_2&\equiv \int_0^\pi\frac{\sin^2f(1+\cos f)}{\left(1 - \frac{f+\sin f}{\pi}\right)^{\frac23}}\;\d f
  \simeq 4.824,\non
  a_3&\equiv \int_0^\pi\frac{\sin^2f\left(1 - \frac{f+\sin f}{\pi}\right)^{\frac23}}{1+\cos f}\;\d f
  \simeq 0.5167,\non
  a_4&\equiv \int_0^\pi\frac{\sin^4f(1+\cos f)}{\left(1 - \frac{f+\sin f}{\pi}\right)^{\frac43}}\;\d f
  \simeq 16.327,
\end{align}
and
\beq
R=\sqrt[3]{N}\tilde{R},\qquad
\tilde{R} = \sqrt[3]{\frac{3\pi\sqrt{2c_6}}{\mu}}.
\eeq
Notice that the pion mass term is linearly proportional to $N$ and
does not affect $N_\star$.
Setting $c_4:=0$, we find the minimum of the leading-order
correction per $N$ as 
\beq
N_\star^{(1,2)} = \sqrt{\frac12+\frac{9\pi^2a_1}{2a_2}}
\simeq 2.197.
\eeq
Explicit checks find that
$M_{\rm LO}(2)/2<M_{\rm LO}(3)/3<M_{\rm LO}(4)/4<M_{\rm LO}(1)<M_{\rm LO}(5)/5$.
We also explicitly find that $M_{\rm LO}(3)<2M_{\rm LO}(2)+M_{\rm LO}(1)$. Setting instead $c_2:=0$, we find the minimum of the leading-order
correction per $N$ as 
\beq
N_\star^{(1,2)} = \sqrt{\frac{2}{1 + \frac{a_4}{9\pi^2a_3}}}
\simeq 1.215.
\eeq
Explicit checks find that $M_{\rm LO}(1)<M_{\rm LO}(N)/N$, for any $N>1$.
In general, $N_\star^{(1,2)}$ is a function of the ratio
$c_4/(c_2\tilde{R}^2)$ and the equation for $N_\star$ reads
\beq
-\frac{3\pi a_1}{N^{4/3}}
+\frac{a_2(2N^2-1)}{3\pi N^{4/3}}
+3\pi a_3 x\left(1 - \frac{2}{N^2}\right)
+\frac{x a_4}{3\pi} = 0,\qquad
x:=\frac{c_4}{c_2\tilde{R}^2}.
\eeq
\begin{figure}[!htp]
  \begin{center}
    \includegraphics[width=0.49\linewidth]{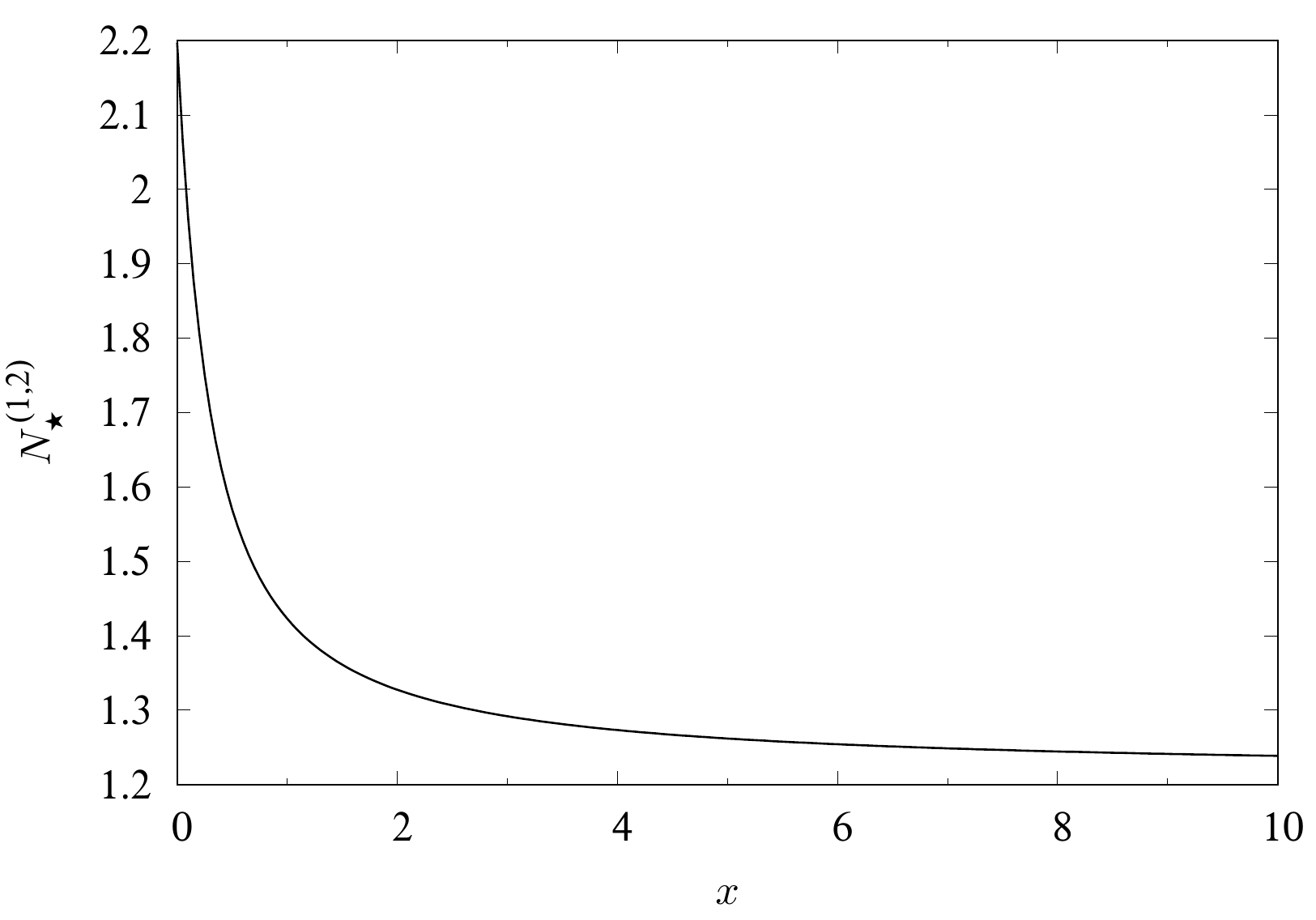}
    \includegraphics[width=0.49\linewidth]{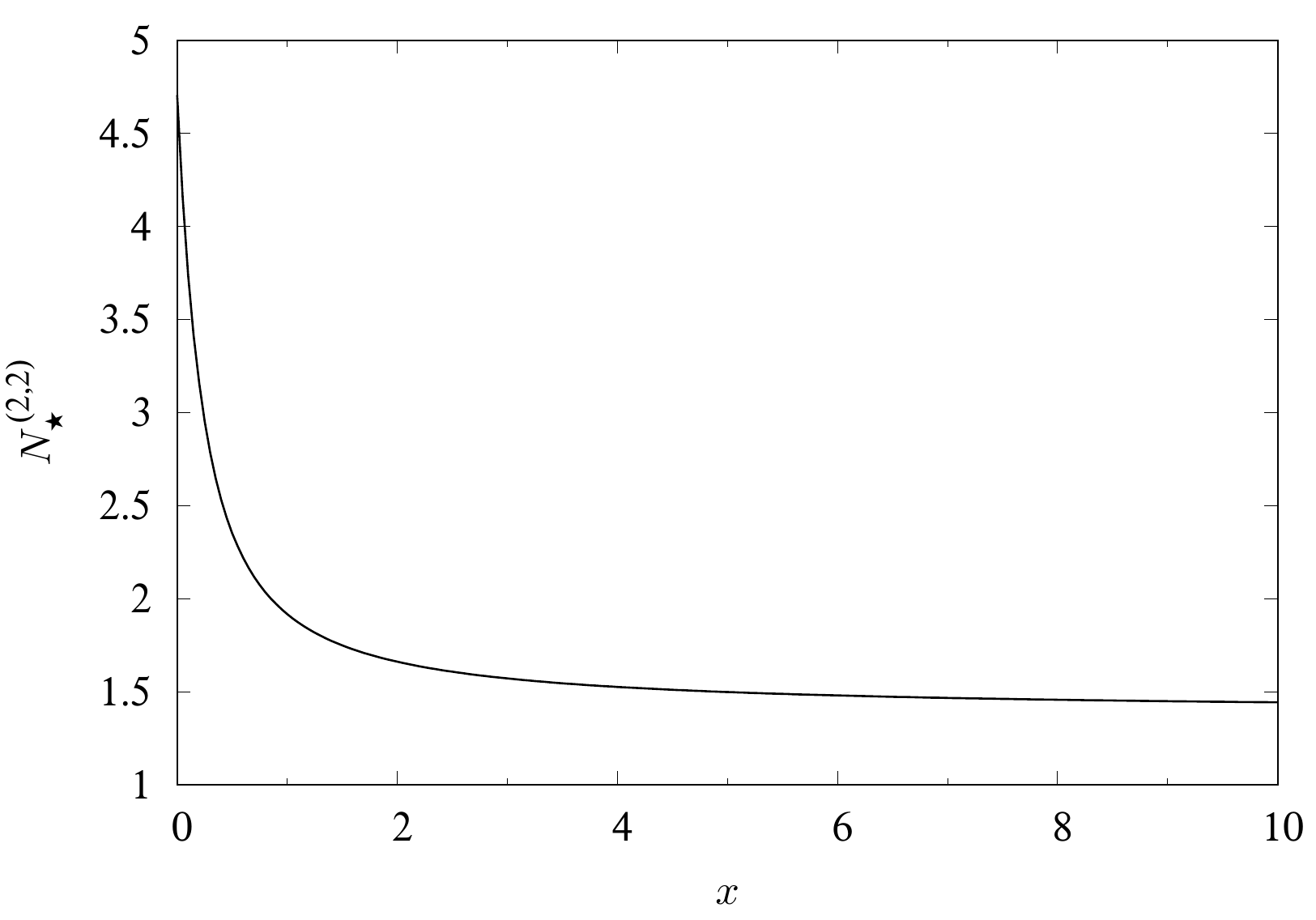}
    \caption{$N_\star^{(1,2)}$ and $N_\star^{(2,2)}$ as functions of
      $x=\frac{c_4}{c_2\tilde{R}^2}$.}
    \label{fig:Nstar}
  \end{center}
\end{figure}
The solution to the equation, namely $N_\star^{(1,2)}$, is shown in
fig.~\ref{fig:Nstar} as a function of $x$. 

Finally, we will consider the case of the potential \eqref{eq:Vsp} with
$(s,p)=(2,2)$, for which we have the Bogomol'nyi mass
\eqref{eq:MBPS22} and BPS solution \eqref{eq:BPSsol22}, and we further
have
\beq
f_r^2 = 9\pi^2\frac{r^4}{R^6}, \qquad
\sin^2f = \sin^2\left(\frac{\pi r^3}{R^3}\right),
\eeq
so now the LO energy is convergent and we can write the LO energy
divided by $2\pi\epsilon$ as
\begin{align}
\frac{M_{\rm LO}^{(2,2)}}{2\pi}
&= \frac{9\pi^2c_2\tilde{R}N^{1/3}}{7}
+ c_2(1+N^2)\Upsilon_1\tilde{R}N^{1/3}
+ c_4(1+N^2)\frac{9\pi^2\Upsilon_2}{\tilde{R}N^{1/3}}
+ \frac{c_4 N^{5/3}\Upsilon_3}{\tilde{R}}\non
&\phantom{=\ }
+ \frac{2 m_\pi^2 N\tilde{R}^3}{3},
\end{align}
where
\begin{align}
  R &= \sqrt[3]{N}\tilde{R}, \qquad
  \tilde{R} = \sqrt[3]{\frac{6\pi\sqrt{c_6}}{\mu}},\\
  \Upsilon_1 &= \int_0^1\sin^2(\pi x^3)\;\d x \simeq 0.29303,\\
  \Upsilon_2 &= \int_0^1\sin^2(\pi x^3)x^4\;\d x \simeq 0.103303,\\
  \Upsilon_3 &= \int_0^1\frac{\sin^4(\pi x^3)}{x^2}\;\d x \simeq 0.370701.
\end{align}
Notice again that the contribution from the pion mass term is linear
in $N$ and hence will not affect $N_\star$.
If we set $c_4:=0$, the LO mass per $N$ has a minimum at
\beq
N_\star^{(2,2)} = \sqrt{\frac12 + \frac{9\pi^2}{14\Upsilon_1}} \simeq 4.707.
\eeq
Explicit checks find that $M_{\rm LO}(5)/5<M_{\rm LO}(4)/4<M_{\rm LO}(6)/6<M_{\rm LO}(7)/7<M_{\rm LO}(3)/3<M_{\rm LO}(8)/8<M_{\rm LO}(9)/9<M_{\rm LO}(2)/2<M_{\rm LO}(10)/10<M_{\rm LO}(11)/11<M_{\rm LO}(12)/12<M_{\rm LO}(13)/13<M_{\rm LO}(14)/14<M_{\rm LO}(15)/15<M_{\rm LO}(1)<M_{\rm LO}(16)/16$.
This means that there are many generalized-restricted harmonic
solutions that have less energy per baryon number than the spherically 
symmetric 1-Skyrmions.
This is quite surprising.

On the other hand, if we set $c_2:=0$, the LO mass per $N$ has a
minimum given by
\beq
N_\star^{(2,2)} = \sqrt{\frac{2}{1+\frac{\Upsilon_3}{9\pi^2\Upsilon_2}}} \simeq 1.386,
\eeq
and explicit checks verify that
$\frac{M_{\rm LO}(N)}{N}<\frac{M_{\rm LO}(N+1)}{N+1}$
for all $N=1,2,\ldots$
In general, $N_\star^{(2,2)}$ is a function of the ratio
$c_4/(c_2\tilde{R}^2)$ and the equation for $N_\star$ reads
\beq
-\frac{3\pi}{7}
+\frac{\Upsilon_1(2N^2-1)}{3\pi}
+\frac{3\pi x\Upsilon_2(N^2-2)}{N^{2/3}}
+\frac{N^{4/3} x\Upsilon_3}{3\pi} = 0, \qquad
x:=\frac{c_4}{c_2\tilde{R}^2}.
\eeq
The solution to this equation, i.e.~$N_\star^{(2,2)}$, is shown in
fig.~\ref{fig:Nstar} as a function of $x$. 

To summarize, we have found that for the potentials \eqref{eq:Vsp}
with $(s,p)=(2,1)$ and $(s,p)=(2,2)$, setting $c_4:=0$, we have
solutions with lower energy per $N$ to leading order in $\epsilon$ for
$N>1$, which means that the spherically symmetric 1-Skyrmion is at
best metastable in the near-BPS limit (for those potentials).
On the other hand, for $c_2:=0$ and $(s,p)=(2,1),(2,2),(1,1),(1,2)$,
we have found that the 1-Skyrmion is the energetically preferred
solution.

In the situation with both $c_2>0$ and $c_4>0$, for the cases
$(s,p)=(1,2)$ and $(s,p)=(2,2)$, the value of $N_{\star}$ rapidly
reaches $\sim 1$ for $c_4\gg c_2R^2$ (see fig.~\ref{fig:Nstar}). As a
consequence, in the following of the paper, we impose the
constraint $c_4\gg c_2R^2$ in order to build stable nuclei made of
$B=1+1+1+\cdots$ Skyrmions.
Note that the addition of $m_\pi>0$ does not change $N_\star$.

\subsection{NLO and \texorpdfstring{N$^2$LO}{N2LO} corrections}

We now consider the next-to-leading order (NLO) and
next-to-next-to-leading order (N$^2$LO) corrections to the
energy, meaning that we have to take into account the corrections of 
order $\mathcal{O}(\epsilon^2)$ and $\mathcal{O}(\epsilon^3)$.
The reason for having to consider the perturbed Lagrangian up to
$\mathcal{O}(\epsilon^3)$ has been discussed in
ref.~\cite{Gudnason:2020tps} for the 2-dimensional case and it is
strictly related with the choice of the compacton-type solution as
the background field. Indeed, since the compacton field is constant
outside its finite domain, all the terms that contain derivatives of
the background field vanish outside said region.
As a result, the first order of the field expansion vanishes outside
the compacton domain and then, iterating the perturbative scheme, all
orders of the expansion vanish too.
In other words, using the compacton solution as the zeroth order of
the field expansion, the ordinary perturbation scheme fails.
To avoid this problem, in ref.~\cite{Gudnason:2020tps}, both the
quadratic order and the third order in $\epsilon$ have been considered
together.
In this way, at the price of harder analytical computations, the
solution of the perturbed field exists even outside the compacton
region and the perturbative method works.
We adopt here the same strategy for the 3-dimensional case.

For the perturbative $\epsilon$-expansion, we will again utilize the
$\Og(4)$ vector field $\Phi=(\Phi_0,\Phi_1,\Phi_2,\Phi_3)$ that is
related to the $\SU(2)$ matrix, $U$ as defined in eqs.~\eqref{eq:U}
and \eqref{eq:Phi}.
In this way, the computations are similar to the 2-dimensional case
(in which the field is parameterized by an $\Og(3)$ vector field).
We will thus perform the perturbative expansion directly in the $\Phi$
field
\beq
\Phi=\varPhi+\delta\Phi,
\eeq
where $\varPhi$ denotes here the BPS background solution and $\dPhi$
is a small perturbation.
In this notation, the Lagrangian \eqref{eq:L} with $\Lag_0$ given by
eq.~\eqref{eq:Vsp} now reads \cite{Gudnason:2015nxa,Gudnason:2017opo}
\begin{align}
  \Lag_6&= \frac{1}{36}\eta_{\mu\mu'}\epsilon^{\mu\nu\rho\sigma}\epsilon^{abcd}\,\Phi^a\p_{\nu} \Phi^b\p_{\rho}\Phi^c\p_{\sigma}\Phi^d\,\epsilon^{\mu'\nu'\rho'\sigma'}\epsilon_{efgh}\Phi^e\p_{\nu'} \Phi^f\p_{\rho'}\Phi^g\p_{\sigma'}\Phi^h \\
  &=-\frac13(\p_\mu\Phi\cdot\p^\nu\Phi)(\p_\nu\Phi\cdot\p^\rho\Phi)(\p_\rho\Phi\cdot\p^\mu\Phi)
  +\frac12(\p_\mu\Phi\cdot\p^\nu\Phi)(\p_\nu\Phi\cdot\p^\mu\Phi)(\p_\rho\Phi\cdot\p^\rho\Phi)\non
  &\phantom{=\ }
  -\frac16(\p_\mu\Phi\cdot\p^\mu\Phi)^3,\\
  \Lag_0 &= -V_{s,p} = -\frac{1}{sp}(1-(\Phi^an^a)^s)^p,\\
  \Lag_2 &=-\frac12(\p_{\mu}\Phi\cdot\p^{\mu}\Phi)\\
  \Lag_4 &=\frac14(\p_{\mu}\Phi\cdot\p^{\nu}\Phi)(\p_{\nu}\Phi\cdot\p^{\mu}\Phi)
  -\frac14(\p_{\mu}\Phi\cdot\p^{\mu}\Phi)^2,\\
  -V_{1,1} &= -(1-\Phi^a n^a),\\
  \Lag_\lambda &=\frac{\lambda}{2}(\Phi\cdot\Phi-1),
\end{align}
where $n^a=\delta^{a0}$ is the vacuum of the theory and we use the
convention $\epsilon^{0123}=1$.

For the NLO and N$^2$LO corrections, we need to calculate the
variation up to third order (in the fields) of the Lagrangian
\eqref{eq:L} (assuming that $\dPhi=\mathcal{O}(\epsilon)$) 
\begin{align}
\Lag^{\rm perturb}[\varPhi,\dPhi] &=\left.\frac{\p\Lag}{\p\lambda\p\Phi^a}\right|\dlambda\,\dPhi^a+\frac{1}{2}\left.\frac{\p\Lag}{\p\lambda\p\Phi^a\p\Phi^b}\right|\dlambda\,\dPhi^a\,\dPhi^b+
\left.\frac{\p\Lag}{\p\Phi^a}\right|\dPhi^a
\non
&\phantom{=\ }
+\left.\frac12\frac{\p^2\Lag}{\p\Phi^a\p\Phi^b}\right|\dPhi^a\dPhi^b
+\left.\frac16\frac{\p^3\Lag}{\p\Phi^a\p\Phi^b\p\Phi^c}\right|\dPhi^a\dPhi^b\dPhi^c
+\left.\frac{\p\Lag}{\p\p_\mu\Phi^a}\right|\p_\mu\dPhi^a 
\non
&\phantom{=\ }
+\left.\frac12\frac{\p^2\Lag}{\p\p_\mu\Phi^a\p\p_\nu\Phi^b}\right|\p_\mu\dPhi^a\p_\nu\dPhi^b
\non
&\phantom{=\ }
+\left.\frac16\frac{\p^3\Lag}{\p\p_\mu\Phi^a\p\p_\nu\Phi^b\p\p_\rho\Phi^c}\right|\p_\mu\dPhi^a\p_\nu\dPhi^b\p_\rho\dPhi^c\non
&= \dlambda\left(\frac12\dPhi\cdot\dPhi + \varPhi\cdot\dPhi\right)
+ \mu^2\Xi
+ \epsilon m_\pi^2\dPhi^0
+ \frac{\lambda_0}{2}\dPhi\cdot\dPhi
-\epsilon J_a^\mu\p_\mu\dPhi^a\non
&\phantom{=\ }
-\frac12V_{a b}^{\mu\nu}\p_\mu\dPhi^a\p_\nu\dPhi^b
-\frac16\Gamma_{a b c}^{\mu\nu\rho}\p_\mu\dPhi^a\p_\nu\dPhi^b\p_\rho\dPhi^c,
\end{align}
where the symbol $|$ means that the expression to the left is
evaluated on the background field $\varPhi$,
and we have defined the symbols
\begingroup
\allowdisplaybreaks
\begin{align}
  \Xi =&\ -\frac{s}{2}(p-1)\big(1-(\varPhi^0)^s\big)^{p-2}(\varPhi^0)^{2s-2}(\dPhi^0)^2
  +\frac12(s-1)\big(1-(\varPhi^0)^s\big)^{p-1}(\varPhi^0)^{s-2}(\dPhi^0)^2\non
  &+\frac16s^2(p-1)(p-2)\big(1-(\varPhi^0)^s\big)^{p-3}(\varPhi^0)^{3s-3}(\dPhi^0)^3\non
  &\qquad-\frac12s(s-1)(p-1)\big(1-(\varPhi^0)^s\big)^{p-2}(\varPhi^0)^{2s-3}(\dPhi^0)^3\non
  &\qquad+\frac16(s-1)(s-2)\big(1-(\varPhi^0)^s\big)^{p-1}(\varPhi^0)^{s-3}(\dPhi^0)^3,\\
  J_a^\mu =&\ c_2\p^\mu\varPhi^a
  -c_4(\p^\mu\varPhi\cdot\p^\nu\varPhi)\p_\nu\varPhi^a
  +c_4(\p_\nu\varPhi\cdot\p^\nu\varPhi)\p^\mu\varPhi^a,\\
  V_{a b}^{\mu\nu} =&\ V_{0a b}^{\mu\nu} + \epsilon V_{1a b}^{\mu\nu},\\
  V_{0a b}^{\mu\nu} =&\
  2c_6(\p_\rho\varPhi\cdot\p_\sigma\varPhi)\p^\rho\varPhi^a\p^\sigma\varPhi^b\eta^{\mu\nu}
  +2c_6(\p^\nu\varPhi\cdot\p_\rho\varPhi)\p^\rho\varPhi^a\p^\mu\varPhi^b
  +2c_6(\p^\mu\varPhi\cdot\p^\nu\varPhi)\p_\rho\varPhi^a\p^\rho\varPhi^b\non&
  +2c_6(\p^\mu\varPhi\cdot\p_\rho\varPhi)\p^\nu\varPhi^a\p^\rho\varPhi^b
  +2c_6(\p^\mu\varPhi\cdot\p_\rho\varPhi)(\p^\nu\varPhi\cdot\p^\rho\varPhi)\delta^{ab}\non&
  -2c_6(\p_\rho\varPhi\cdot\p^\rho\varPhi)\p_\sigma\varPhi^a\p^\sigma\varPhi^b\eta^{\mu\nu}
  -2c_6(\p_\rho\varPhi\cdot\p^\rho\varPhi)\p^\nu\varPhi^a\p^\mu\varPhi^b\non&
  -2c_6(\p^\mu\varPhi\cdot\p^\nu\varPhi)(\p_\rho\varPhi\cdot\p^\rho\varPhi)\delta^{ab}
  -4c_6(\p^\mu\varPhi\cdot\p_\rho\varPhi)\p^\rho\varPhi^a\p^\nu\varPhi^b\non&
  -4c_6(\p^\nu\varPhi\cdot\p_\rho\varPhi)\p^\mu\varPhi^a\p^\rho\varPhi^b
  -c_6(\p_\rho\varPhi\cdot\p_\sigma\varPhi)(\p^\rho\varPhi\cdot\p^\sigma\varPhi)\eta^{\mu\nu}\delta^{ab}\non&
  +4c_6(\p_\rho\varPhi\cdot\p^\rho\varPhi)\p^\mu\varPhi^a\p^\nu\varPhi^b
  +c_6(\p_\rho\varPhi\cdot\p^\rho\varPhi)(\p_\sigma\varPhi\cdot\p^\sigma\varPhi)\eta^{\mu\nu}\delta^{ab}
  ,\\
  V_{1a b}^{\mu\nu} =&\ c_2\eta^{\mu\nu}\delta^{a b}
  +c_4(\p_\rho\varPhi\cdot\p^\rho\varPhi)\eta^{\mu\nu}\delta^{a b}
  +2c_4\p^\mu\varPhi^a\p^\nu\varPhi^b
  -c_4(\p^\mu\varPhi\cdot\p^\nu\varPhi)\delta^{a b}
  -c_4\p^\mu\varPhi^b\p^\nu\varPhi^a\non&
  -c_4\p_\rho\varPhi^a\p^\rho\varPhi^b\eta^{\mu\nu},\\
  \Gamma_{a b c}^{\mu\nu\rho} =&\ 
  8c_6\p^\mu\varPhi^a\p^\nu\varPhi^b\p^\rho\varPhi^c
  +2c_6\p^\rho\varPhi^a\p^\mu\varPhi^b\p^\nu\varPhi^c
  +2c_6\p^\nu\varPhi^a\p^\rho\varPhi^b\p^\mu\varPhi^c
  -4c_6\p^\nu\varPhi^a\p^\mu\varPhi^b\p^\rho\varPhi^c\non&
  -4c_6\p^\rho\varPhi^a\p^\nu\varPhi^b\p^\mu\varPhi^c
  -4c_6\p^\mu\varPhi^a\p^\rho\varPhi^b\p^\nu\varPhi^c\non&
  +c_6\eta^{\mu\nu}\big(2\p^\rho\varPhi^a\p_\sigma\varPhi^b\p^\sigma\varPhi^c
  +2\p_\sigma\varPhi^a\p^\rho\varPhi^b\p^\sigma\varPhi^c
  -4\p_\sigma\varPhi^a\p^\sigma\varPhi^b\p^\rho\varPhi^c\big)\non&
  +c_6\eta^{\mu\rho}\big(2\p^\nu\varPhi^a\p_\sigma\varPhi^b\p^\sigma\varPhi^c
  +2\p_\sigma\varPhi^a\p^\sigma\varPhi^b\p^\nu\varPhi^c
  -4\p_\sigma\varPhi^a\p^\nu\varPhi^b\p^\sigma\varPhi^c\big)\non&
  +c_6\eta^{\nu\rho}\big(2\p_\sigma\varPhi^a\p^\mu\varPhi^b\p^\sigma\varPhi^c
  +2\p_\sigma\varPhi^a\p^\sigma\varPhi^b\p^\mu\varPhi^c
  -4\p^\mu\varPhi^a\p_\sigma\varPhi^b\p^\sigma\varPhi^c\big)\non&
  +c_6\delta^{ab}\big(2(\p^\nu\varPhi\cdot\p^\rho\varPhi)\p^\mu\varPhi^c
  +2(\p^\mu\varPhi\cdot\p^\rho\varPhi)\p^\nu\varPhi^c
  -4(\p^\mu\varPhi\cdot\p^\nu\varPhi)\p^\rho\varPhi^c\big)\non&
  +c_6\delta^{ac}\big(2(\p^\mu\varPhi\cdot\p^\nu\varPhi)\p^\rho\varPhi^b
  +2(\p^\nu\varPhi\cdot\p^\rho\varPhi)\p^\mu\varPhi^b
  -4(\p^\mu\varPhi\cdot\p^\rho\varPhi)\p^\nu\varPhi^b\big)\non&
  +c_6\delta^{bc}\big(2(\p^\mu\varPhi\cdot\p^\nu\varPhi)\p^\rho\varPhi^a
  +2(\p^\mu\varPhi\cdot\p^\rho\varPhi)\p^\nu\varPhi^a
  -4(\p^\nu\varPhi\cdot\p^\rho\varPhi)\p^\mu\varPhi^a\big)\non&
  +4c_6\delta^{ab}\eta^{\mu\nu}\big((\p_\sigma\varPhi\cdot\p^\sigma\varPhi)\p^\rho\varPhi^c
  -(\p^\rho\varPhi\cdot\p_\sigma\varPhi)\p^\sigma\varPhi^c\big)\non&
  +2c_6\delta^{ab}\eta^{\mu\rho}\big((\p^\nu\varPhi\cdot\p_\sigma\varPhi)\p^\sigma\varPhi^c
  -(\p_\sigma\varPhi\cdot\p^\sigma\varPhi)\p^\nu\varPhi^c\big)\non&
  +2c_6\delta^{ab}\eta^{\nu\rho}\big((\p^\mu\varPhi\cdot\p_\sigma\varPhi)\p^\sigma\varPhi^c
  -(\p_\sigma\varPhi\cdot\p^\sigma\varPhi)\p^\mu\varPhi^c\big)\non&
  +2c_6\delta^{ac}\eta^{\mu\nu}\big((\p^\rho\varPhi\cdot\p_\sigma\varPhi)\p^\sigma\varPhi^b
  -(\p_\sigma\varPhi\cdot\p^\sigma\varPhi)\p^\rho\varPhi^b\big)\non&
  +4c_6\delta^{ac}\eta^{\mu\rho}\big((\p_\sigma\varPhi\cdot\p^\sigma\varPhi)\p^\nu\varPhi^b
  -(\p^\nu\varPhi\cdot\p_\sigma\varPhi)\p^\sigma\varPhi^b\big)\non&
  +2c_6\delta^{ac}\eta^{\nu\rho}\big((\p^\mu\varPhi\cdot\p_\sigma\varPhi)\p^\sigma\varPhi^b
  -(\p_\sigma\varPhi\cdot\p^\sigma\varPhi)\p^\mu\varPhi^b\big)\non&
  +2c_6\delta^{bc}\eta^{\mu\nu}\big((\p^\rho\varPhi\cdot\p_\sigma\varPhi)\p^\sigma\varPhi^a
  -(\p_\sigma\varPhi\cdot\p^\sigma\varPhi)\p^\rho\varPhi^a\big)\non&
  +2c_6\delta^{bc}\eta^{\mu\rho}\big((\p^\nu\varPhi\cdot\p_\sigma\varPhi)\p^\sigma\varPhi^a
  -(\p_\sigma\varPhi\cdot\p^\sigma\varPhi)\p^\nu\varPhi^a\big)\non&
  +4c_6\delta^{bc}\eta^{\nu\rho}\big((\p_\sigma\varPhi\cdot\p^\sigma\varPhi)\p^\mu\varPhi^a
  -(\p^\mu\varPhi\cdot\p_\sigma\varPhi)\p^\sigma\varPhi^a\big).
\end{align}
\endgroup
In the perturbed Lagrangian we have consistently expanded the Lagrange
multiplier $\lambda$ as $\lambda\to\lambda_0+\dlambda$, where
$\lambda_0$ is the Lagrange multiplier solution of the background BPS
model
\begin{align}
  \lambda_0 =&\ 
  -2c_6(\p^\nu\varPhi\cdot\p^\rho\varPhi)(\p_\rho\varPhi\cdot\p^\mu\varPhi)(\p_\mu\p_\nu\varPhi\cdot\varPhi)
  +c_6(\p_\nu\varPhi\cdot\p^\rho\varPhi)(\p_\rho\varPhi\cdot\p^\nu\varPhi)(\p^2\varPhi\cdot\varPhi)\non&
  +2c_6(\p_\rho\varPhi\cdot\p^\rho\varPhi)(\p^\mu\varPhi\cdot\p^\nu\varPhi)(\p_\mu\p_\nu\varPhi\cdot\varPhi)
  -c_6(\p_\mu\varPhi\cdot\p^\mu\varPhi)^2(\p^2\varPhi\cdot\varPhi)\non&
  -\mu^2\big(1-(\varPhi^0)^s\big)^{p-1}(\varPhi^0)^s.
\end{align}
The role of $\dlambda$ is to ensure that the norm of the unit four-vector $\Phi$ does not change up to the accuracy of the perturbation order. Indeed, the equation of motion for the perturbed Lagrangian with respect to $\dlambda$ gives 
\beq\label{eq:lambda_cons}
\left(\frac12\dPhi^2 + \varPhi\cdot\dPhi\right) = 0.
\eeq
In order to solve the above constraint equation, it will prove
convenient to use differential forms with $\dPhi=\dPhi^a\d y^a$ a 1-form
on a 4-dimensional space in which the target space is embedded, and a
natural Ansatz is to take $\dPhi=*(\Delta\wedge\Phi)+\omega$ with
$\omega$ a 1-form (to be determined), since the first term is
transverse to $\Phi$ by construction.
$\Delta$ is a 2-form, which will parametrize the tangent directions to
the target space, as we will see later.
Computing the terms in eq.~\eqref{eq:lambda_cons}, we have
\begin{align}
\frac12(\dPhi,\dPhi) &= \frac12|\Delta\wedge\Phi|^2 + \frac12|\omega|^2,\label{eq:dPhi_sq}\\
(\Phi,\dPhi) &= (\Phi,\omega),
\end{align}
with $(~,~)$ the inner product.
The cross terms in eq.~\eqref{eq:dPhi_sq} vanish due to antisymmetry:
\beq
\frac12\int *(\Delta\wedge\Phi)\wedge *\omega + \frac12\int\omega\wedge\Delta\wedge\Phi
=-\frac12\int \omega\wedge\Delta\wedge\Phi + \frac12\int\omega\wedge\Delta\wedge\Phi
=0,
\eeq
since $*^2=-1$ for a 1-form in 4 dimensions.
Writing out eq.~\eqref{eq:lambda_cons}, we have
\begin{align}
&\frac12\int \Delta\wedge\Phi\wedge *(\Delta\wedge\Phi)
+\int\Phi\wedge *\omega + \frac12\int \omega\wedge *\omega\non
&=\int\Phi\wedge\left(\frac12\Delta\wedge *(\Delta\wedge\Phi)
+*\omega\right) + \frac12\int \omega\wedge *\omega = 0.
\end{align}
Setting the parenthesis to zero yields the 1-form solution
\beq
\dPhi = *(\Delta\wedge\Phi) + \frac12*\Delta\wedge *(\Delta\wedge\Phi).
\eeq
which is consistent, because $|\omega|^2$ is of order $\Delta^4$
and hence $\epsilon^4$.
Writing out the components of $\dPhi$, we get
\beq\label{eq:ansatz}
\dPhi^a=\epsilon^{abcd}\Delta_{bc}\varPhi^d
+\frac{1}{2}\epsilon^{abcd}\Delta_{bc}\epsilon^{defg}\Delta_{ef}\varPhi^g.
\eeq
The solution can also be viewed as due to the standard Gram-Schmidt
orthonormalization algorithm to second order.
The norm of the vector field $\Phi$ is therefore  
\begin{equation}
  \Phi^a\Phi^a=1+\mathcal{O}(\epsilon^4),
\end{equation}
as we request.
Moreover, by using the Ansatz \eqref{eq:ansatz}, it is clear that the
contribution to the energy of the terms multiplied by $\dlambda$ will
be of order $\mathcal{O}(\epsilon^5)$ and then we can neglect them in
the final calculation of the total energy.
The job of $\dlambda$ was indeed just to cast the form of the
perturbation as found in eq.~\eqref{eq:ansatz}.

\subsection{Axially symmetric perturbations}\label{sec:axial_perturb}

It will prove convenient to define the following basis vectors
\begin{align}
\Phi_r &=
\begin{pmatrix}
  -\sin f(r)\\
  \cos f(r)\sin(\theta)\cos(N\varphi)\\
  \cos f(r)\sin(\theta)\sin(N\varphi) \\
  \cos f(r)\cos(\theta)
\end{pmatrix},\quad
\Phi_\theta =
\begin{pmatrix}
  0\\
  \cos(\theta)\cos(N\varphi)\\
  \cos(\theta)\sin(N\varphi) \\
  -\sin(\theta)
\end{pmatrix},\quad
\Phi_\varphi =
\begin{pmatrix}
  0\\
  -\sin(N\varphi)\\
  \cos(N\varphi) \\
  0
\end{pmatrix},
\end{align}
in terms of the background field solution with axial symmetry
\beq
\varPhi =
\begin{pmatrix}
  \cos f(r)\\
  \sin f(r)\sin(\theta)\cos(N\varphi)\\
  \sin f(r)\sin(\theta)\sin(N\varphi)\\
  \sin f(r)\cos(\theta)
\end{pmatrix},
\eeq
for which the perturbation tensor for axially symmetric perturbations
are pointed in the direction of the tensor product of the $\theta$ and
$\varphi$ directions, hence we have
\beq
\Delta_{a b}^r = -\Phi_a^\theta\Phi_b^\varphi \,\df(r),
\eeq
and therefore the perturbation field for axially symmetric
perturbations reads
\beq
\dPhi = 
\begin{pmatrix}
  -\sin f - \frac12\cos f\df\\
  \sin\theta\cos(N\varphi)\left(\cos f - \frac12\sin f\df\right)\\
  \sin\theta\sin(N\varphi)\left(\cos f - \frac12\sin f\df\right)\\
  \cos\theta\left(\cos f - \frac12\sin f\df\right)
\end{pmatrix}\df.
\eeq
Writing out the total field, we have
\begin{align}
  \Phi &= \varPhi+\dPhi\non
  &=
  \begin{pmatrix}
    \cos f\\
    \sin f\sin(\theta)\cos(N\varphi)\\
    \sin f\sin(\theta)\sin(N\varphi)\\
    \sin f\cos(\theta)
  \end{pmatrix}
  +
  \begin{pmatrix}
    -\sin f\\
    \cos f\sin\theta\cos(N\varphi)\\
    \cos f\sin\theta\sin(N\varphi)\\
    \cos f\cos\theta
  \end{pmatrix}\df
  -
  \begin{pmatrix}
    \cos f\\
    \sin f\sin\theta\cos(N\varphi)\\
    \sin f\sin\theta\sin(N\varphi)\\
    \sin f\cos\theta
  \end{pmatrix}\frac{\df^2}{2}\non
  &\simeq
  \begin{pmatrix}
    \cos(f + \df)\\
    \sin(f + \df)\sin(\theta)\cos(N\varphi)\\
    \sin(f + \df)\sin(\theta)\sin(N\varphi)\\
    \sin(f + \df)\cos(\theta)
  \end{pmatrix}
  + \mathcal{O}(\df^3).
\end{align}
It is hence clear that the perturbation preserves the length of the
field $\Phi$, as any change in the function $f$ does not change the
length of the vector field $\Phi$.

Restricting to a radial perturbation in the profile function,
$\df=\df(r)$, we can write the perturbation energy as
\beq
\calE^{\rm perturb}(f,\df) =
\calE_2^{\rm perturb}(f,\df)
+\calE_{3,\rm quad}^{\rm perturb}(f,\df)
+\calE_{3,\rm cubic}^{\rm perturb}(f,\df),
\eeq
with
\begin{align}
  \calE_2^{\rm perturb}(f,\df) =&\ 
  \frac{\mu^2}{2}\cos^{s-2}f\big(1-\cos^sf\big)^{p-2}
  \big(1 - s + (s p - 1)\cos^sf\big)\sin^2(f)\df^2\non&
  +\frac{\mu^2}{2}\cos^sf\big(1-\cos^sf\big)^{p-1}\df^2
  -\frac{2c_6 N^2}{r^4}\sin^2f(4\sin^2f-3)f_r^2\df^2\non&
  +\epsilon m_\pi^2\sin(f)\df
  +\epsilon c_2 f_r \df_r
  +\frac{\epsilon c_2(1+N^2)}{2r^2}\sin(2f)\df\non&
  +\frac{\epsilon c_4(1+N^2)}{2r^2}\sin(2f)f_r^2\df
  +\frac{\epsilon c_4 N^2}{r^4}\sin^2f\sin(2f)\df\non&
  +\frac{\epsilon c_4(1+N^2)}{r^2}\sin^2(f)f_r\df_r
  +\frac{c_6 N^2}{r^4}\sin^4(f)\df_r^2\non&
  +\frac{4c_6 N^2}{r^4}\sin^2f\sin(2f)f_r\df\df_r,
\label{eq:E2_axial}
\end{align}
for the NLO terms,
\begin{align}
  \calE_{3,\rm quad}^{\rm perturb}(f,\df) =&\
  \frac{\epsilon m_\pi^2}{2}\cos(f)\df^2
  +\frac{\epsilon c_2}{2}\df_r^2
  +\frac{\epsilon c_2(1+N^2)}{2r^2}\cos(2f)\df^2\non&
  -\frac{\epsilon c_4 N^2}{r^4}\sin^2f(4\sin^2f - 3)\df^2
  +\frac{\epsilon c_4(1+N^2)}{2r^2}\cos(2f)f_r^2\df^2\non&
  +\frac{\epsilon c_4(1+N^2)}{2r^2}\sin^2(f)\df_r^2
  +\frac{\epsilon c_4(1+N^2)}{r^2}\sin(2f)f_r\df\df_r,
\label{eq:E3q_axial}
\end{align}
for the NNLO terms quadratic in $\df$ and
\begin{align}
  \calE_{3,\rm cubic}^{\rm perturb}(f,\df) =&\
  \frac{\mu^2}{2}\cos^{s-1}f\big(1-\cos^sf\big)^{p-2}
  \big(1 - s + (s p - 1)\cos^sf\big)\sin(f)\df^3\non&
  +\frac{\mu^2}{6}\cos^{s-3}f\big(1-\cos^sf\big)^{p-3}
  \big(2 - 3s + s^2 + (s-1)(4 + (1-3p)s)\cos^sf\non&
  \qquad\qquad\qquad\qquad\qquad\qquad\quad
  +(sp-2)(sp-1)\cos^{2s}f\big)\sin^3(f)\df^3\non&
  -\frac{c_6 N^2}{r^4}(5\sin^2f - 2)\sin(2f)f_r^2\df^3
  -\frac{3c_6 N^2}{r^4}\sin^2(f)(5\sin^2f-4)f_r\df^2\df_r\non&
  +\frac{2c_6 N^2}{r^4}\sin^2f\sin(2f)\df\df_r^2,
\label{eq:E3c_axial}
\end{align}
for the NNLO terms cubic in $\df$.

Outside the support of the compacton, $f=f_r=0$ and hence the
perturbation energy reduces to
\beq
\calE^{\rm perturb,\ outside}(f,\df) =
\epsilon c_2\left(\frac12\df_r^2 + \frac{1+N^2}{2}\frac{\df^2}{r^2}
+ \frac{m_\pi^2}{2c_2}\df^2\right).
\eeq
The problem simplifies for spherical symmetry for which $N=1$, since
the boundary of the compacton becomes a sphere of radius $R$, hence
simplifying drastically the boundary conditions for the outside
perturbations.
The corresponding equation of motion is the modified spherical Bessel
equation
\beq
r^2\df_{rr} + 2r\df_r - 2\df - \frac{m_\pi^2 r^2}{c_2}\df = 0,
\eeq
which in turn has the analytic solution being the first modified
spherical Bessel function of the second kind
\beq
\df = \alpha k_1\left(\tfrac{m_\pi r}{\sqrt{c_2}}\right)
= \alpha e^{-\frac{m_\pi r}{\sqrt{c_2}}}\left(\frac{\sqrt{c_2}}{m_\pi r} + \frac{c_2}{m_\pi^2 r^2}\right),
\qquad
\alpha > 0.
\label{eq:df_outside_sol}
\eeq
The perturbation outside of the compacton is thus a free massive boson
with the mass of the pion, as is expected on physical grounds.

In order to perform numerical calculations with the axially symmetric
Ansatz and $\df=\df(r)$, we actually need to pick a BPS background for
which the leading order energy correction is minimized for $N=1$, and
it is furthermore needed that the derivative of the BPS solution is
finite at the compacton radius; these constraints leave us only with
the two cases $(s,p)=(1,2)$ and $(s,p)=(2,2)$, both with $c_4\gg c_2R^2$
(since we need $N_\star\approx 1$ for the spherically symmetric
solution to be a minimizer of the energy functional, see
fig.~\ref{fig:Nstar}). 
On the other hand, it is necessary that $c_2\neq 0$, in order for the
tail of the perturbation to exist outside of the compacton.
The reason that it is necessary to have a finite derivative of the BPS
profile function at the compacton radius ($f_r(R)$) is that we will
have to impose a cusp condition on the perturbative computation, which
becomes nontrivial if the condition has to cancel an infinite negative
derivative.
In this paper, we will consider only the case of the BPS potential
$(s,p)=(1,2)$, whereas we leave the case $(s,p)=(2,2)$ for a future
work.

\subsubsection{\texorpdfstring{$(s,p)=(1,2)$}{(s,p)=(1,2)}}

Considering the potential \eqref{eq:Vsp} with $(s,p)=(1,2)$, which has
Bogomol'nyi mass \eqref{eq:MBPS12} and BPS solution
\eqref{eq:BPSsol2}, we can analytically determine the derivative of
the BPS solution at the compacton radius $R$:
\beq
f_r(R) = -\frac{3\pi}{2R},
\eeq
and hence we need to impose the following condition on the
perturbation field
\beq
\df_r(R^-) - \df_r(R^+) = \frac{3\pi}{2R},
\eeq
with $R$ given in eq.~\eqref{eq:Rsol2}.
Using the analytic solution \eqref{eq:df_outside_sol}, we have
\begin{align}
  \df_r(R^-) &= \frac{3\pi}{2R}
  -\frac{\alpha e^{-\frac{m_\pi R}{\sqrt{c_2}}}}{R}
  \left(1 + \frac{2\sqrt{c_2}}{m_\pi R} + \frac{2c_2}{m_\pi^2 R^2}\right)\non
  &= \frac{3\pi}{2R} - \frac{\df(R^-)}{R}
  \left(1 + \frac{m_\pi R}{\sqrt{c_2}} + \frac{1}{1 + \frac{m_\pi R}{\sqrt{c_2}}}\right),
\label{eq:cusp_condition}
\end{align}
which is a Robin-type of boundary condition. 

\begin{figure}[!htp]
  \begin{center}
    \mbox{\subfloat[$\frac{m_\pi}{\sqrt{c_2}}=1$]{\includegraphics[width=0.49\linewidth]{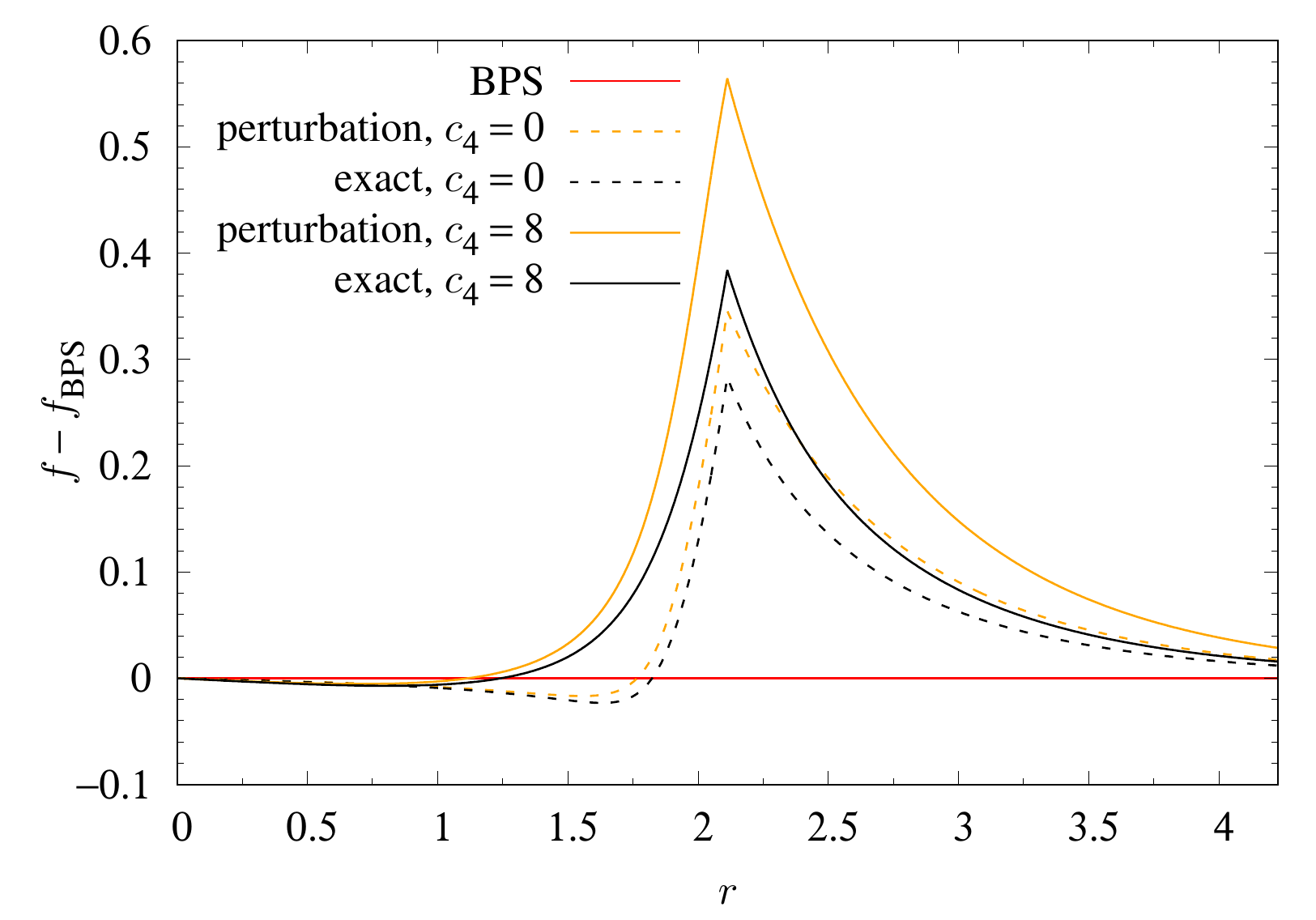}}
      \subfloat[$\frac{m_\pi}{\sqrt{c_2}}=2$]{\includegraphics[width=0.49\linewidth]{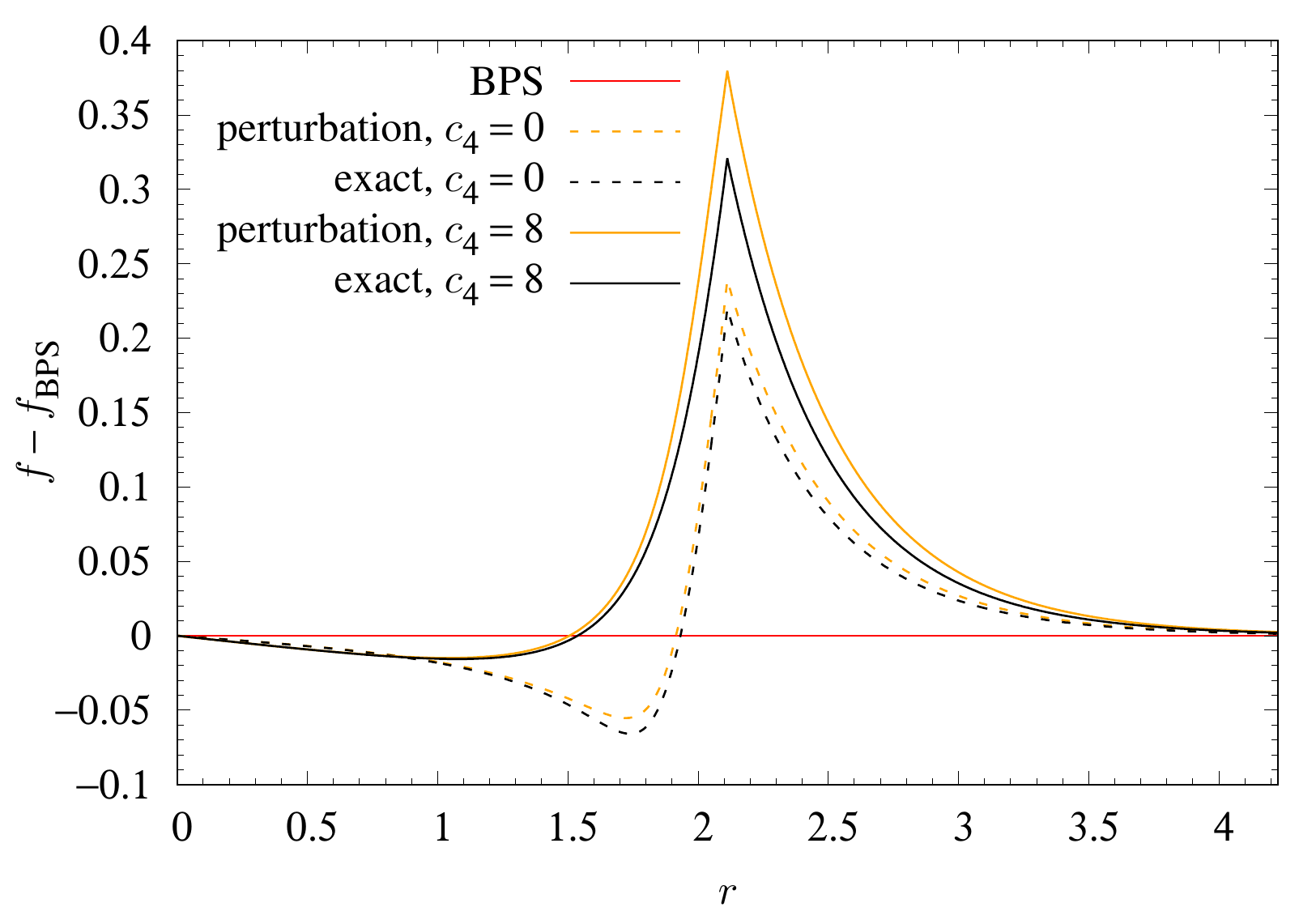}}}
    \mbox{\subfloat[$\frac{m_\pi}{\sqrt{c_2}}=3$]{\includegraphics[width=0.49\linewidth]{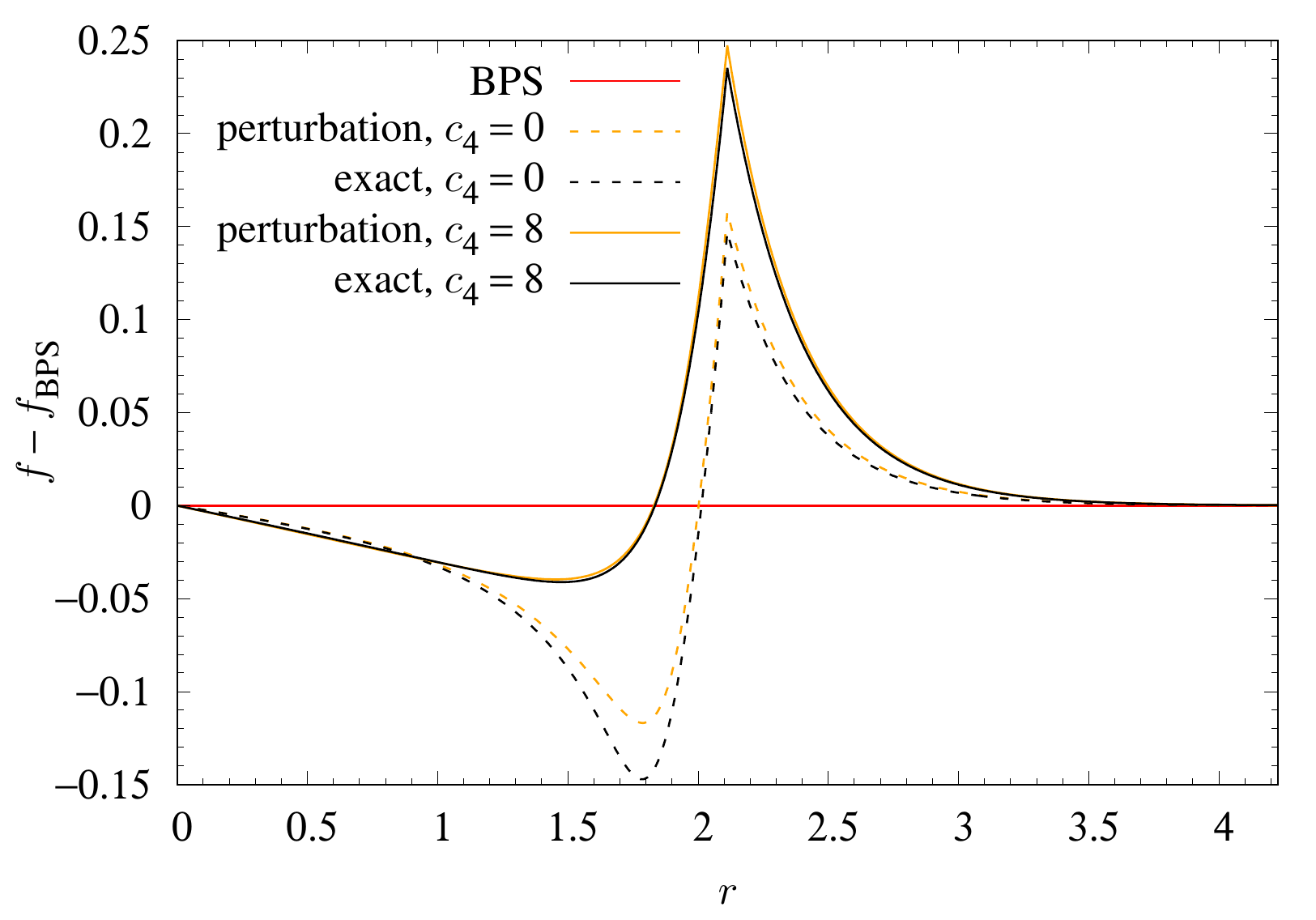}}
      \subfloat[$\frac{m_\pi}{\sqrt{c_2}}=3$]{\includegraphics[width=0.49\linewidth]{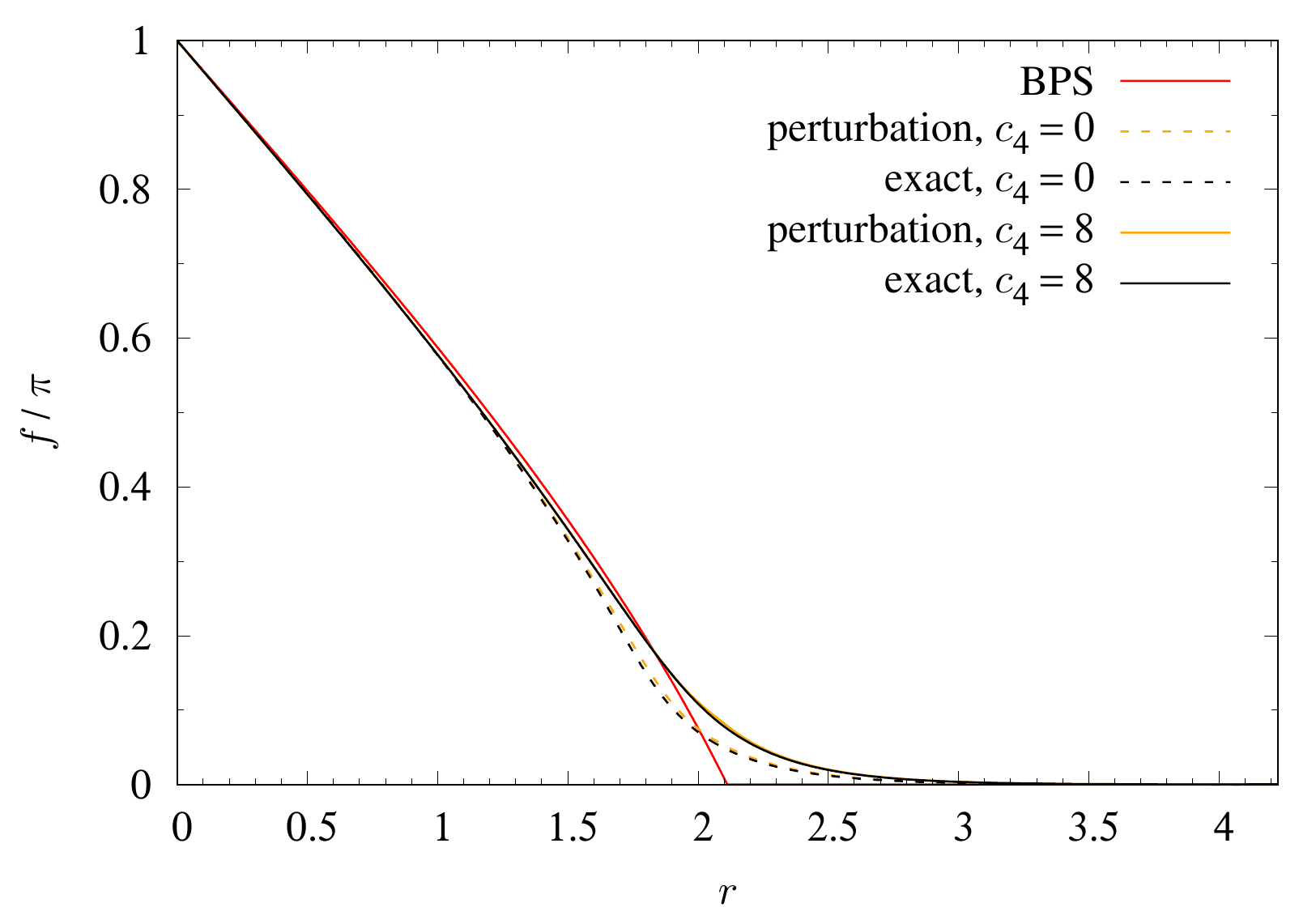}}}
    \caption{The profile function $f=f_{\rm BPS}+\df$ at N$^2$LO using
      linearized equation of motion in the BPS-Skyrme model with
      $(s,p)=(1,2)$, for $c_4=0,8$ and $m_\pi=1,2,3$.
      The other parameters of the model has been set as: $c_2=1$,
      $c_6=\frac12$, $\mu=1$ and hence the compacton radius $R=(3\pi)^{\frac13}\simeq2.112$.
    }
    \label{fig:fsol12}
  \end{center}
\end{figure}

In fig.~\ref{fig:fsol12} is shown the profile function $f(r)$ for the
$N=1$ spherically symmetric Skyrmion with $\epsilon=0.01$ for various
values of $m_\pi/\sqrt{c_2}$.
The perturbative scheme (orange curves) is compared to the exact
numerical results (black curves).
The cusp in the perturbative solution is imposed by using the
condition \eqref{eq:cusp_condition}.
For $c_4=0$ the perturbative scheme captures well the true solution
for $m_\pi/\sqrt{c_2}=2$, but not quite yet for large $c_4=8$, whereas
it works well for large $c_4=8$ with $m_\pi/\sqrt{c_2}=3$ being
slightly large.
We found in the previous section that $c_4\gg c_2R^2$ is
necessary for the spherically symmetric Skyrmion to be the true
minimizer of the energy functional, which is the reason for choosing
$c_4=8\gg(3\pi)^{2/3}$ for $c_6=1/2$ and $\mu=1$.

From fig.~\ref{fig:fsol12}, we recognize that the perturbative method
seems to better approximate the exact near-BPS solution when
increasing the pion mass $m_{\pi}$.
A possible explanation for that behavior is the
following. The perturbative expansion of the Skyrme field \eqref{expu}
obviously works in the hypothesis of $\dU\ll 1$, so that the
truncation of the series is justified by neglecting the smaller and
smaller higher orders. Therefore, the smaller the difference
between the BPS background and the exact solution is, the smaller 
$\dU$ is required to be.
In this work, the BPS background is a compacton so a suppressed
near-BPS tail (obtained with large $m_{\pi}$) should increase the
accuracy of the perturbative method.
Moreover, the linearization of the equation of motion
operated at the NLO+N$^2$LO is better justified for $\dU$ very small,
i.e.~when the tail is well suppressed by a large $m_{\pi}$. 

\begin{figure}[!tp]
  \begin{center}
    \mbox{\includegraphics[width=0.49\linewidth]{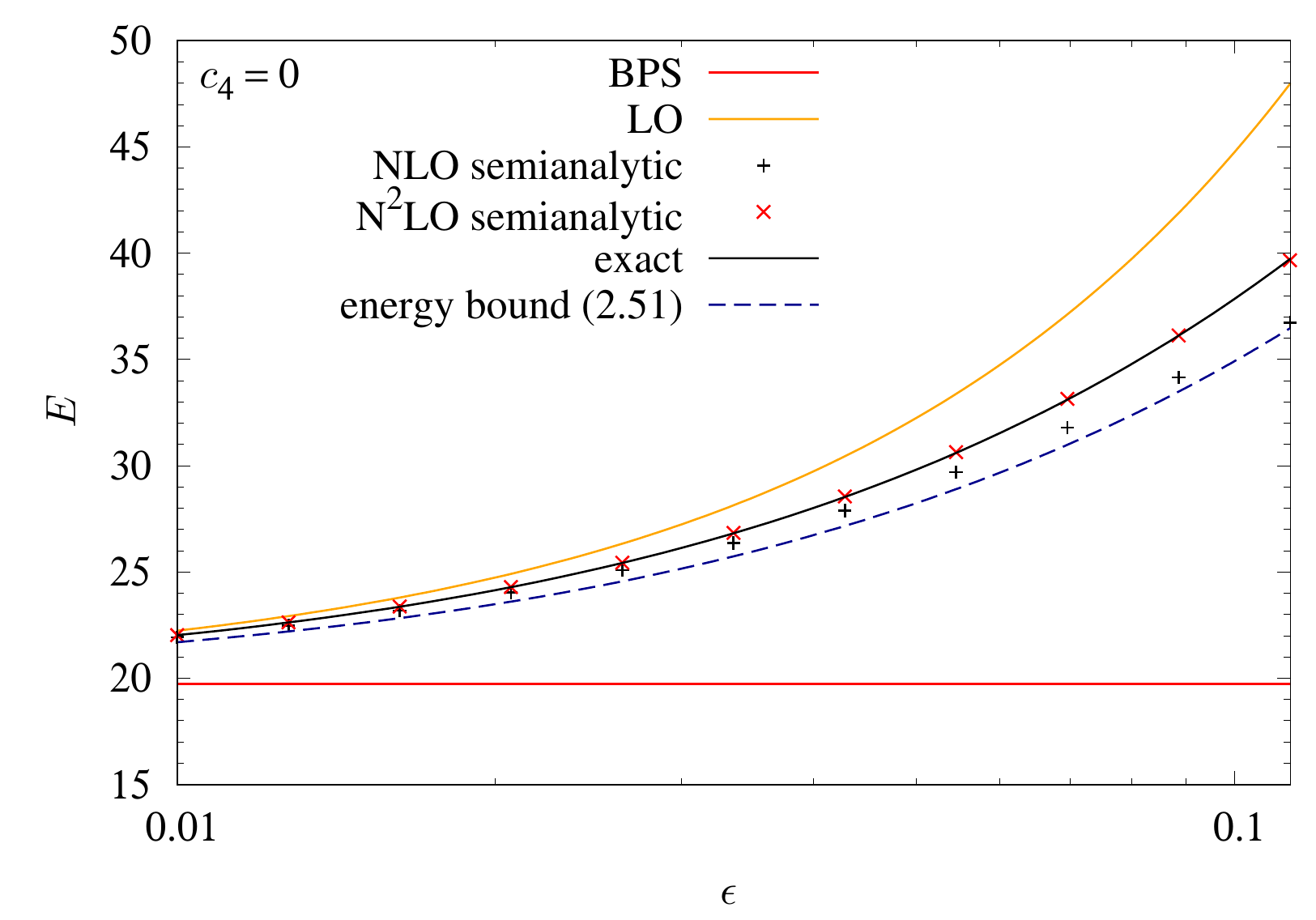}
      \includegraphics[width=0.49\linewidth]{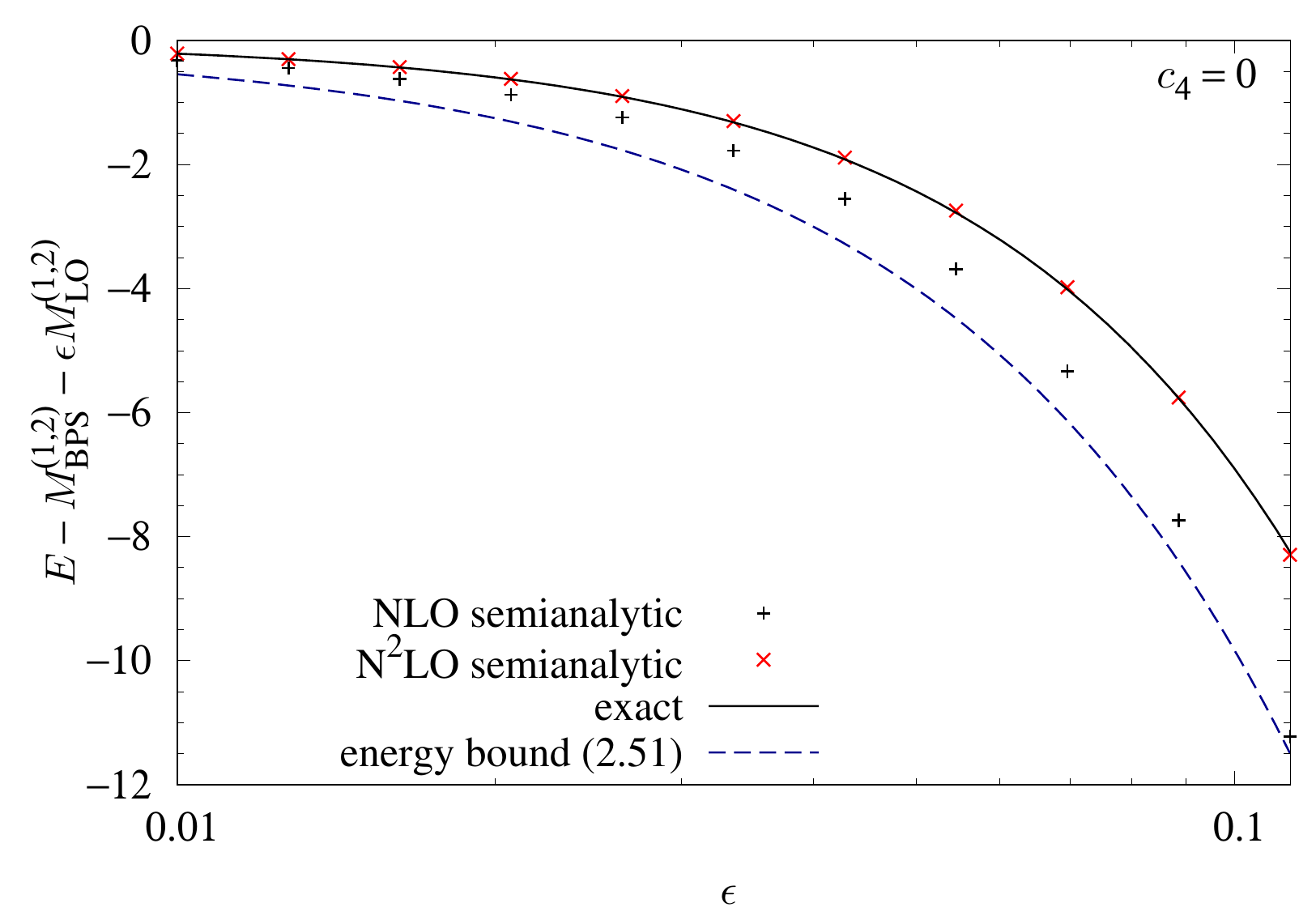}}
    \mbox{\includegraphics[width=0.49\linewidth]{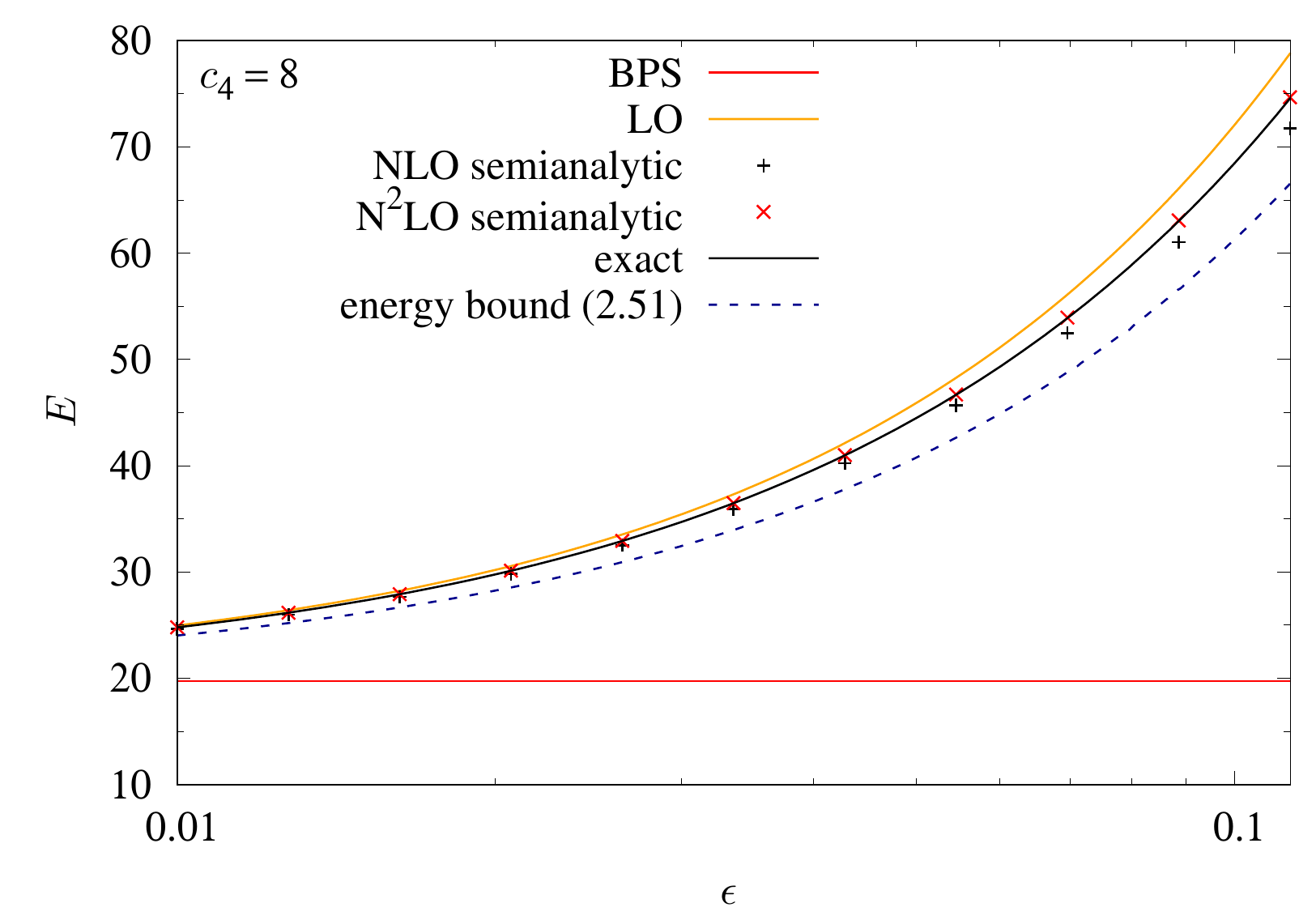}
      \includegraphics[width=0.49\linewidth]{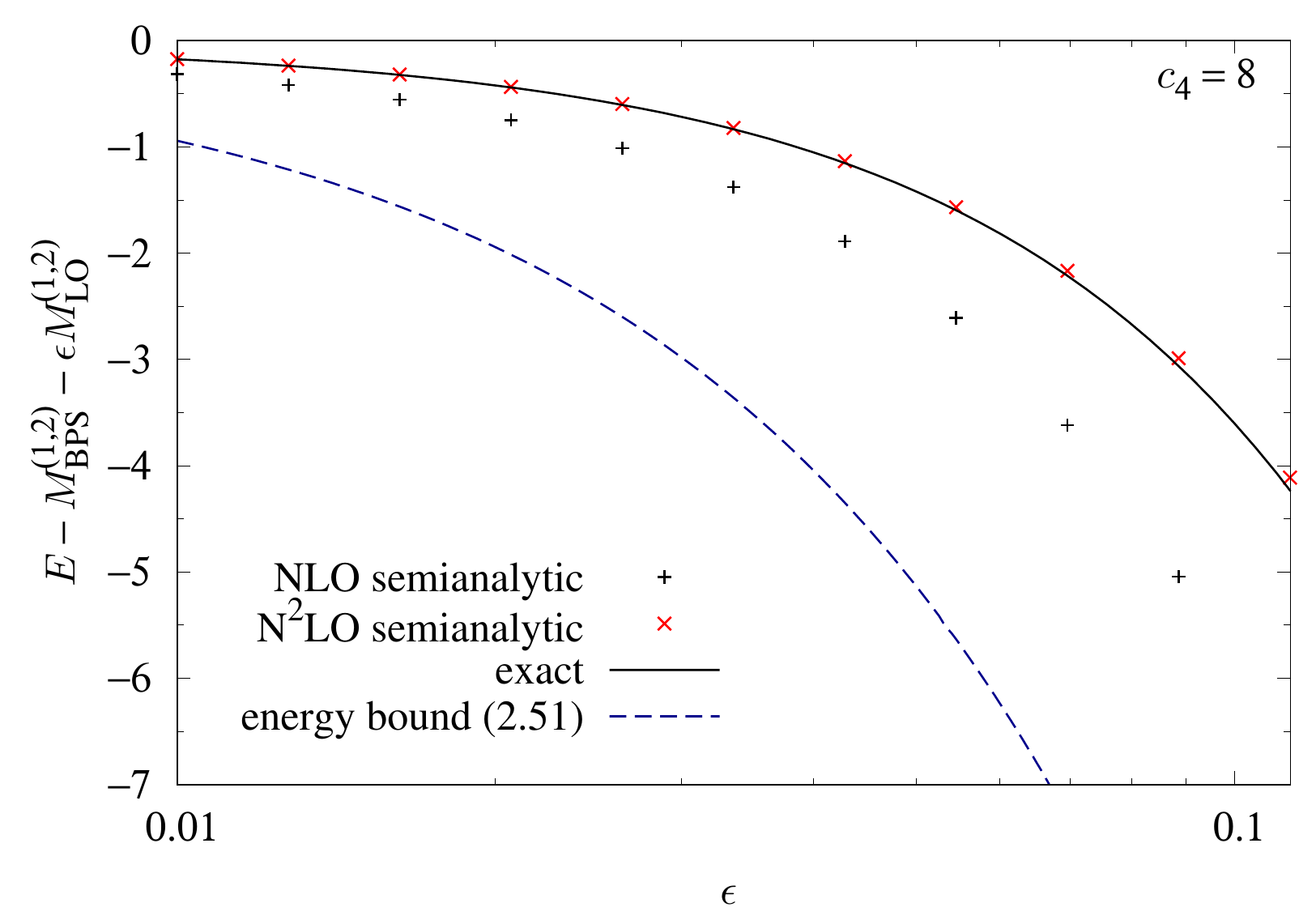}}
    \caption{The mass of the Skyrmion in the perturbative
      $\epsilon$-expansion as a function of $\epsilon$ on a logarithmic scale.
      The red solid line represents the BPS bound, the orange solid
      line is the LO correction to the energy, the black pluses are
      NLO corrections and finally the red crosses are N$^2$LO
      corrections.
      For comparison, the solid black line shows the exact ODE
      calculation.
      We also show the energy bound \eqref{eq:bound} with a
      blue-dashed line.
      The top row shows the case of $c_4=0$ and the bottom row shows
      $c_4=8$.
      In this figure $c_2=1$, $c_6=\tfrac12$, $\mu=1$, $m_\pi=3$,
      and $(s,p)=(1,2)$.
    }
    \label{fig:en_delta}
  \end{center}
\end{figure}

We are now ready to compare the energies of the exact numerical
calculations with those of the perturbative scheme.
The result is shown in fig.~\ref{fig:en_delta}.
The NLO correction to the energy is calculated using
eq.~\eqref{eq:E2_axial} which contributes with $\epsilon^2$ to the
energy and the N$^2$LO correction is calculated using the sum of
eqs.~\eqref{eq:E3q_axial} and \eqref{eq:E3c_axial} contributing of
order $\epsilon^3$.

Fitting the NLO and N$^2$LO corrections to the energy, we can write an
approximate formula for the energy in the perturbative scheme (for
$N=1$, $c_2=1$, $m_\pi=3$, $c_6=\tfrac12$ and $\mu=1$):
\begin{align}
  E(\epsilon) &= M_{\rm BPS}^{(1,2)} + \epsilon M_{\rm LO}^{(1,2)}
  + \epsilon^2 M_{\rm NLO}^{(1,2)} + \epsilon^3 M_{\rm N^2LO}^{(1,2)}
  \non
  &= 2\pi^2
  + \epsilon\left(39.79 + 5.43c_4\right)
  + \epsilon^2\left(-331.39 + 24.91c_4\right)
  + \epsilon^3\left(229.87 - 20.36c_4\right).
\end{align}
The energy for different values of $c_6$ and $\mu$ (with modified
values of $m_\pi$ and $c_2$) can be recovered by a scaling argument of
length and energy scales.

\section{Binding energies}\label{sec:binding}

In order to calculate the binding energy between two $B=1$ Skyrmions,
we need to write down the energy with generic fluctuations turned on
\begin{align}
\Delta_{ab} &= \Delta_{ab}^r + \Delta_{ab}^\theta + \Delta_{ab}^\varphi\non
&= -\Phi_a^\theta\Phi_b^\varphi \df(\bx)
-\Phi_a^\varphi\Phi_b^r \dtheta(\bx)
-\Phi_a^r\Phi_b^\theta \dvarphi(\bx),\label{eq:Delta_general}
\end{align}
 and with $\df$ not being restricted to being dependent only
on the radial coordinate.

Geometrically, there is the direction $\Phi$ and only three tangent
directions ($\Phi^r$, $\Phi^\theta$ and $\Phi^\varphi$), since the
target space is a 3-sphere.
The tensor fluctuation \eqref{eq:Delta_general} is the most general
nonvanishing tensor that can be constructed out of tensor products of
these vector directions (i.e.~$\Phi$, $\Phi^r$, $\Phi^\theta$ and
$\Phi^\varphi$); that is, any inclusion of $\Phi$ gives no
contribution to $\dPhi$ of Eq.~\eqref{eq:ansatz}.

\subsection{General fluctuation energy}\label{sec:general_energy}

The NLO + N$^2$LO energy density is given by
\begin{equation}
  \mathcal{E}^{\rm perturb}(f,\df,\dtheta,\dvarphi)
  = \mathcal{E}_2^{\rm perturb}(f,\df,\dtheta,\dvarphi)
  + \mathcal{E}_{3,\rm quad}^{\rm perturb}(f,\df,\dtheta,\dvarphi)
  + \mathcal{E}_{3,\rm cubic}^{\rm perturb}(f,\df,\dtheta,\dvarphi),
\end{equation}
with
\begin{align}
  \mathcal{E}_2^{\rm perturb}
  &= \epsilon X_{0\epsilon}^a\Omega^a
  + \sum_a\epsilon X_{1\epsilon}^a\bxhat_a^i\p_i\Omega^a
  + \Omega^aY_0^{ab}\Omega^b
  + \Omega^aY_1^{abi}\p_i\Omega^b
  + \p_i\Omega^aY_2^{abij}\p_j\Omega^b,\label{eq:E2general}\\
  \mathcal{E}_{3,\rm quad}^{\rm perturb} &= \epsilon\Omega^aY_{0\epsilon}^{ab}\Omega^b
  +\epsilon\Omega^aY_{1\epsilon}^{abi}\p_i\Omega^b
  +\epsilon\p_i\Omega^aY_{2\epsilon}^{abij}\p_j\Omega^b,\\
  \mathcal{E}_{3,\rm cubic}^{\rm perturb} &= Z_0^{abc}\Omega^a\Omega^b\Omega^c
  + \p_i\Omega^aZ_1^{abci}\Omega^b\Omega^c
  + \p_i\Omega^a\p_j\Omega^bZ_2^{abcij}\Omega^c\non&\phantom{=\ }
  + \p_i\Omega^a\p_j\Omega^bZ_3^{abcijk}\p_k\Omega^c,
\end{align}
$i,j,k=1,2,3$ the spatial indices, $a,b,c=1,2,3$ the 3-vector indices,
the definitions
\begin{align}
\Omega^a &=
\begin{pmatrix}
  \df\\\dtheta\\\dvarphi
\end{pmatrix},\quad
\bxhat_1^i =
\begin{pmatrix}
  \sin\theta\cos\varphi\\\sin\theta\sin\varphi\\\cos\theta
\end{pmatrix},\quad
\bxhat_2^i =
\begin{pmatrix}
  \cos\theta\cos\varphi\\\cos\theta\sin\varphi\\-\sin\theta
\end{pmatrix},\quad
\bxhat_3^i =
\begin{pmatrix}
  -\sin\varphi\\\cos\varphi\\0
\end{pmatrix},
\end{align}
and the tensors
\begin{align}
X_{0\epsilon}^a &=
\begin{pmatrix}
  \frac{c_2}{r^2}\sin(2f)
  +\frac{c_4}{r^2}\sin(2f)f_r^2
  +\frac{c_4}{r^4}\sin(2f)\sin^2f
  +m_\pi^2\sin f\\
  \cot\theta\sin f\left(\frac{c_2}{r^2}
  +\frac{c_4}{r^4}\sin^2f
  +\frac{c_4}{r^2}f_r^2\right)\\
  0
\end{pmatrix},\label{eq:X0epsilon}\\
X_{1\epsilon}^a &= 
\begin{pmatrix}
  \left(c_2 + \frac{2c_4}{r^2}\sin^2f\right)f_r\\
  \frac1r\left(c_2 + c_4f_r^2 + \frac{c_4}{r^2}\sin^2f\right)\sin f\\
  \frac1r\left(c_2 + c_4f_r^2 + \frac{c_4}{r^2}\sin^2f\right)\sin f
\end{pmatrix},\label{eq:X1epsilon}
\end{align}
for the linear terms
\begingroup
\allowdisplaybreaks
\begin{align}
Y_0^{ab} &=
\begin{pmatrix}
  Y_0^{11} & \frac{3c_6}{2r^4}\cot\theta\sin f\sin(2f)f_r^2 & 0\\
  \frac{3c_6}{2r^4}\cot\theta\sin f\sin(2f)f_r^2 & Y_0^{22} & 0\\
  0 & 0 & Y_0^{33}
\end{pmatrix},\\
Y_0^{11} &= -\frac{\mu^2}{4}\cos^{s-2}f(1-\cos^sf)^{p-2}
\left(-2+2s\sin^2f+\cos^sf(2-2sp\sin^2f)\right)\non&\phantom{=\ }
+\frac{2c_6}{r^4}(3 - 4\sin^2f)\sin^2(f)f_r^2,\\
Y_0^{22} &= \frac{\mu^2}{2}\cos^sf(1-\cos^sf)^{p-1}
+\frac{c_6}{r^4}(\cot^2\theta - 2\sin^2f)\sin^2(f)f_r^2,\\
Y_0^{33} &= \frac{\mu^2}{2}\cos^sf(1-\cos^sf)^{p-1}
-\frac{2c_6}{r^4}\sin^4(f)f_r^2,\\
Y_1^{abi} &=
\frac{c_6}{r^3}s^2_f f_r
\begin{pmatrix}
  \frac{4}{r}s_{2f}\bxhat_1^i & 6c_ff_r\bxhat_2^i & 6c_f f_r\bxhat_3^i\\
  \frac{4}{r}s_f(\bxhat_2^i + s_\theta^{-1}\delta^{i3}) & (\frac{s_{2f}}{r}+\frac{4f_r}{t^{2}_\theta})\bxhat_1^i - \frac{4f_r}{s_\theta t_\theta}\delta^{i3} & \frac{2f_r}{t_\theta}\bxhat_3^i\\
  0 & 0 & (\frac{s_{2f}}{r}+\frac{2f_r}{t^{2}_\theta})\bxhat_1^i - \frac{2f_r}{s_\theta t_\theta}\delta^{i3}
\end{pmatrix},\\
Y_2^{abij} &= \frac{c_6}{r^2}s^2_f
\begin{pmatrix}
  \frac{s^2_f}{r^2}\bxhat_1^i\bxhat_1^j & \frac{s_ff_r}{r}(\bxhat_1^i\bxhat_2^j + \epsilon^{ijk}\bxhat_3^k) & \frac{s_ff_r}{r}(\bxhat_1^i\bxhat_3^j - \epsilon^{ijk}\bxhat_2^k)\\
  \frac{s_ff_r}{r}(\bxhat_2^i\bxhat_1^j - \epsilon^{ijk}\bxhat_3^k) & f_r^2\bxhat_2^i\bxhat_2^j & f_r^2(\bxhat_2^i\bxhat_3^j+\epsilon^{ijk}\bxhat_1^k)\\
  \frac{s_ff_r}{r}(\bxhat_3^i\bxhat_1^j + \epsilon^{ijk}\bxhat_2^k) & f_r^2(\bxhat_3^i\bxhat_2^j-\epsilon^{ijk}\bxhat_1^k) & f_r^2\bxhat_3^i\bxhat_3^j
\end{pmatrix},
\end{align}
\endgroup
for the quadratic NLO terms
\begingroup
\allowdisplaybreaks
\begin{align}
Y_{0\epsilon}^{11} &= \frac{c_2}{r^2}\cos(2f)
+\frac{c_4}{r^2}\cos(2f)f_r^2
+\frac{c_4}{r^4}\sin^2f(3-4\sin^2f)
+\frac{m_\pi^2}{2}\cos f,\\
Y_{0\epsilon}^{22} &= \frac{c_2}{2r^2}(\cos2f+\cot^2\theta)
-\frac{c_2}{2}f_r^2
+\frac{c_4}{2r^4}\sin^2f(\cos(2f)+\cot^2\theta)
\non&\phantom{=\ }
-\frac{c_4}{2r^2}(3\sin^2f - \cot^2\theta)f_r^2
+\frac{m_\pi^2}{2}\cos f,\\
Y_{0\epsilon}^{33} &= \frac{c_2}{2r^2}(\cos2f+\cot^2\theta)
-\frac{c_2}{2}f_r^2
+\frac{c_4}{2r^4}\sin^2f\cos2f
-\frac{c_4}{2r^2}(3\sin^2f - \cot^2\theta)f_r^2
\non&\phantom{=\ }
+\frac{m_\pi^2}{2}\cos f\\
Y_{0\epsilon}^{ab} &=
\begin{pmatrix}
  Y_{0\epsilon}^{11} & \frac{c_2c_f}{2r^2t_\theta} + \frac{c_4c_f}{2r^4t_\theta}(r^2f_r^2 + 3s^2_f) & 0\\
  \frac{c_2c_f}{2r^2t_\theta} + \frac{c_4c_f}{2r^4t_\theta}(r^2f_r^2 + 3s^2_f) & Y_{0\epsilon}^{22} & 0\\
  0 & 0 & Y_{0\epsilon}^{33}
\end{pmatrix},\\
Y_{1\epsilon}^{abi} &= \frac{c_2}{r}c_f
\begin{pmatrix}
  0 & \bxhat_2^i & \bxhat_3^i\\
  -\bxhat_2^i & 0 & \frac{1}{t_\theta c_f}\bxhat_3^i\\
  -\bxhat_3^i & -\frac{1}{t_\theta c_f}\bxhat_3^i & 0
\end{pmatrix}\non
\mathop+\frac{c_4}{r^2}&s_f
\begin{pmatrix}
  4c_ff_r\bxhat_1^i & \left(\frac{3s_{2f}}{2r} + \frac{rf_r^2}{t_f}\right)\bxhat_2^i & \left(\frac{3c_{2f}}{2r} + \frac{rf_r^2}{t_f}\right)\bxhat_3^i \\
  \left(2f_r - \frac{s_{2f}}{2r}\right)\bxhat_2^i + \frac{2f_r}{s_\theta}\delta^{i3} & \left(\frac{2s_f}{rt_\theta^2} + c_ff_r\right)\bxhat_1^i - \frac{2c_\theta s_f}{rs_\theta^2}\delta^{i3} & \cot\theta\left(\frac{s_f}{r}+\frac{rf_r^2}{s_f}\right)\bxhat_3^i\\
  -\frac{s_{2f}}{2r}\bxhat_3^i & -\frac{rf_r^2}{t_\theta s_f}\bxhat_3^i & \left(\frac{s_f}{rt_\theta^2} + c_ff_r\right)\bxhat_1^i - \frac{c_\theta s_f}{rs_\theta^2}\delta^{i3}
\end{pmatrix},\\
Y_{2\epsilon}^{abij} &= \frac{c_2}{2}
\begin{pmatrix}
  \delta^{ij} & 0 & 0\\
  0 & \delta^{ij} & 0\\
  0 & 0 & \delta^{ij}
\end{pmatrix}\non&
\mathop+\frac{c_4}{2}
\begin{pmatrix}
  \frac{s_f^2}{r^2}(\delta^{ij} + \bxhat_1^i\bxhat_1^j) & \frac{s_ff_r}{r}(\bxhat_1^i\bxhat_2^j + \epsilon^{ijk}\bxhat_3^k) & \frac{s_ff_r}{r}(\bxhat_1^i\bxhat_3^j - \epsilon^{ijk}\bxhat_2^k)\\
  \frac{s_ff_r}{r}(\bxhat_2^i\bxhat_1^j - \epsilon^{ijk}\bxhat_3^k) & \frac{s_f^2}{r^2}\bxhat_{!3}^i\bxhat_{!3}^j + f_r^2\bxhat_{!1}^i\bxhat_{!1}^j & \frac{s_f^2}{r^2}(\bxhat_2^i\bxhat_3^j + \epsilon^{ijk}\bxhat_1^k)\\
  \frac{s_ff_r}{r}(\bxhat_3^i\bxhat_1^j + \epsilon^{ijk}\bxhat_2^k) & \frac{s_f^2}{r^2}(\bxhat_3^i\bxhat_2^j - \epsilon^{ijk}\bxhat_1^k) & \frac{s_f^2}{r^2}\bxhat_{!2}^i\bxhat_{!2}^j + f_r^2\bxhat_{!1}^i\bxhat_{!1}^j
\end{pmatrix},
\end{align}
\endgroup
for the quadratic N$^2$LO terms and
\begingroup
\allowdisplaybreaks
\begin{align}
  Z_0^{111} &= \frac{\mu^2}{6}\cos^{s-3}f(1 - \cos^2f)^{p-3}
  \sin f \Big((s-1)(-3 + (1 + s)\sin^2f)
  \non&\phantom{=\ }\quad
  + (sp-1)\cos^{2s}f(-3 + (sp+1)\sin^2f)
  \non&\phantom{=\ }\quad
  + \cos^sf(3(s(p+1)-2) + (2 + s^2(1 - 3p))\sin^2f)\Big)
  \non&\phantom{=\ }
  +\frac{c_6}{r^4}\sin(2f)(2 - 5\sin^2f)f_r^2,\\
  Z_0^{222} &= -\frac{3c_6}{r^4}\cot\theta\sin^3(f)f_r^2,\\
  Z_0^{112} &= \frac{3c_6}{4r^4}\cot\theta(3\sin(3f) - \sin f)f_r^2,\\
  Z_0^{221} &= \frac{\mu^2}{2}\cos^{s-1}f(1-\cos^sf)^{p-2}(1-s+(sp-1)\cos^sf)\sin f
  \non&\phantom{=\ }
  +\frac{c_6}{r^4}\sin(2f)(\cot^2\theta-3\sin^2f)f_r^2,\\
  Z_0^{331} &= \frac{\mu^2}{2}\cos^{s-1}f(1-\cos^sf)^{p-2}(1-s+(sp-1)\cos^sf)\sin f
  -\frac{3c_6}{r^4}\sin^2f\sin(2f)f_r^2,\\
  Z_0^{332} &= -\frac{3c_6}{r^4}\cot\theta\sin^3(f)f_r^2,\\
  Z_1^{1bci} &= \frac{c_6}{r^4}s_f^2f_r
  \begin{pmatrix}
    3(4 - 5s^2_f)\bxhat_1^i & 6c_f(\bxhat_2^i + \csc\theta\delta^{i3}) & 0\\
    6c_f(\bxhat_2^i + \csc\theta\delta^{i3}) & \left(\frac{2}{t_\theta^2} - 3s^2_f\right)\bxhat_1^i & 0\\
    0 & 0 & -3s_f^2\bxhat_1^i
  \end{pmatrix},\\
  Z_1^{2bci} &= \frac{c_6}{r^3}s_ff_r
  \begin{pmatrix}
    3(2-3s_f^2)f_r\bxhat_2^i & c_f\left(\frac{3s_{2f}}{2r} + \frac{4f_r}{t_\theta^2}\right)\bxhat_1^i  & 0\\
    c_f\left(\frac{3s_{2f}}{2r} + \frac{4f_r}{t_\theta^2}\right)\bxhat_1^i & \left(\frac{2s_{2f}}{r} + \frac{2f_r}{t_\theta^2} - 3s_f^2f_r\right)\bxhat_2^i & 0\\
    0 & 0 & -3s_f^2f_r\bxhat_2^i
  \end{pmatrix}\non
  &\phantom{=\ }
  +\frac{2c_6}{r^3}\frac{s_{2f}f_r}{s_\theta}
  \begin{pmatrix}
    0 & -\frac{f_r}{t_\theta} & 0\\
    -\frac{f_r}{t_\theta} & \frac{s_f}{r} & 0\\
    0 & 0 & 0
  \end{pmatrix}
  \delta^{i3},\\
  Z_1^{3bci} &= \frac{c_6}{r^3}s_ff_r
  \begin{pmatrix}
    3(2-3s_f^2)f_r\bxhat_3^i & \frac{2c_ff_r}{t_\theta}\bxhat_3^i & c_f\left(\frac{3s_{2f}}{2r}+\frac{2f_r}{t_\theta^2}\right)\bxhat_1^i\\
    \frac{2c_ff_r}{t_\theta}\bxhat_3^i & -3s_f^2f_r\bxhat_3^i & \left(\frac{s_{2f}}{r} + \frac{f_r}{t_\theta^2}\right)\bxhat_2^i\\
    c_f\left(\frac{3s_{2f}}{2r}+\frac{2f_r}{t_\theta^2}\right)\bxhat_1^i & \left(\frac{s_{2f}}{r} + \frac{f_r}{t_\theta^2}\right)\bxhat_2^i & -3s_f^2f_r\bxhat_3^i\\
  \end{pmatrix}\non&\phantom{=\ }
  +\frac{c_6}{r^3}\frac{s_{2f}f_r}{s_\theta}
  \begin{pmatrix}
    0 & 0 & -\frac{f_r}{t_\theta}\\
    0 & 0 & \frac{s_f}{r}\\
    -\frac{f_r}{t_\theta} & \frac{s_f}{r} & 0
  \end{pmatrix}\delta^{i3},\\
  Z_2^{ab1ij} &= \frac{c_6}{r^2}s_{2f}
  \begin{pmatrix}
    \frac{2s_f^2}{r^2}\bxhat_1^i\bxhat_1^j & \frac{3s_ff_r}{2r}(\bxhat_1^i\bxhat_2^j + \epsilon^{ijk}\bxhat_3^k) & \frac{3s_ff_r}{2r}(\bxhat_1^i\bxhat_3^j - \epsilon^{ijk}\bxhat_2^k)\\
    \frac{3s_ff_r}{2r}(\bxhat_2^i\bxhat_1^j - \epsilon^{ijk}\bxhat_3^k) & f_r^2\bxhat_2^i\bxhat_2^j & f_r^2(\bxhat_2^i\bxhat_3^j + \epsilon^{ijk}\bxhat_1^k)\\
    \frac{3s_ff_r}{2r}(\bxhat_3^i\bxhat_1^j + \epsilon^{ijk}\bxhat_2^k) & f_r^2(\bxhat_3^i\bxhat_2^j - \epsilon^{ijk}\bxhat_1^k) & f_r^2\bxhat_3^i\bxhat_3^j
  \end{pmatrix},\\
  Z_2^{ab2ij} &= \frac{c_6}{r^2}s_f
  \begin{pmatrix}
    \frac{2s_f^2}{r^2}(\bxhat_2^i\bxhat_1^j + s_\theta^{-1}\delta^{i3}\bxhat_1^j) & Z_2^{122ij} & Z_2^{132ij}\\
    Z_2^{212ij} & \frac{s_{2f}f_r}{r}(\bxhat_2^i\bxhat_1^j - \frac12\epsilon^{ijk}\bxhat_3^k) + \frac{2f_r^2}{t_\theta}\bxhat_2^i\bxhat_2^j & Z_2^{232ij}\\
    Z_2^{312ij} & Z_2^{322ij} & 0
  \end{pmatrix},\\
  Z_2^{122ij} &= Z_2^{212ji} = \frac{2s_ff_r}{r}\bxhat_2^i\bxhat_2^j + \frac{s_fs_{2f}}{2r^2}\bxhat_1^i\bxhat_1^j + \frac{2s_ff_r}{rt_\theta}\epsilon^{ijk}\bxhat_3^k + \frac{2s_ff_r}{rs_\theta}\delta^{i3}\bxhat_2^j,\\
  Z_2^{132ij} &= Z_2^{312ji} = \frac{s_ff_r}{r}(\bxhat_2^i\bxhat_3^j+\epsilon^{ijk}\bxhat_1^k-s_\theta^{-1}\bxhat_3^i\delta^{j3}+2s_\theta^{-1}\delta^{i3}\bxhat_3^j),\\
  Z_2^{232ij} &= Z_2^{322ji} = \frac{f_r^2}{t_\theta}(\bxhat_2^i\bxhat_3^j + \epsilon^{ijk}\bxhat_1^k) + \frac{s_{2f}f_r}{2r}(\bxhat_1^i\bxhat_3^j - \epsilon^{ijk}\bxhat_2^k),\\
  Z_2^{ab3ij} &= \frac{c_6}{r^2}s_f
  \begin{pmatrix}
    0 & 0 & Z_2^{133ij}\\
    0 & 0 & Z_2^{233ij}\\
    Z_2^{313ij} & Z_2^{323ij} & \frac{2f_r^2}{t_\theta}(\bxhat_2^i\bxhat_3^j - \frac{t_\theta}{2}\epsilon^{ijk}\bxhat_2^k) + \frac{s_{2f}f_r}{r}(\bxhat_1^i\bxhat_3^j + \frac12\epsilon^{ijk}\bxhat_2^k) - \frac{f_r^2}{s_\theta}\epsilon^{ij3}
  \end{pmatrix},\\
  Z_2^{133ij} &= Z_2^{313ji} = \frac{s_ff_r}{r}(\bxhat_2^i\bxhat_2^j + t_\theta^{-1}\epsilon^{ijk}\bxhat_3^k + s_\theta^{-1}\delta^{i3}\bxhat_2^j) + \frac{s_{2f}s_f}{2r^2}\bxhat_1^i\bxhat_1^j,\\
  Z_2^{233ij} &= Z_2^{323ji} = \frac{f_r^2}{t_\theta}\bxhat_2^i\bxhat_2^j + \frac{s_{2f}f_r}{2r}(\bxhat_1^i\bxhat_2^j + \epsilon^{ijk}\bxhat_3^k),\\
  Z_3^{112ijk} &= \frac{4c_6}{r^3}s_f^3
  \left(\bxhat_1^i\epsilon^{jkl}\bxhat_3^l + \frac12\epsilon^{ijl}\bxhat_3^l\bxhat_1^k\right),\\
  Z_3^{113ijk} &= \frac{4c_6}{r^3}s_f^3
  \left(-\bxhat_1^i\epsilon^{jkl}\bxhat_2^l - \frac12\epsilon^{ijl}\bxhat_2^l\bxhat_1^k\right),\\
  Z_3^{221ijk} &= \frac{4c_6}{r^2}s_f^2f_r
  \left(-\bxhat_2^i\epsilon^{jkl}\bxhat_3^l - \frac12\epsilon^{ijl}\bxhat_3^l\bxhat_2^k\right),\\
  Z_3^{123ijk} &= \frac{4c_6}{r^2}s_f^2f_r
  \left(\bxhat_1^i\epsilon^{jkl}\bxhat_1^l + \frac12\epsilon^{ijl}\bxhat_1^l\bxhat_1^k + \epsilon^{ijl}\bxhat_3^l\bxhat_3^k + \frac12\bxhat_3^i\epsilon^{jkl}\bxhat_3^l\right),\\
  Z_3^{223ijk} &= \frac{4c_6}{r}s_ff_r^2
  \left(\bxhat_2^i\epsilon^{jkl}\bxhat_1^l + \frac12\epsilon^{ijl}\bxhat_1^l\bxhat_2^k\right),\\
  Z_3^{332ijk} &= \frac{4c_6}{r}s_ff_r^2
  \left(-\bxhat_3^i\epsilon^{jkl}\bxhat_1^l - \frac12\epsilon^{ijl}\bxhat_1^l\bxhat_3^k\right),
\end{align}
\endgroup
for the cubic N$^2$LO terms.

The equations of motion for the general fluctuations in Cartesian
coordinates read
\begin{align}
  \p_i\left(
  \Omega^b Y_1^{bai}
  +\epsilon\Omega^b Y_{1\epsilon}^{bai}
  +2Y_2^{abij}\p_j\Omega^b
  +2\epsilon Y_{2\epsilon}^{abij}\p_j\Omega^b\right)&\non\qquad
  \mathop-2Y_0^{ab}\Omega^b
  -2\epsilon Y_{0\epsilon}^{ab}\Omega^b
  -Y_1^{abi}\p_i\Omega^b
  -\epsilon Y_{1\epsilon}^{abi}\p_i\Omega^b
  &= \epsilon X_{0\epsilon}^a - \epsilon\p_i(X_{1\epsilon}^a\bxhat_a^i),
  \label{eq:eom_general_fluc}
\end{align}
($a$ not summed over).
We have used the short-hand notation for the trigonometric functions
\beq
s_\theta=\sin\theta,\quad
s_f=\sin f,\quad
c_\theta=\cos\theta,\quad
c_f=\cos f,\quad
t_\theta=\tan\theta,
\eeq
and so on, and we have used the following short-hand index summation
rule
\beq
\bxhat_{!1}^i\bxhat_{!1}^j := \sum_{a\neq 1}\bxhat_a^i\bxhat_a^j
=\delta^{ij} - \bxhat_1^i\bxhat_1^j,
\eeq
and similarly for other excluded directions.

\subsection{Spherical symmetry}

We will now show that if we restrict to the spherically symmetric
$B=1$ Skyrmion and impose $\df=\df(r)$, the fluctuations $\dtheta$ and
$\dvarphi$ decouple and are solved by their trivial solution. 

First we notice that there are, seemingly, source terms for the
fluctuation field $\dtheta$ in eqs.~\eqref{eq:X0epsilon} and
\eqref{eq:X1epsilon}.
However, using the identity
\beq
\p_iX = \bxhat_1^i\p_rX + \frac1r\bxhat_2^i\p_\theta X + \frac{1}{r\sin\theta}\bxhat_3^i\p_\varphi X,
\label{eq:id_deriv}
\eeq
the linear terms in the energy density \eqref{eq:E2general} for the 
fluctuations $\dtheta$ and $\dvarphi$ read 
\beq
\epsilon\sin f
\left(\frac{c_2}{r^2} + \frac{c_4}{r^4}\sin^2f + \frac{c_4}{r^2}f_r^2\right)
\left(\cot\theta\dtheta + \p_\theta\dtheta + \p_\varphi\dvarphi\right).
\eeq
Now including the integration measure, we have
\begin{align}
&\epsilon\int \d r\d\theta\dvarphi
\sin f
\left(c_2 + \frac{c_4}{r^2}\sin^2f + c_4f_r^2\right)
\left(\cos\theta\dtheta + \sin\theta\p_\theta\dtheta + \sin\theta\p_\varphi\dvarphi\right)\non
&=\epsilon\int \d r\d\theta\dvarphi\;
\p_\theta\left[
\sin f
\left(c_2 + \frac{c_4}{r^2}\sin^2f + c_4f_r^2\right)
\sin\theta\dtheta
\right]\non
&\phantom{=\ }
+\epsilon\int \d r\d\theta\dvarphi\;
\p_\varphi\left[
\sin\theta\sin f
\left(c_2 + \frac{c_4}{r^2}\sin^2f + c_4f_r^2\right)
\dvarphi
\right],
\end{align}
which are clearly total derivatives and hence do no contribute to the
equations of motion for the fluctuation fields.
In particular, this means that the sources, i.e.~the right-hand side
of the equation of motion \eqref{eq:eom_general_fluc} take the form
\beq
\epsilon
\begin{pmatrix}
  *\\0\\0
\end{pmatrix},
\eeq
thus are only turning on the fluctuation $\df=\Omega^1$. 

Next, we will show that only the non-radial derivatives of $\df$ turn
on the fluctuations $\dtheta$ and $\dvarphi$.
Starting with the non-derivative terms in the equations of motion
\eqref{eq:eom_general_fluc}, we observe that
\beq
\p_i(Y_1^i)^{\rm T} - 2Y_0 =
\begin{pmatrix}
  * & * & 0\\
  0 & * & 0\\
  0 & 0 & *
\end{pmatrix},\qquad
\p_i(Y_{1\epsilon}^i)^{\rm T} - 2Y_{0\epsilon} =
\begin{pmatrix}
  * & * & 0\\
  0 & * & 0\\
  0 & 0 & *
\end{pmatrix},
\eeq
where we treat the tensors as matrices in $ab$: $a=1,2,3$ being the row
and the equation index and $b=1,2,3$ being the column and field index.
It is thus clear -- at this stage -- that a nonvanishing $\dtheta$
affects the equation of motion for $\df$, but a nonvanishing $\df$
does not affect the equations of motion for $\dtheta$ and $\dvarphi$:
it does not act as a source for the latter fluctuation fields.

Considering now the one-derivative terms of the equations of motion,
we find
\begin{align}
\left[(Y_1^i)^{\rm T} - Y_1^i + 2\p_jY_2^{ji}\right]\bxhat_1^i = 
\begin{pmatrix}
  * & * & 0\\
  0 & * & 0\\
  0 & 0 & *
\end{pmatrix},\qquad
\left[(Y_{1\epsilon}^i)^{\rm T} - Y_{1\epsilon}^i + 2\p_jY_{2\epsilon}^{ji}\right]\bxhat_1^i = 
\begin{pmatrix}
  * & * & 0\\
  0 & * & 0\\
  0 & 0 & *
\end{pmatrix},\non
\left[(Y_1^i)^{\rm T} - Y_1^i + 2\p_jY_2^{ji}\right]\bxhat_2^i = 
\begin{pmatrix}
  0 & * & 0\\
  * & * & 0\\
  0 & 0 & *
\end{pmatrix},\qquad
\left[(Y_{1\epsilon}^i)^{\rm T} - Y_{1\epsilon}^i + 2\p_jY_{2\epsilon}^{ji}\right]\bxhat_2^i = 
\begin{pmatrix}
  0 & * & 0\\
  * & * & 0\\
  0 & 0 & *
\end{pmatrix},\non
\left[(Y_1^i)^{\rm T} - Y_1^i + 2\p_jY_2^{ji}\right]\bxhat_3^i = 
\begin{pmatrix}
  0 & 0 & *\\
  0 & 0 & 0\\
  * & * & 0
\end{pmatrix},\qquad
\left[(Y_{1\epsilon}^i)^{\rm T} - Y_{1\epsilon}^i + 2\p_jY_{2\epsilon}^{ji}\right]\bxhat_3^i = 
\begin{pmatrix}
  0 & 0 & *\\
  0 & 0 & *\\
  * & * & 0
\end{pmatrix},
\end{align}
from which we can see that a nonvanishing radial derivative of the
fluctuation $\p_r\df$ does not turn on the fluctuations $\dtheta$ and
$\dvarphi$ (see the first line), whereas a nonvanishing $\p_\theta\df$
acts as a source for $\dtheta$ (see the second line) and a
nonvanishing $\p_\varphi\df$ acts as a source for $\dvarphi$,
recalling the identity \eqref{eq:id_deriv} (see the third line).

Finally, we need to consider the double-derivatives of the fluctuation
fields in the equations of motion and we find
\begin{equation}
(Y_2^{ij} + Y_2^{ji})\bxhat_k^i\bxhat_l^j \propto \delta^{ka}\delta^{lb} + \delta^{kb}\delta^{la},\qquad
(Y_{2\epsilon}^{ij} + Y_{2\epsilon}^{ji})\bxhat_k^i\bxhat_l^j \propto
\delta^{kl}\delta^{ab} + \delta^{ka}\delta^{lb} + \delta^{kb}\delta^{la},
\end{equation}
(distinguishing only vanishing and nonvanishing elements of the
tensor and with $ab$ being the matrix indices) and hence it is clear
again that the only sources for the fluctuations $\dtheta$ and
$\dvarphi$ are $\p_r\p_\theta\df$ and $\p_r\p_\varphi\df$.

This completes the proof that $\df=\df(r)$ does not turn on the
fluctuations $\dtheta$ or $\dvarphi$ and their equations of motion are
homogeneous and satisfied by $\dtheta=\dvarphi=0$, which is compatible
with their boundary conditions at spatial infinity.
Hence, without a nonspherical fluctuation field $\df$, the angular
fluctuation fields remain turned off.

\subsection{Boundary conditions}

\begin{figure}[!htp]
  \centering
  \includegraphics[width=0.5\linewidth]{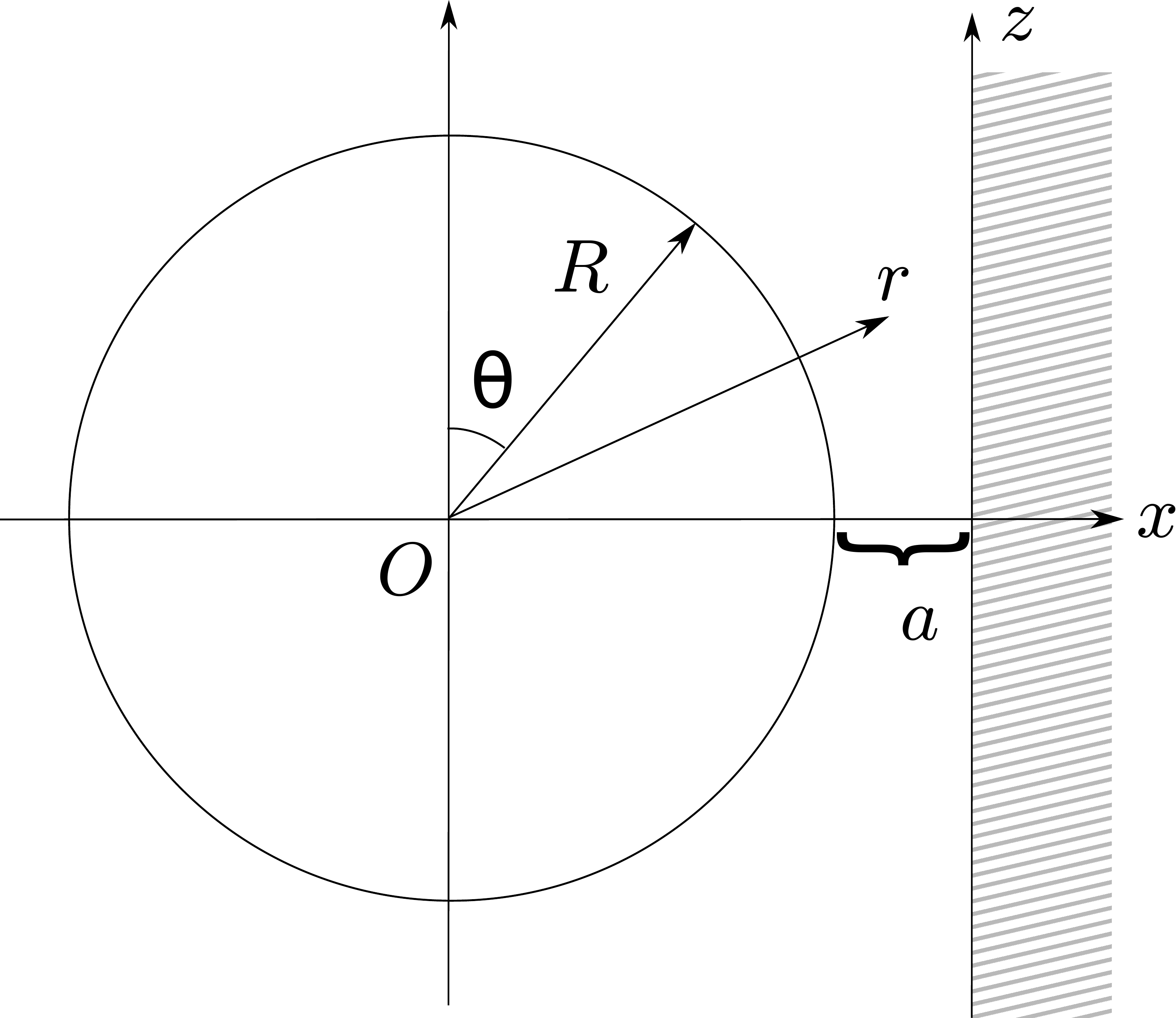}
  \caption{Setup for computation of the binding energy between two
    $B=1$ Skyrmions using the mirror trick and certain gluing
    conditions at $x=0$ and the $B=1$ Skyrmion is placed at $x=-R-a$.
    The figure is shown at $y=\varphi=0$. }
  \label{fig:BC}
\end{figure}

The equation of motion \eqref{eq:eom_general_fluc} must be accompanied
by suitable boundary conditions: the cusp condition at the boundary of
the $B=1$ compacton, the fluctuations must vanish at spatial infinity
and finally, we will solve the problem of binding energies by
performing a mirror trick similar to the problem in two dimensions
\cite{Gudnason:2020tps}, see fig.~\ref{fig:BC}.

A problem is that the best coordinates for imposing the cusp condition
on the boundary of the compacton are the spherical coordinates with
origin $O$, whereas the best coordinates for imposing the gluing
conditions at $x=0$ are Cartesian coordinates.
Here, we will utilize the fact that we are only solving the linearized
equation of motion \eqref{eq:eom_general_fluc} and hence, we can find
the solution using the superposition of two solutions:
\beq
\df = \df^{\rm rad} + \df^{\rm glue},
\label{eq:split}
\eeq
which is the sum of the radial fluctuation of
sec.~\ref{sec:axial_perturb}, $\df^{\rm rad}$, that solves the cusp
condition on the boundary of the compacton and a new fluctuation field
$\df^{\rm glue}$ that is only subject to the gluing condition as well
as the boundary condition at spatial infinity.

The solution for the $B=1$ Skyrmion that obeys the cusp condition
$\df^{\rm rad}$ is exactly the fluctuation found in
sec.~\ref{sec:axial_perturb}; the new fluctuation field subject to the
gluing conditions can thus be calculated in Cartesian coordinates,
which is why we have written the energy of
sec.~\ref{sec:general_energy} in Cartesian coordinates.
Notice that the field $\df^{\rm glue}$ experiences a smooth
background, since the cusp condition and the termination of the BPS
solution on the compacton boundary add up to a smooth total field.
We will now discuss the gluing condition in more details next.

\subsection{Gluing condition}

We will assume that the two $B=1$ Skyrmions should be placed in their
attractive channel, which is dictated by viewing the Skyrmions as
triplets of dipoles, which by the kinetic term provides and attractive
channel \cite{Schroers:1993yk}\footnote{See also
  Ref.~\cite{Gudnason:2020arj} for an explicit numerical computation
  of the interaction potential.}.
The gluing conditions for the attractive channel of two $B=1$
Skyrmions are
\beq
\begin{pmatrix}
  \p_x\Phi^0\\
  \p_x\Phi^1\\
  \Phi^2\\
  \p_x\Phi^2
\end{pmatrix}
= 0,
\label{eq:gluing_condition_Phi}
\eeq
where 3 conditions are Neumann and one is Dirichlet, which is
necessary for computing the gluing with Skyrmions in the attractive
channel (e.g.~all Neumann condition would lead to a
Skyrmion-anti-Skyrmion pair).
In order to impose the correct boundary conditions on the fluctuation
fields, we first write out the total field
\begin{align}
  \Phi &= \varPhi + \dPhi\non
  &= \varPhi + \Phi_r\df + \Phi_\theta\dtheta + \Phi_\varphi\dvarphi
  -\frac12\varPhi(\df^2 + \dtheta^2 + \dvarphi^2).
\end{align}
Since the gluing condition should be applied outside the domain of the
compacton, we need to set $f=0$ of the background solution, for which
the total field reads
\begin{align}
  \Phi 
  &=
  \begin{pmatrix}
    1\\0\\0\\0
  \end{pmatrix}
  \left(1 - \frac{\df^2}{2} - \frac{\dtheta^2}{2} - \frac{\dvarphi^2}{2}\right) +
  \begin{pmatrix}
    0\\
    \sin\theta\cos\varphi\\
    \sin\theta\sin\varphi\\
    \cos\theta
  \end{pmatrix}
  \df +
  \begin{pmatrix}
    0\\
    \cos\theta\cos\varphi\\
    \cos\theta\sin\varphi\\
    -\sin\theta
  \end{pmatrix}
  \dtheta +
  \begin{pmatrix}
    0\\
    -\sin\varphi\\
    \cos\varphi\\
    0
  \end{pmatrix}
  \dvarphi.
\end{align}
Using the identity \eqref{eq:id_deriv}, we obtain the following gluing
conditions for the fluctuation fields
\begin{align}
\df\df_x + \dtheta\dtheta_x + \dvarphi\dvarphi_x &= 0,\\
\sin\theta\sin\varphi\df
+\cos\theta\sin\varphi\dtheta
+\cos\varphi\dvarphi &= 0,\\
\frac1r\cos\theta\cos\varphi(\cos\theta\df - \sin\theta\dtheta)
+\sin\theta\df_x
+\cos\theta\dtheta_x
-\tan\varphi\dvarphi_x &= 0,\\
\frac1r\cos\varphi(\sin\theta\df + \cos\theta\dtheta)
-\df_x
+\tan\theta\dtheta_x &= 0.
\end{align}
Since this is a complicated mixture of a nonlinear boundary condition
and a Robin-type boundary condition on the fluctuation fields, we will
solve the linear Robin-part of the boundary condition and verify
a posteriori that the quadratic part is approximately satisfied.
Using the discrete $x$-derivative to order $h_x^2$ with $h_x$ being
the lattice spacing, the solution reads
\begingroup
\allowdisplaybreaks
\begin{align}
  \df_0^{\rm glue} &=
  \frac{3r^2(\cos^2\theta + \sin^2\theta\cos^2\varphi)}{4h_x^2\cos^2\theta\cos^4\varphi + 9r^2}
  (4\df_{-1}^{\rm glue} - \df_{-2}^{\rm glue})
  \non&\phantom{=\ }
  +\frac{r\cos\theta(2h_x\cos^3\varphi - 3r\sin\theta\sin^2\varphi)}{4h_x^2\cos^2\theta\cos^4\varphi + 9r^2}(4\dtheta_{-1} - \dtheta_{-2})
  \non&\phantom{=\ }
  -\frac{r\sin(2\varphi)(2h_x\cos^2\theta\cos\varphi + 3r\sin\theta)}{2(4h_x^2\cos^2\theta\cos^4\varphi + 9r^2)}(4\dvarphi_{-1} - \dvarphi_{-2})
  \non&\phantom{=\ }
  +\frac{4h_x^2\cos^2\theta\cos^4\varphi + 9r^2\sin^2\theta\sin^2\varphi}{4h_x^2\cos^2\theta\cos^4\varphi + 9r^2}\df^{\rm rad}
  \non&\phantom{=\ }
  +\frac{6h_xr^2\sin\theta\cos\varphi(\cos^2\theta + \sin^2\theta\cos^2\varphi)}{4h_x^2\cos^2\theta\cos^4\varphi + 9r^2}\df_r^{\rm rad},\label{eq:glue1}\\
  \dtheta_0 &=
  \frac{r(2h_x\cos^3\varphi + 3r\sin\theta\sin^2\varphi)}{4h_x^2\cos\theta\cos^4\varphi + 9r^2}
  (4\df_{-1}^{\rm glue} - \df_{-2}^{\rm glue})
  \non&\phantom{=\ }
  -\frac{3r^2(2h_x\cos^4\theta\cos^3\varphi + 3r\sin\theta\cos^2\varphi + 2h_x\cos^2\theta\sin^2\theta\cos\varphi + 3r\sin^3\theta\sin^2\varphi)}{(4h_x^2\cos^2\theta\cos^4\varphi + 9r^2)(2h_x\cos^2\theta + 3r\sin\theta)}
  \non&\phantom{=\ }
  \quad\times(4\dtheta_{-1} - \dtheta_{-2})
  \non&\phantom{=\ }
  +\frac{r\sin(2\varphi)(12h_xr\cos(2\theta)\cos\theta\cos\varphi - \sin(2\theta)(4h_x^2\cos^2\theta\cos^2\varphi - 9r^2)}{4(4h_x^2\cos^2\theta\cos^4\varphi + 9r^2)(2h_x\cos^2\theta\cos\varphi + 3r\sin\theta)}
  \non&\phantom{=\ }
  \quad\times(4\dvarphi_{-1} - \dvarphi_{-2})
  \non&\phantom{=\ }
  +\frac{3r\cos\theta(4h_x^2c_\theta^2c_\varphi^4 + 6h_xrc_\theta^2s_\theta c_\varphi + 9r^2s_\theta^2s_\varphi^2 + 6h_xrs_\theta^3c_\varphi^3)}{(4h_x^2\cos^2\theta\cos^4\varphi + 9r^2)(2h_x\cos^2\theta\cos\varphi + 3r\sin\theta)}\df^{\rm rad}
  \non&\phantom{=\ }
  -\frac{h_xr\sin(2\theta)\cos\varphi(2h_x\cos^4\varphi + 3r\sin\theta\sin^2\varphi)}{4h_x^2\cos^2\theta\cos^4\varphi + 9r^2}\df_r^{\rm rad},\\
  \dvarphi_{0} &=
  -\frac{r\sin(2\varphi)(2h_x\cos^2\theta\cos\varphi - 3r\sin\theta)}{2(4h_x^2\cos^2\theta\cos^4\varphi + 9r^2)}
  (4\df_{-1}^{\rm glue} - \df_{-2}^{\rm glue})
  \non&\phantom{=\ }
  -\frac{2r\cos\theta\sin(2\varphi)(3r + 2h_x\sin\theta\cos\varphi)}{3(h_x^2 + 12r^2) + h^2(3\cos(2\theta) + 2\cos^2\theta(4\cos(2\varphi) + \cos(4\varphi)))}
  (4\dtheta_{-1} - \dtheta_{-2})
  \non&\phantom{=\ }
  +\frac{3r^2\sin^2\varphi}{4h_x^2\cos^2\theta\cos^4\varphi + 9r^2}
  (4\dvarphi_{-1} - \dvarphi_{-2})
  \non&\phantom{=\ }
  +\frac{6r\sin(2\varphi)(3r\sin\theta - 2h_x\cos^2\theta\cos\varphi)}{3(h_X^2 + 12r^2) + h_x^2(3\cos(2\theta) + 2\cos^2\theta(4\cos(2\varphi) + \cos(4\varphi)))}\df^{\rm rad}
  \non&\phantom{=\ }
  +\frac{4h_xr\cos\varphi\sin(2\varphi)\sin\theta(2h_x\cos^2\theta\cos\varphi - 3r\sin\theta)}{3(h_x^2 + 12r^2) + h_x^2(3\cos(2\theta) + 2\cos^2\theta(4\cos(2\varphi) + \cos(4\varphi)))}\df_r^{\rm rad}.\label{eq:glue3}
\end{align}
\endgroup
It is simple to check that the gluing condition is regular for
$3r>2h_x$, which is always true for the gluing condition since $r>R$,
where $R=1\gg h_x$ in numerical calculations.
The subscripts on the fluctuation fields correspond to lattice indices
in the $x$-direction, i.e.~$\dtheta_0$ corresponds to $\dtheta(x=0)$,
$\dtheta_{-1}=\dtheta(x=-h_x)$ and $\dtheta_{-2}=\dtheta(x=-2h_x)$ and
similarly for the other fluctuation fields.
The coordinate system for the spherical polar coordinates is
\beq
x+R+a+\i y = r\sin\theta e^{\i\varphi}, \qquad
z = r\cos\theta.
\eeq
With the gluing conditions in hand, we are now ready to perform
numerical computations of the binding energies.

\subsection{Numerical results}

We will now compute the binding energies numerically within the
semianalytic $\epsilon$-expansion up to N$^2$LO,
i.e.~$\mathcal{O}(\epsilon^3)$.
We choose the potential $(s,p)=(1,2)$ which corresponds to $(1-\cos f)^2$
because this potential allows for the spherically symmetric Skyrmion
being stable within the axially symmetric Ansatz for $c_2=1$ and
$c_4=8$ (or generically any $c_4\gg c_2R^2$). We further fix the
parameters of the numerical calculation by setting $c_6=\frac12$ and
$\mu=1$, which yields a compacton radius, $R=(3\pi)^{\frac13}$.
Now, in order for the cusp condition to be sufficient for the
$\epsilon$-expansion scheme to capture the true Skyrmion solution, we
need $m_\pi\gg 1$ and chose $m_\pi=3$ as in
sec.~\ref{sec:axial_perturb}. 

\def\widthone{0.133}
\begin{figure}[!htp]
\centering
\subfloat[]{\mbox{
\includegraphics[width=\widthone\linewidth]{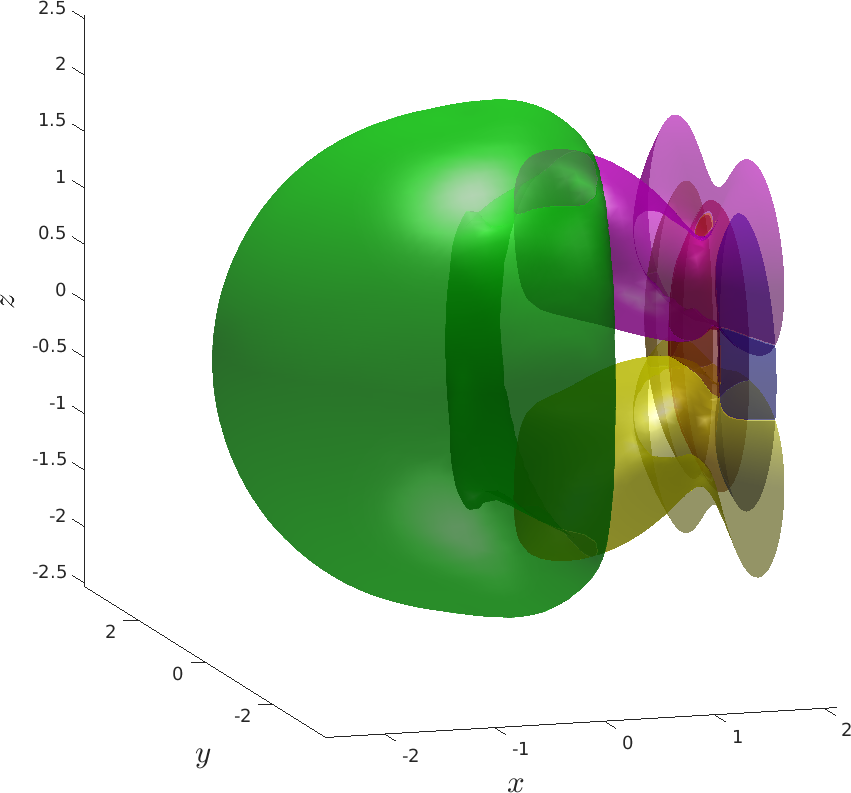}
\includegraphics[width=\widthone\linewidth]{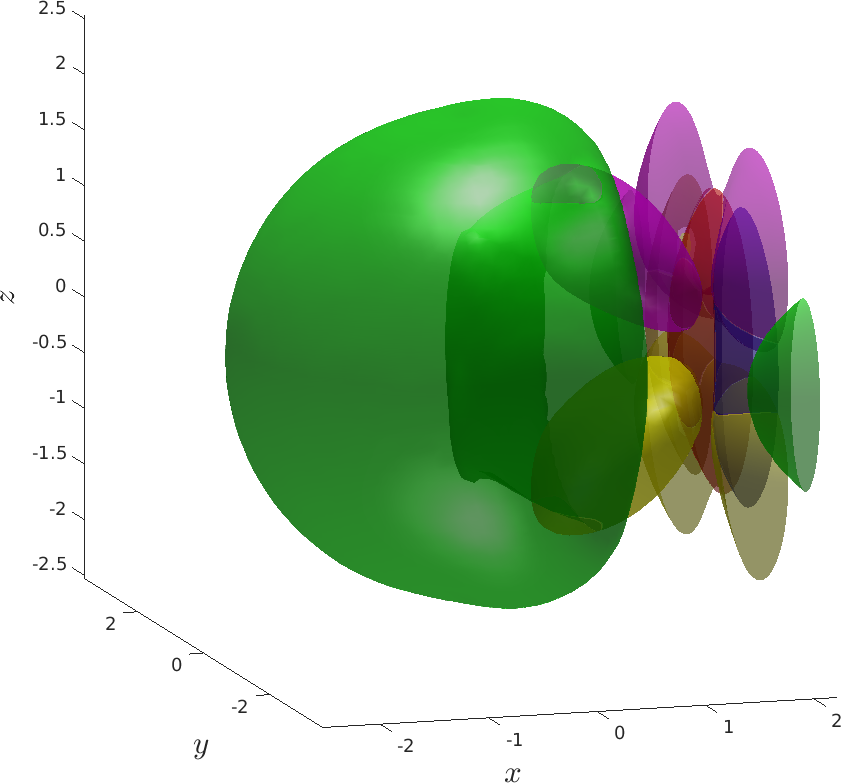}
\includegraphics[width=\widthone\linewidth]{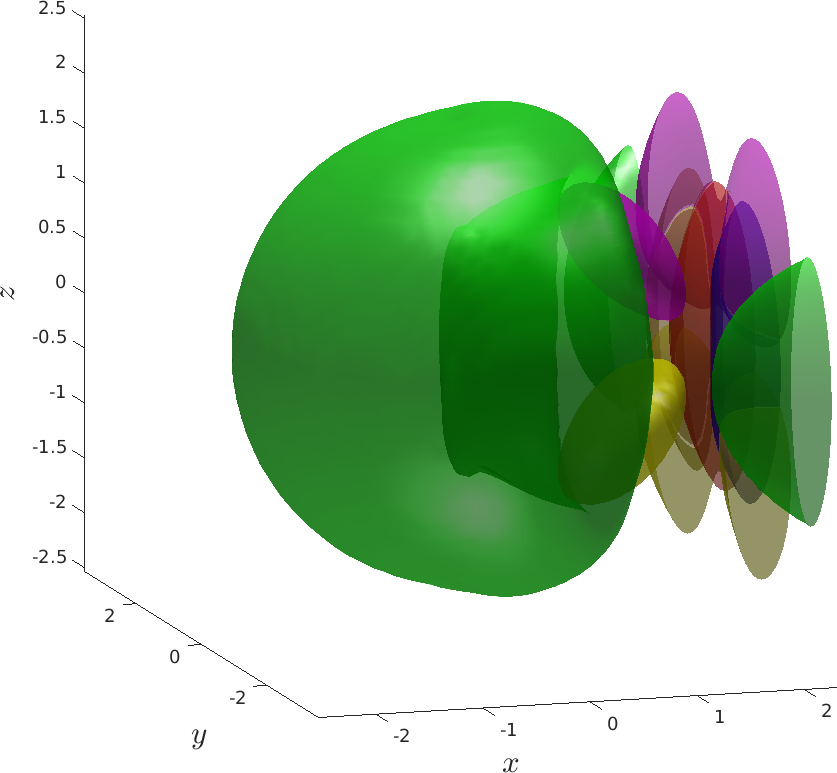}
\includegraphics[width=\widthone\linewidth]{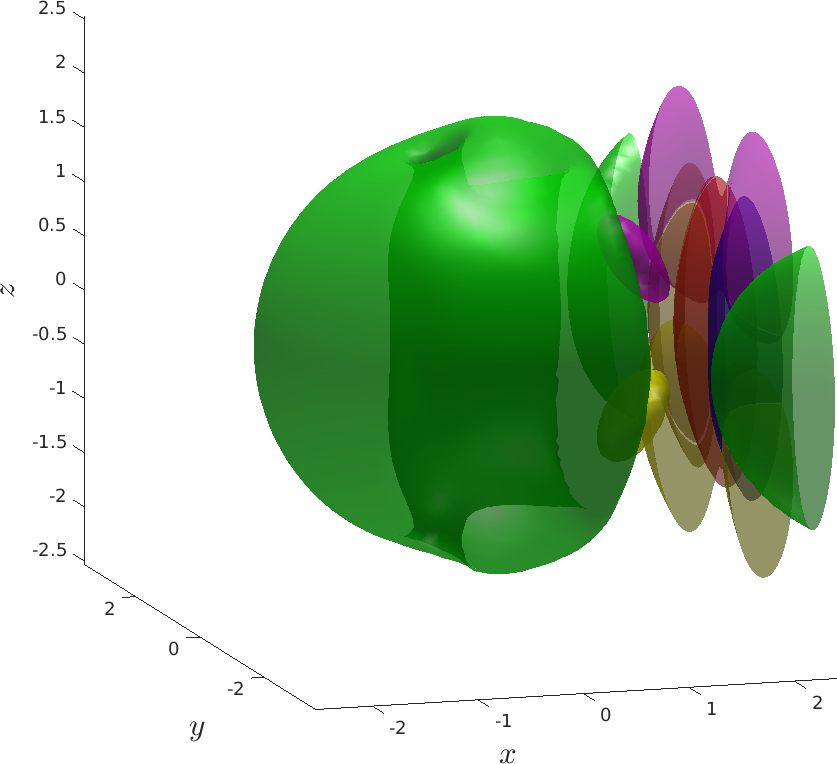}
\includegraphics[width=\widthone\linewidth]{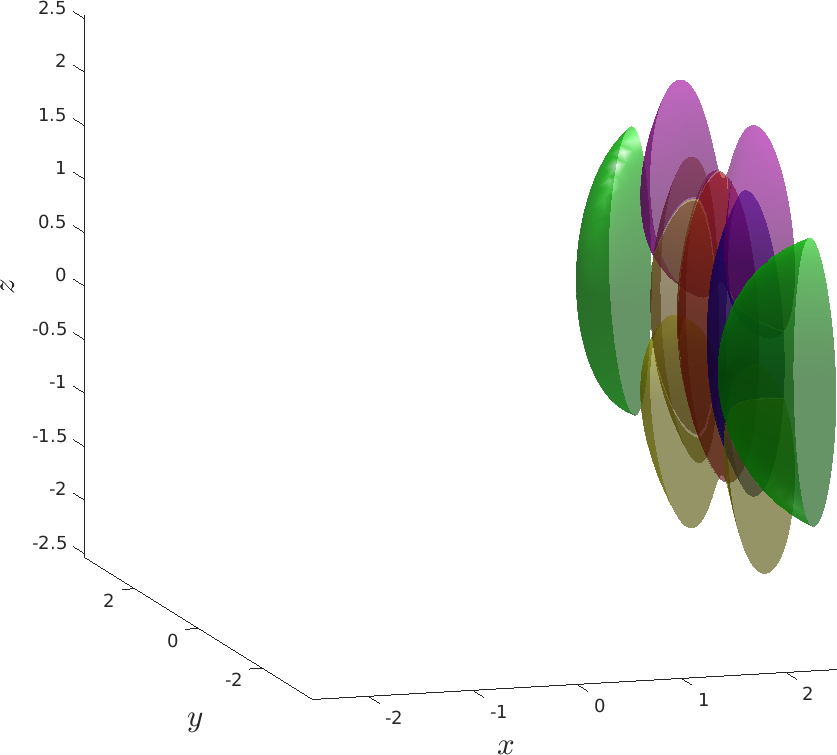}
\includegraphics[width=\widthone\linewidth]{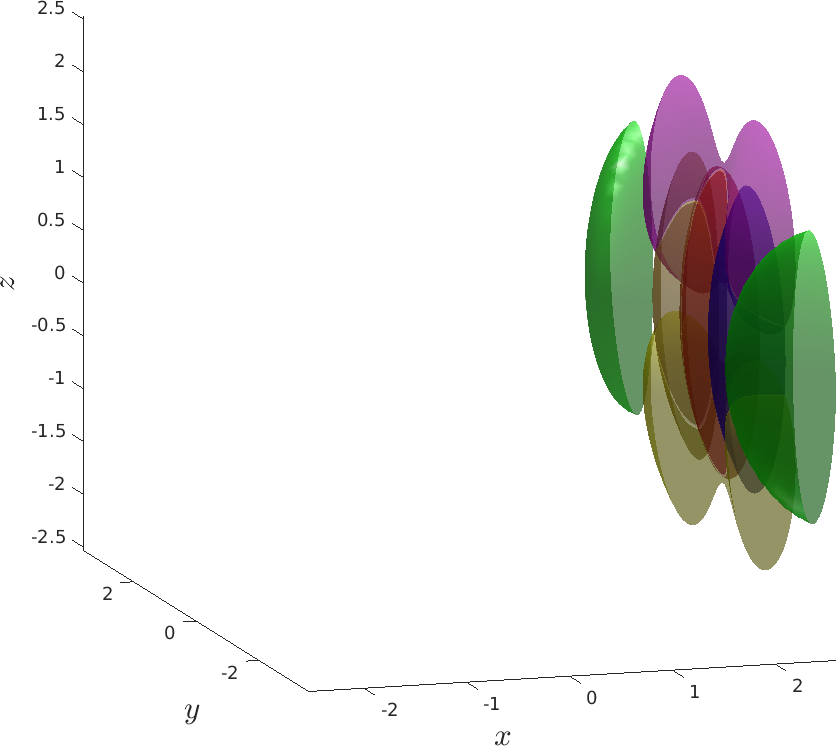}
\includegraphics[width=\widthone\linewidth]{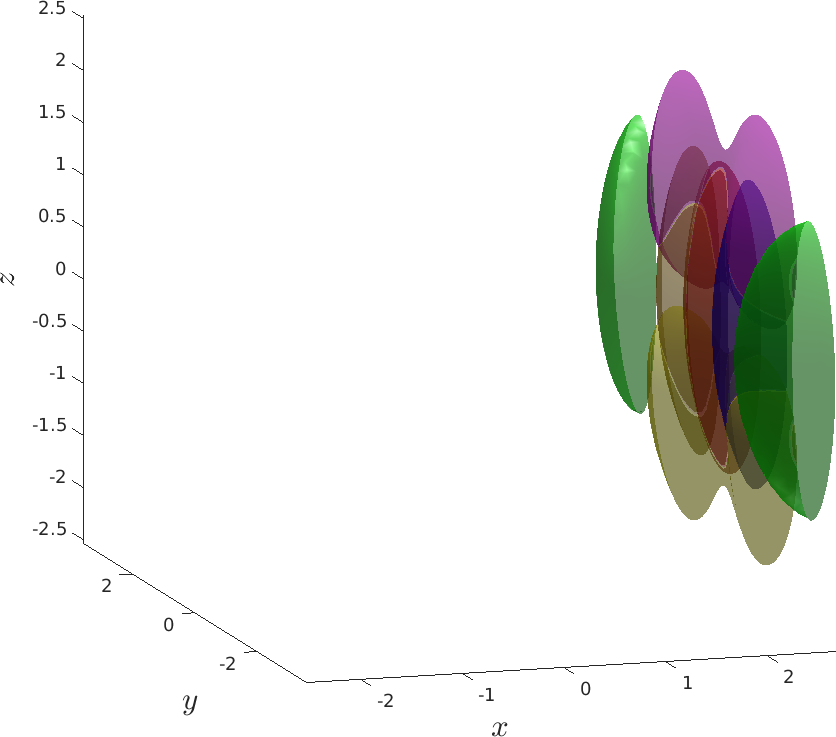}
}}\\
\subfloat[]{\mbox{
\includegraphics[width=\widthone\linewidth]{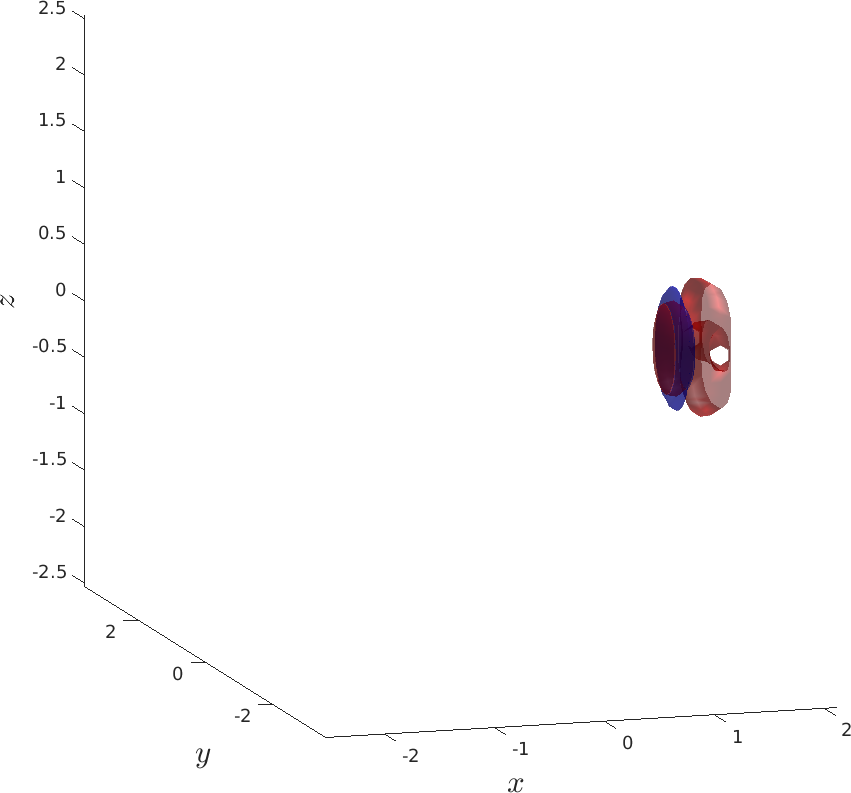}
\includegraphics[width=\widthone\linewidth]{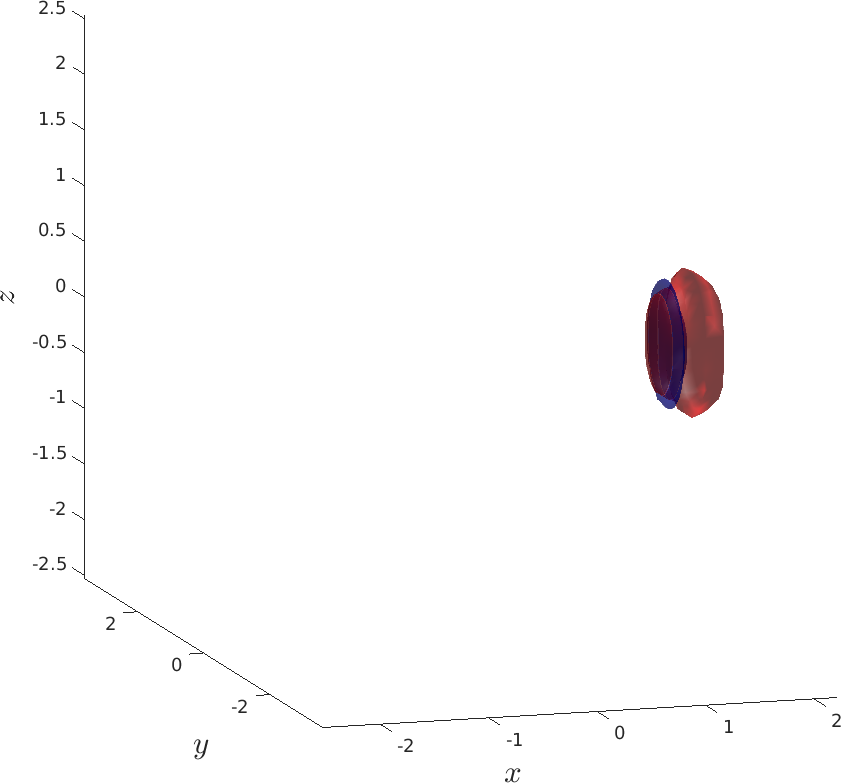}
\includegraphics[width=\widthone\linewidth]{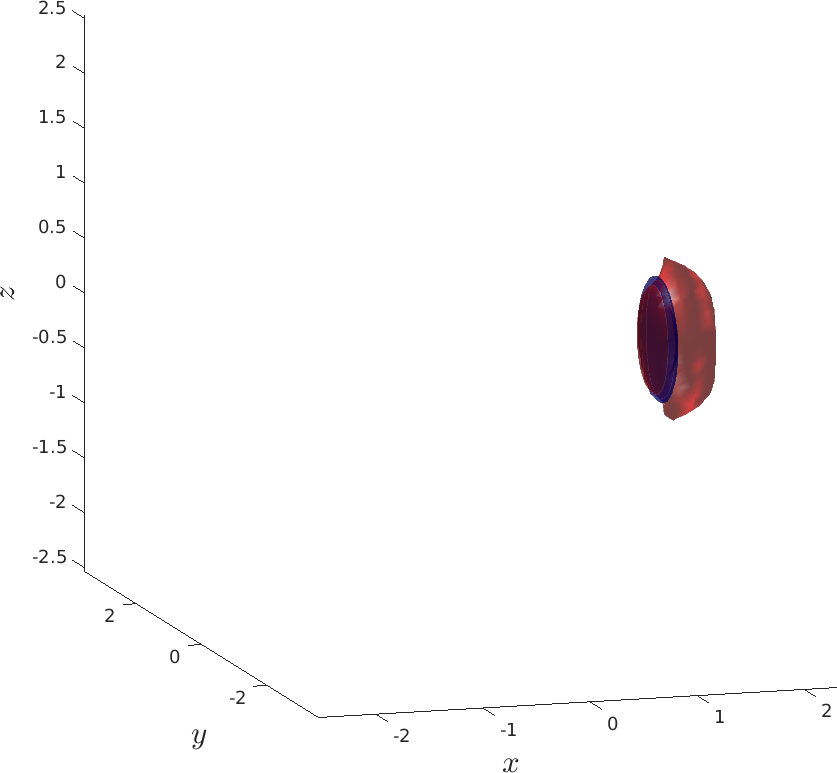}
\includegraphics[width=\widthone\linewidth]{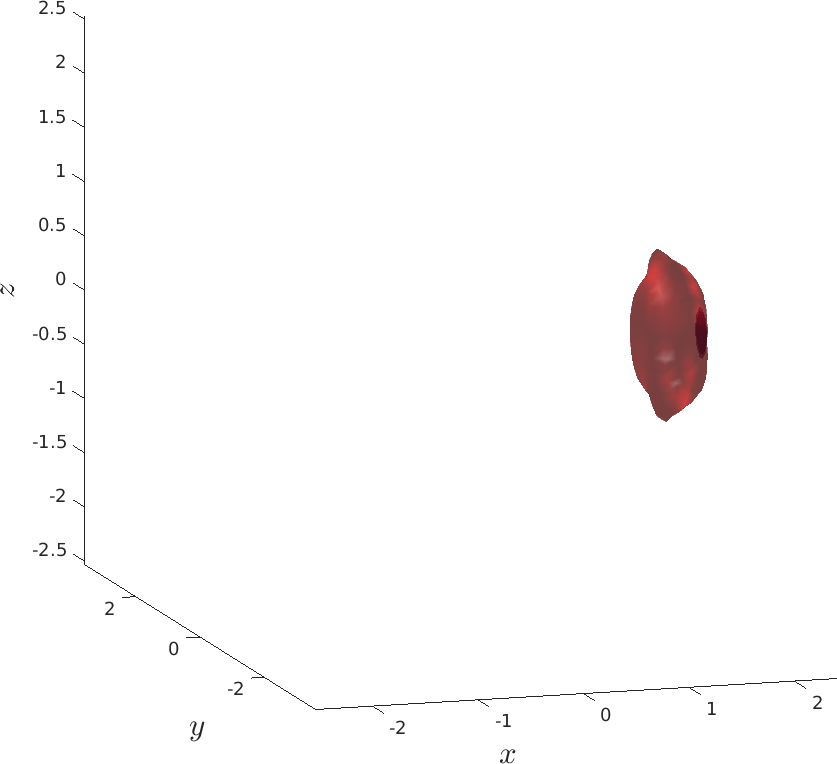}
\includegraphics[width=\widthone\linewidth]{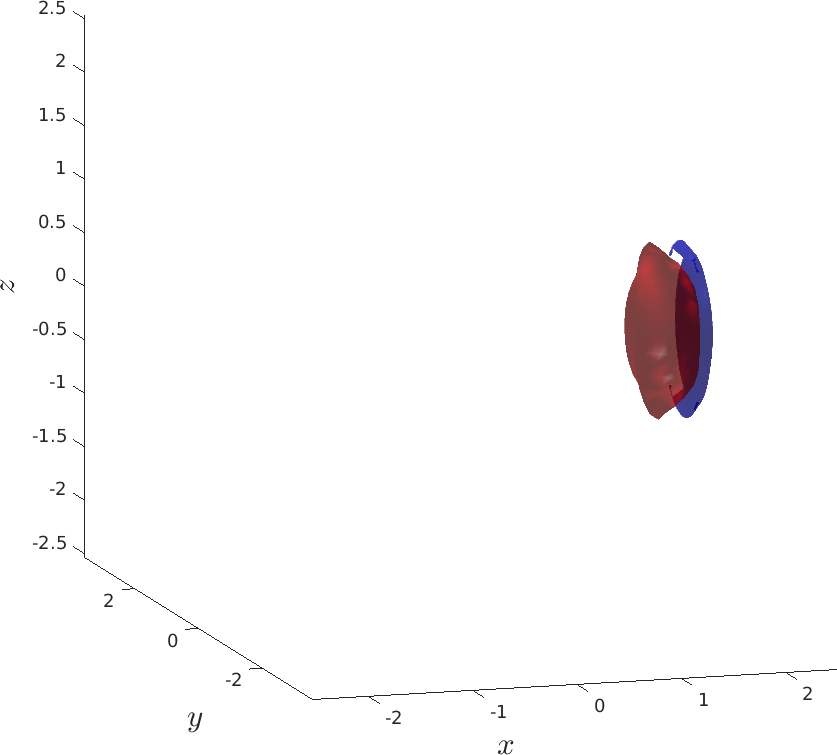}
\includegraphics[width=\widthone\linewidth]{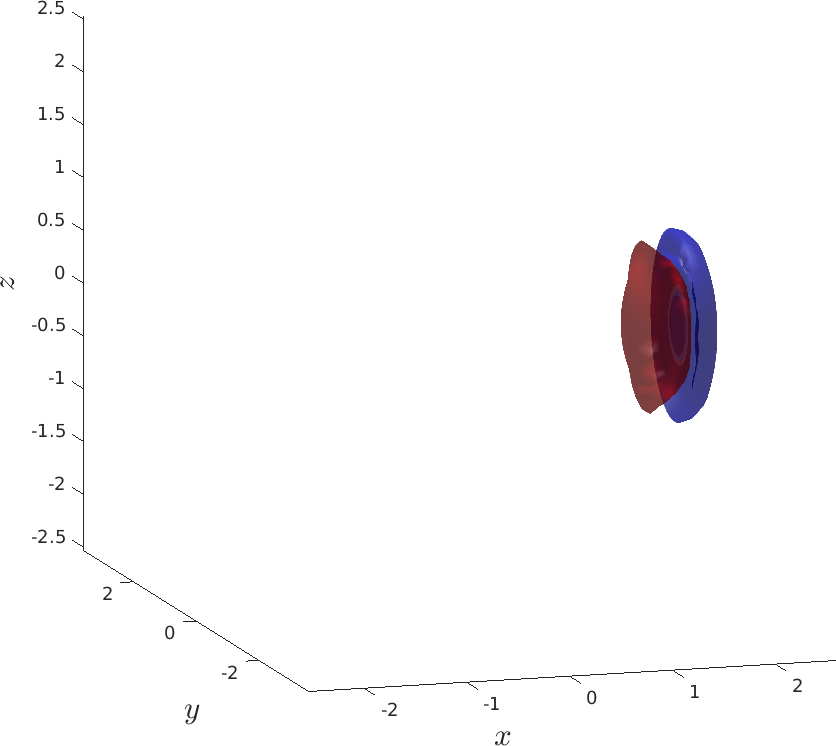}
\includegraphics[width=\widthone\linewidth]{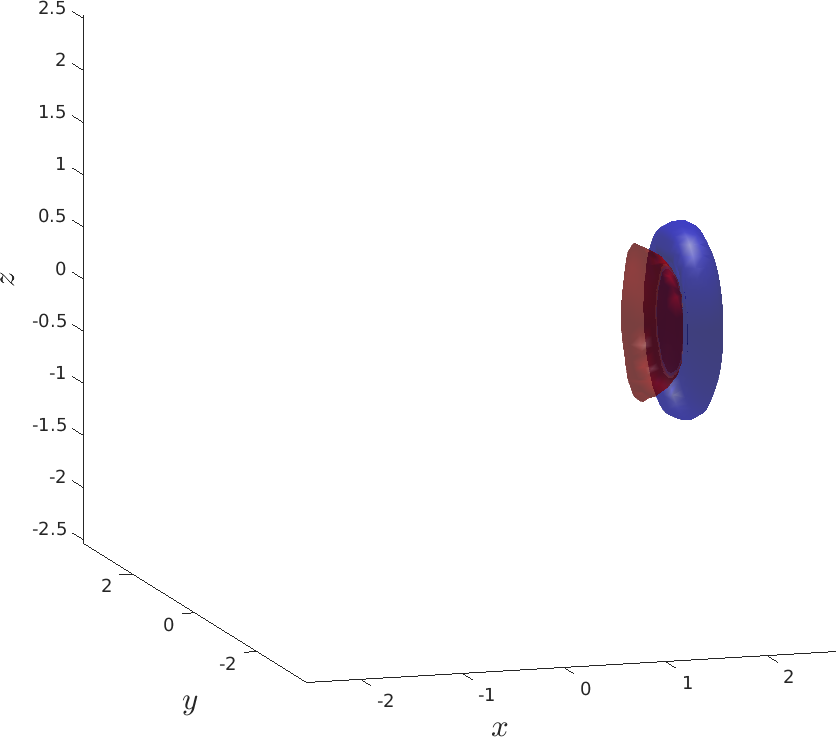}}}
\caption{Numerical solution for the fluctuation fields $\df^{\rm glue}$, 
  $\dtheta$ and $\dvarphi$ for two $B=1$ Skyrmions (showing only the
  left-hand side of the bound state) for $\epsilon=0.01$ and
  separation distances $2a=(0,1,2,3,4,5,6)2h_x$, with $h_x\simeq 0.094$.
  (a) shows the field solutions with constant
  isosurfaces at $\frac14$ of their respective maximal values, where
  the column corresponds to the separation distance.
  (b) shows the corresponding NLO (red) and N$^2$LO (blue)
  perturbation energies at $\frac14$ of their respective minima
  (notice that their contributions are negative).
  The color coding of (a) is that positive
  $\df^{\rm glue}$ is shown with red, negative $\df^{\rm glue}$ with
  green, positive $\dtheta$ with yellow, negative $\dtheta$ with
  magenta, positive $\dvarphi$ with blue and negative $\dvarphi$ with
  orange isosurfaces.
  In this figure $c_2=1$, $c_4=8$, $c_6=\tfrac12$, $\mu=1$,
  $m_\pi=3$, $R=(3\pi)^{\frac13}$ and $(s,p)=(1,2)$.
}
\label{fig:binding001}
\end{figure}

\def\widthtwo{0.133}
\begin{figure}[!htp]
\centering
\subfloat[]{\mbox{
\includegraphics[width=\widthtwo\linewidth]{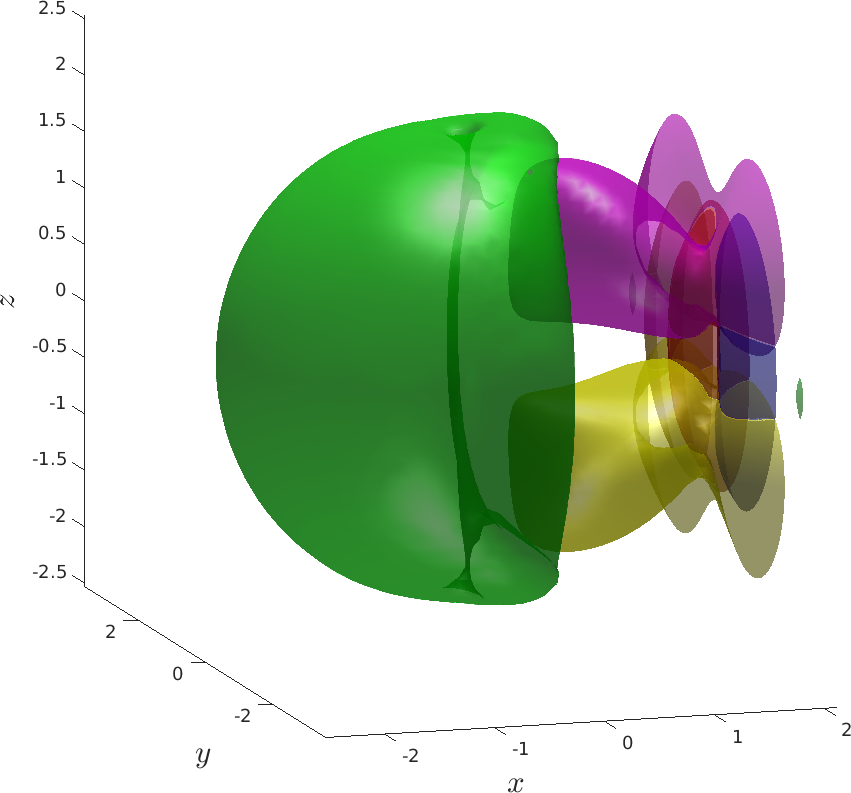}
\includegraphics[width=\widthtwo\linewidth]{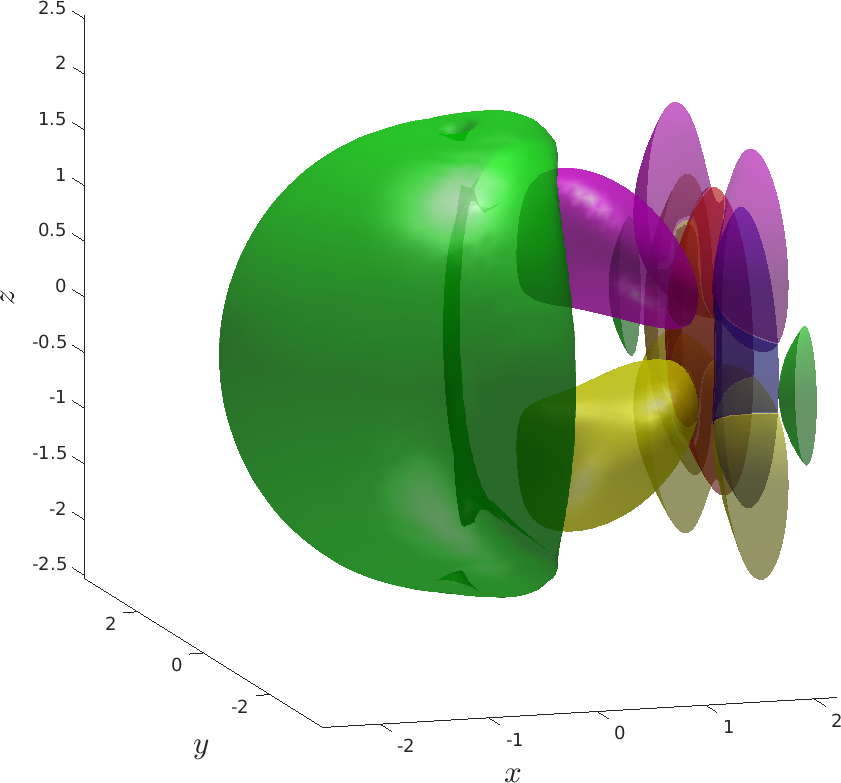}
\includegraphics[width=\widthtwo\linewidth]{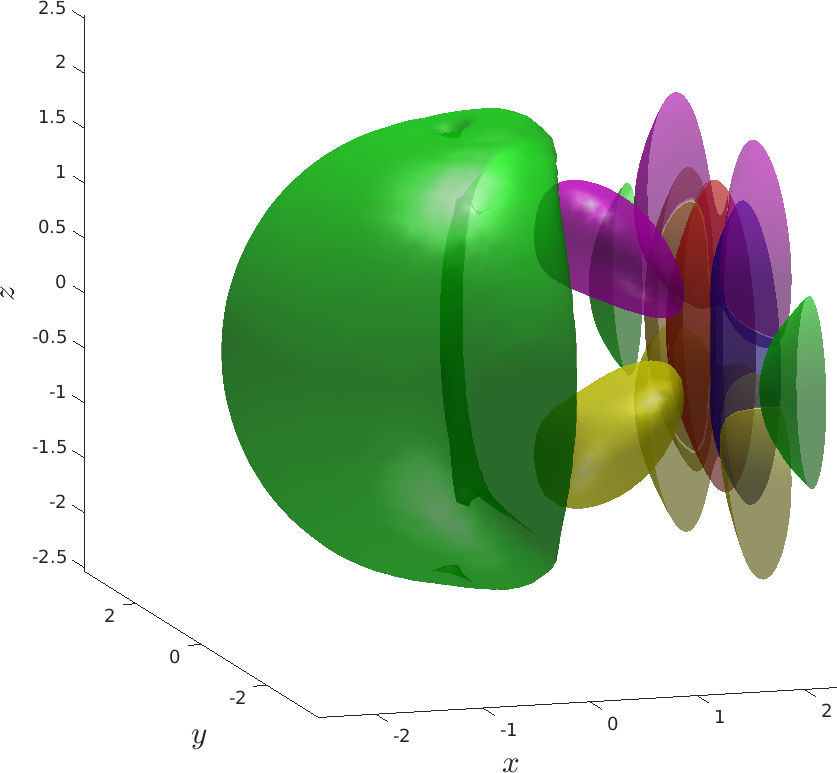}
\includegraphics[width=\widthtwo\linewidth]{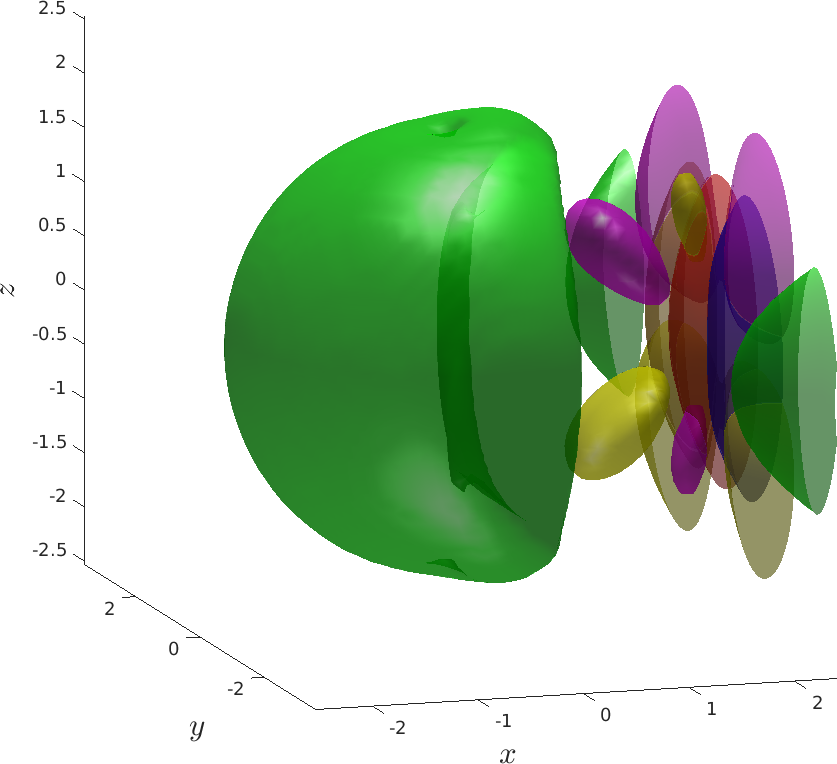}
\includegraphics[width=\widthtwo\linewidth]{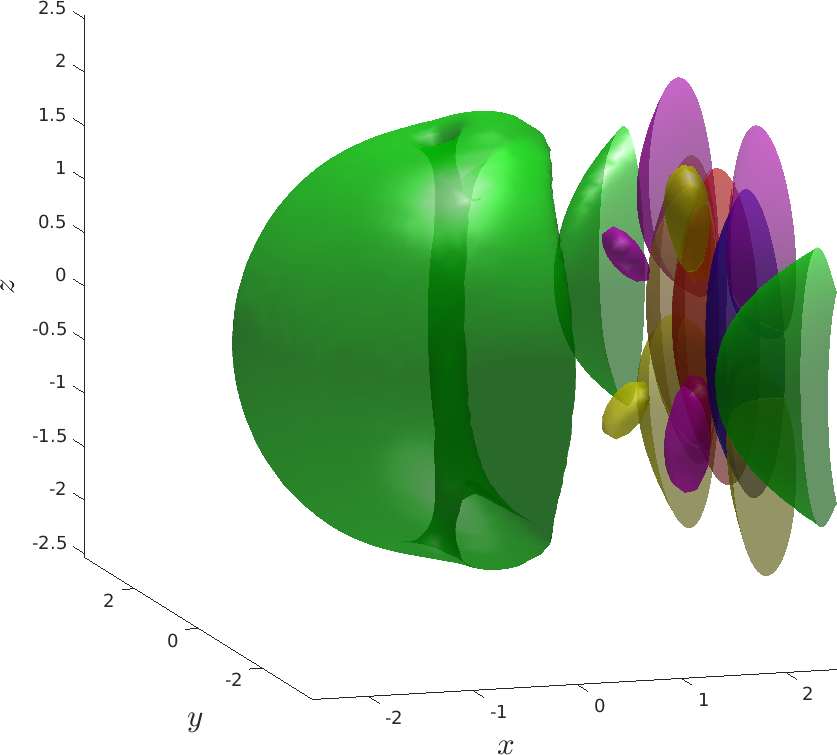}
\includegraphics[width=\widthtwo\linewidth]{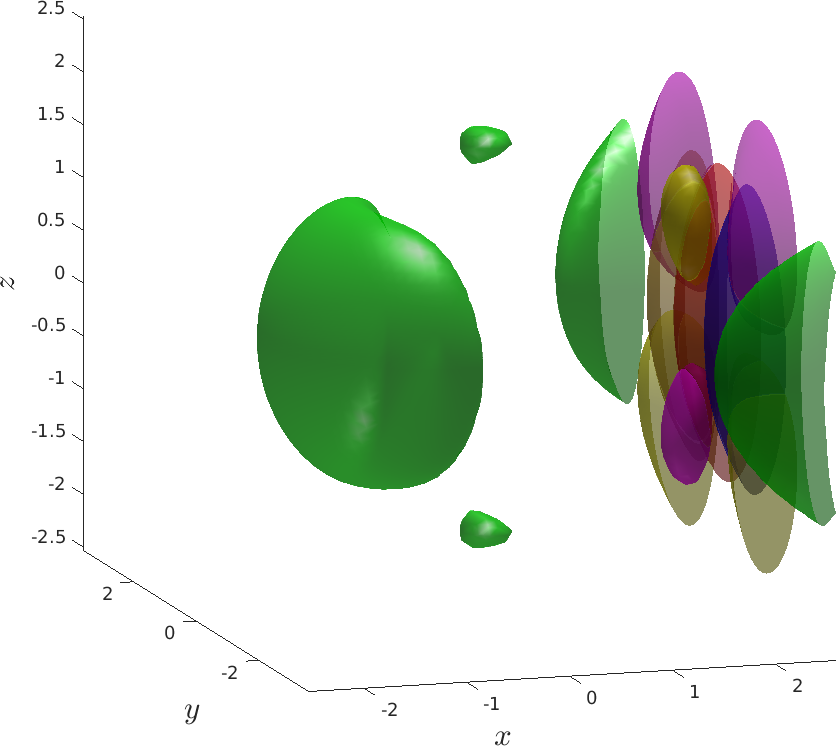}
\includegraphics[width=\widthtwo\linewidth]{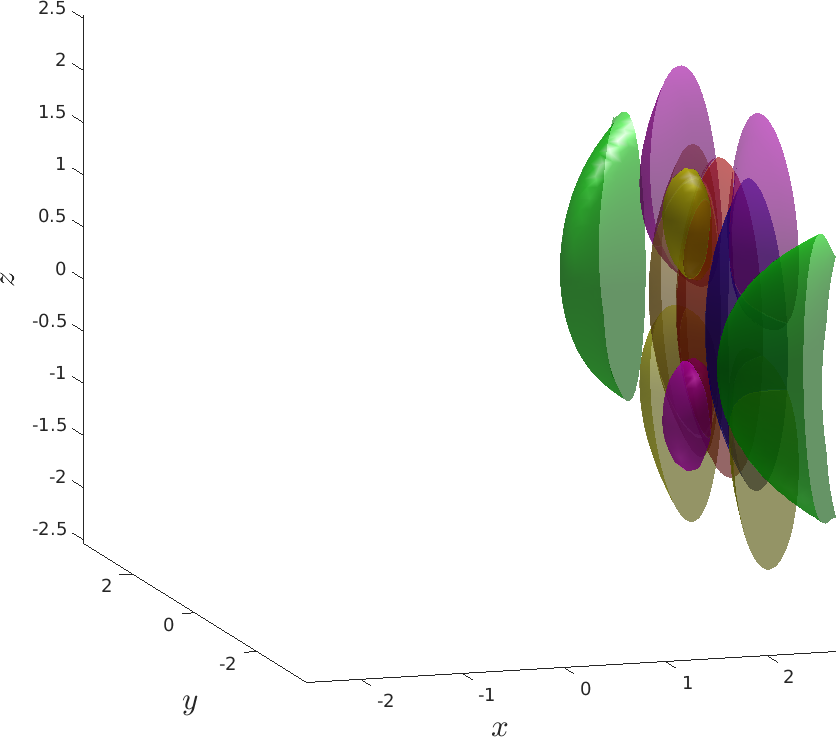}}}\\
\subfloat[]{\mbox{
\includegraphics[width=\widthtwo\linewidth]{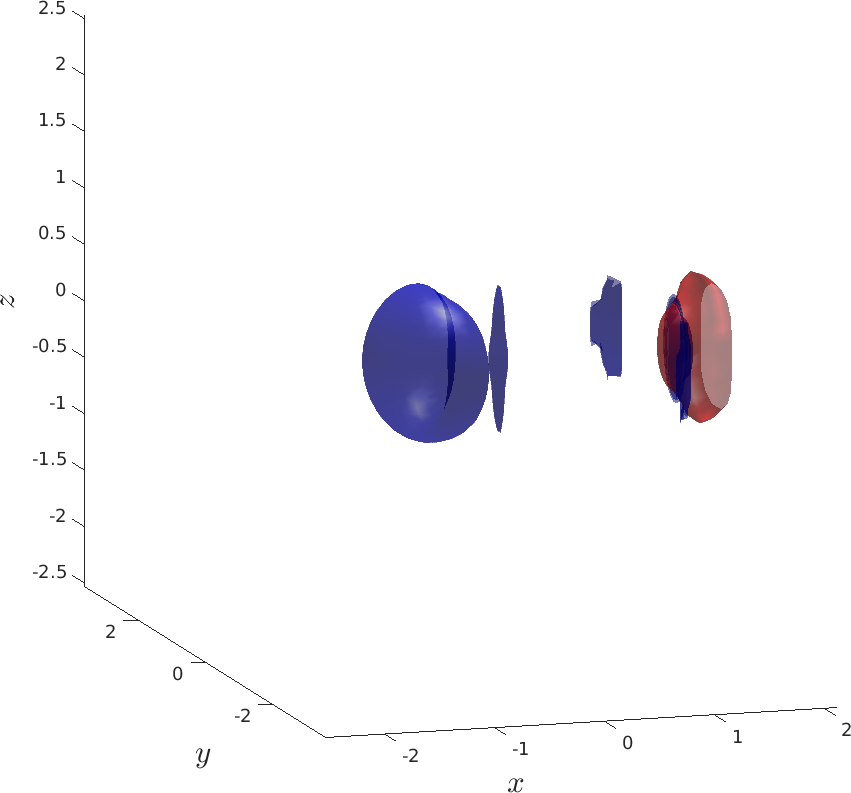}
\includegraphics[width=\widthtwo\linewidth]{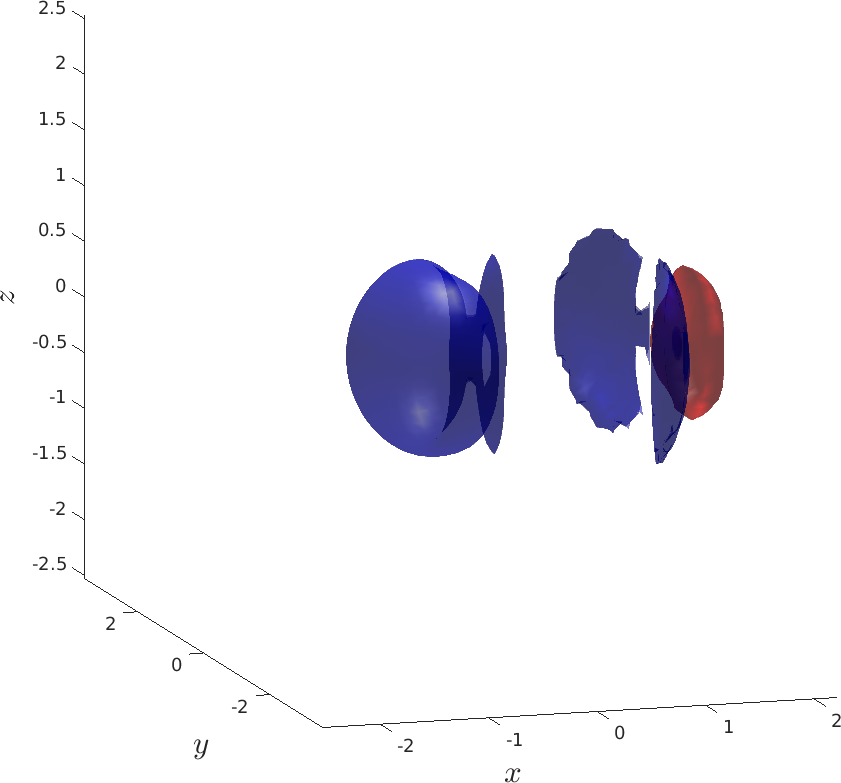}
\includegraphics[width=\widthtwo\linewidth]{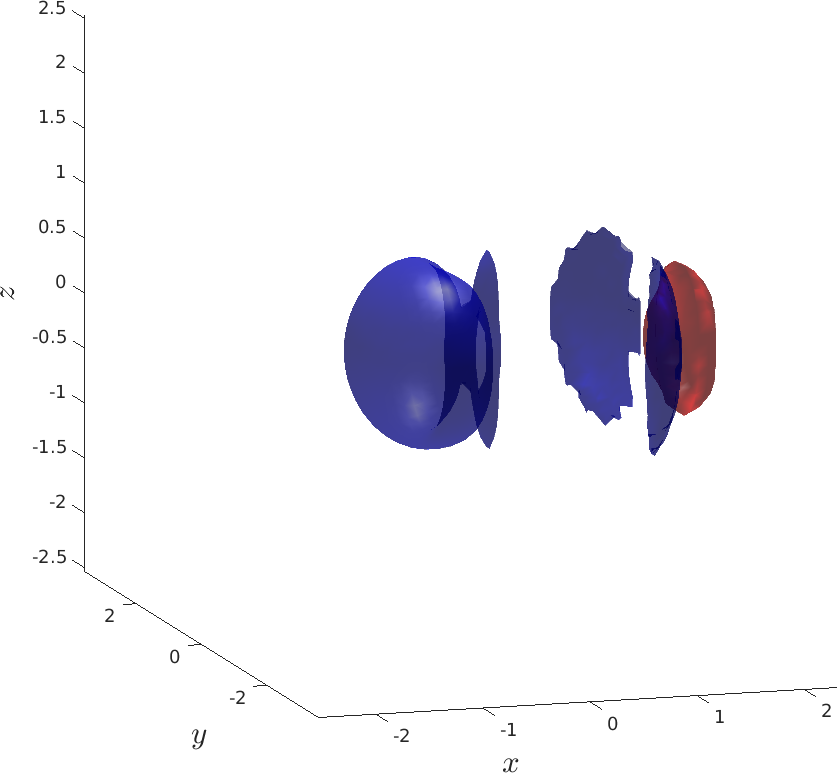}
\includegraphics[width=\widthtwo\linewidth]{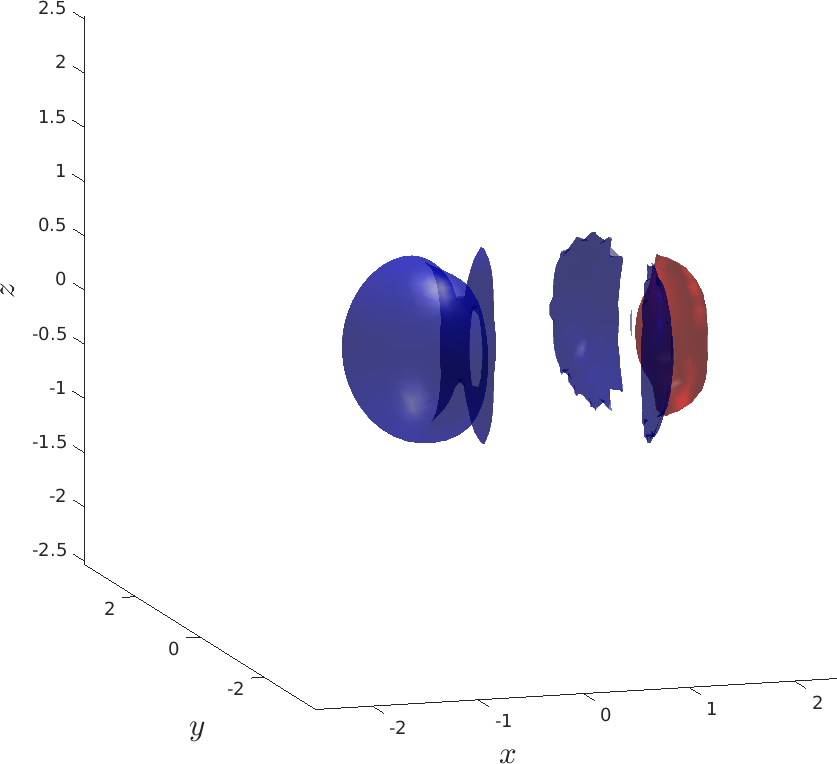}
\includegraphics[width=\widthtwo\linewidth]{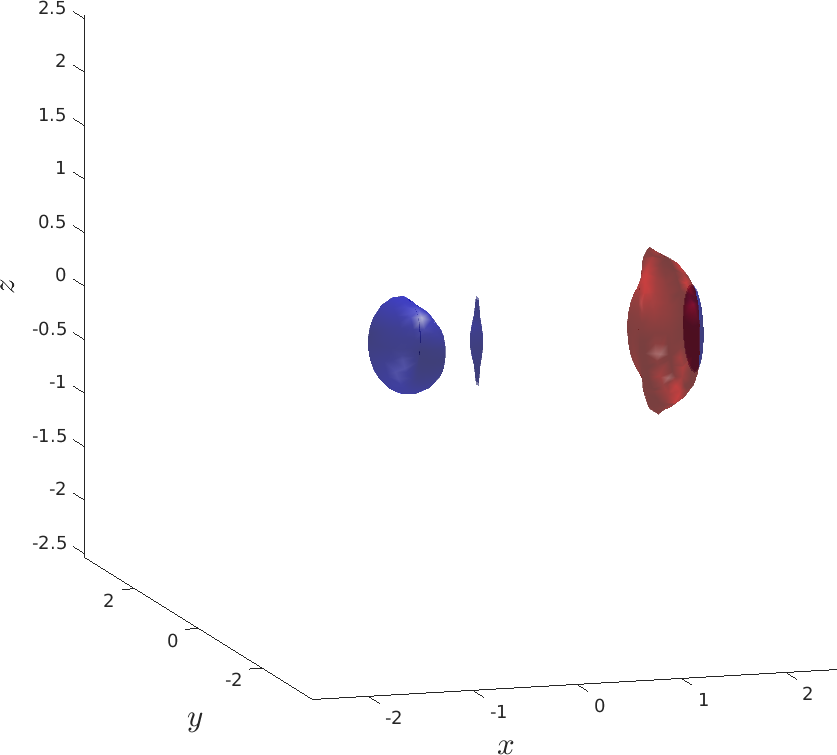}
\includegraphics[width=\widthtwo\linewidth]{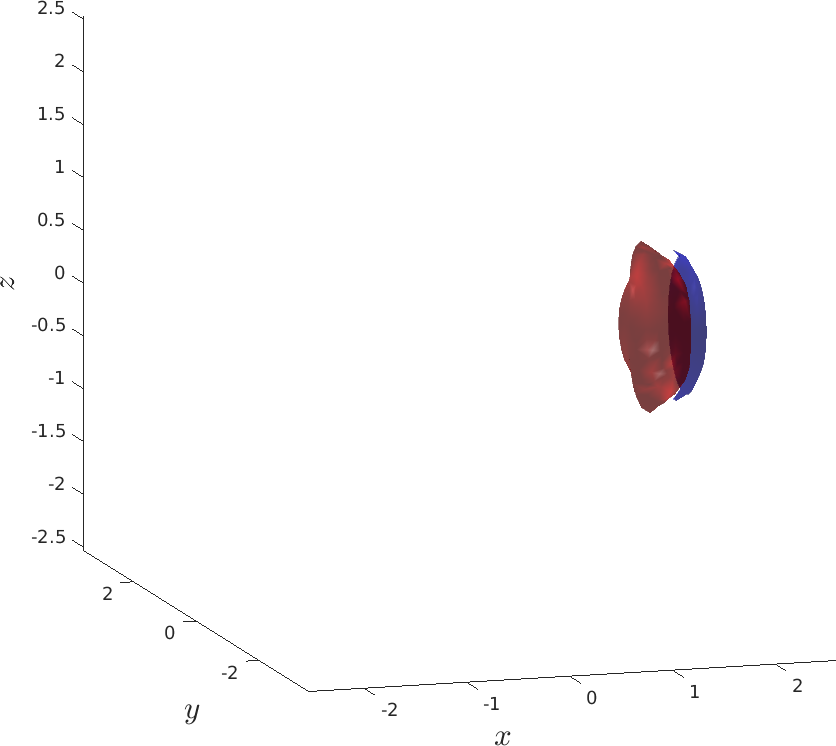}
\includegraphics[width=\widthtwo\linewidth]{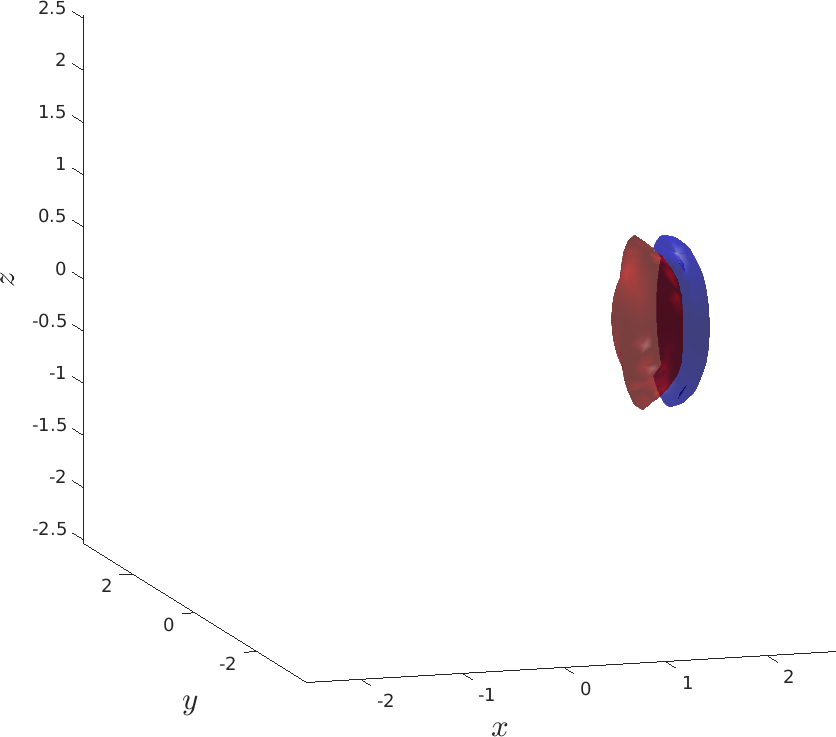}}}
\caption{Numerical solution for the fluctuation fields $\df^{\rm glue}$, 
  $\dtheta$ and $\dvarphi$ for two $B=1$ Skyrmions (showing only the
  left-hand side of the bound state) for $\epsilon=0.0428$ and
  separation distances $2a=(0,1,2,3,4,5,6)2h_x$, with $h_x\simeq 0.094$.
  For further details, see the caption of fig.~\ref{fig:binding001}.
}
\label{fig:binding00428}
\end{figure}

\begin{figure}[!htp]
\centering
\mbox{
\includegraphics[width=0.32\linewidth]{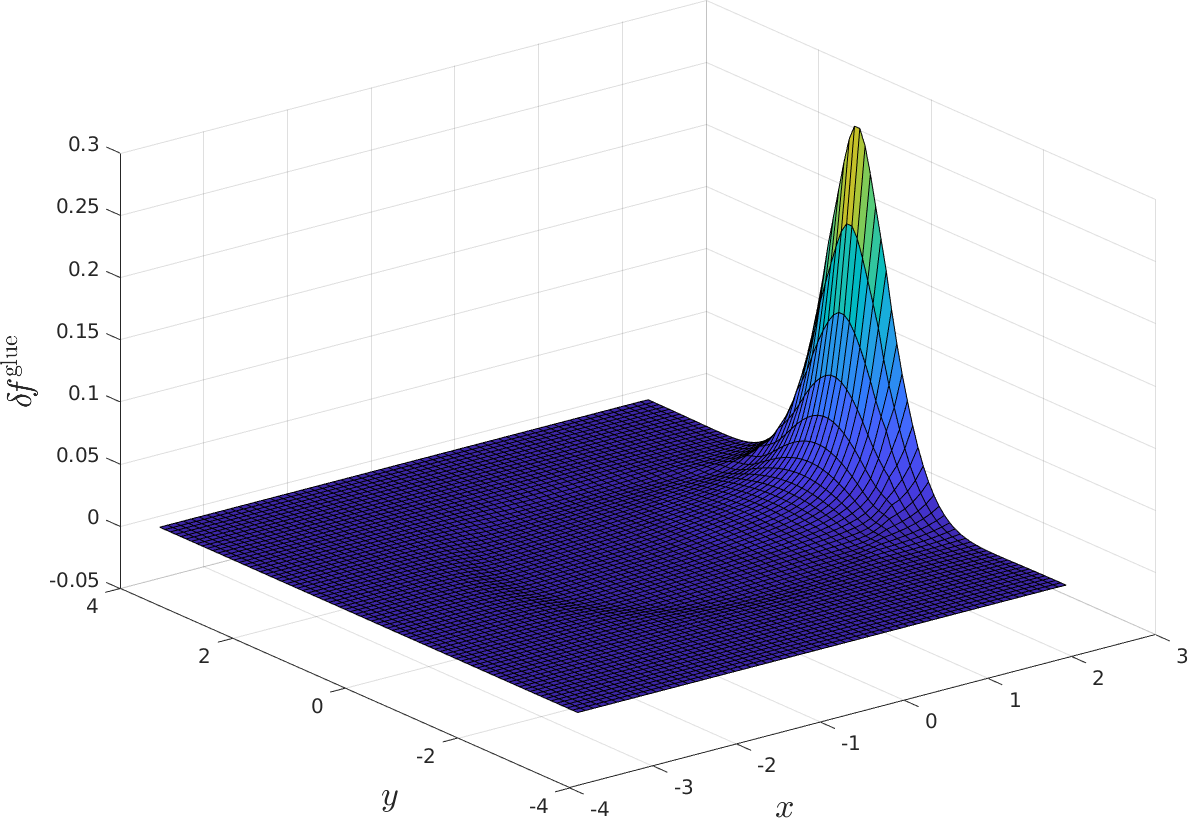}
\includegraphics[width=0.32\linewidth]{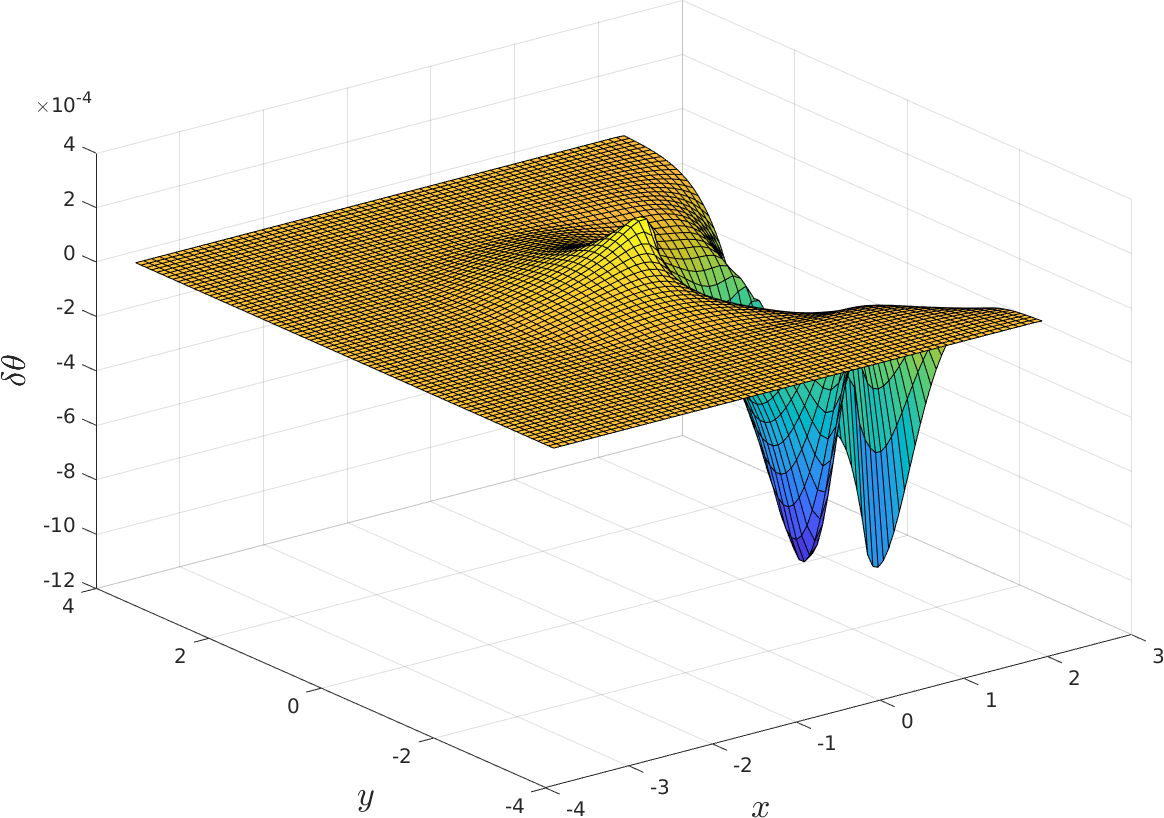}
\includegraphics[width=0.32\linewidth]{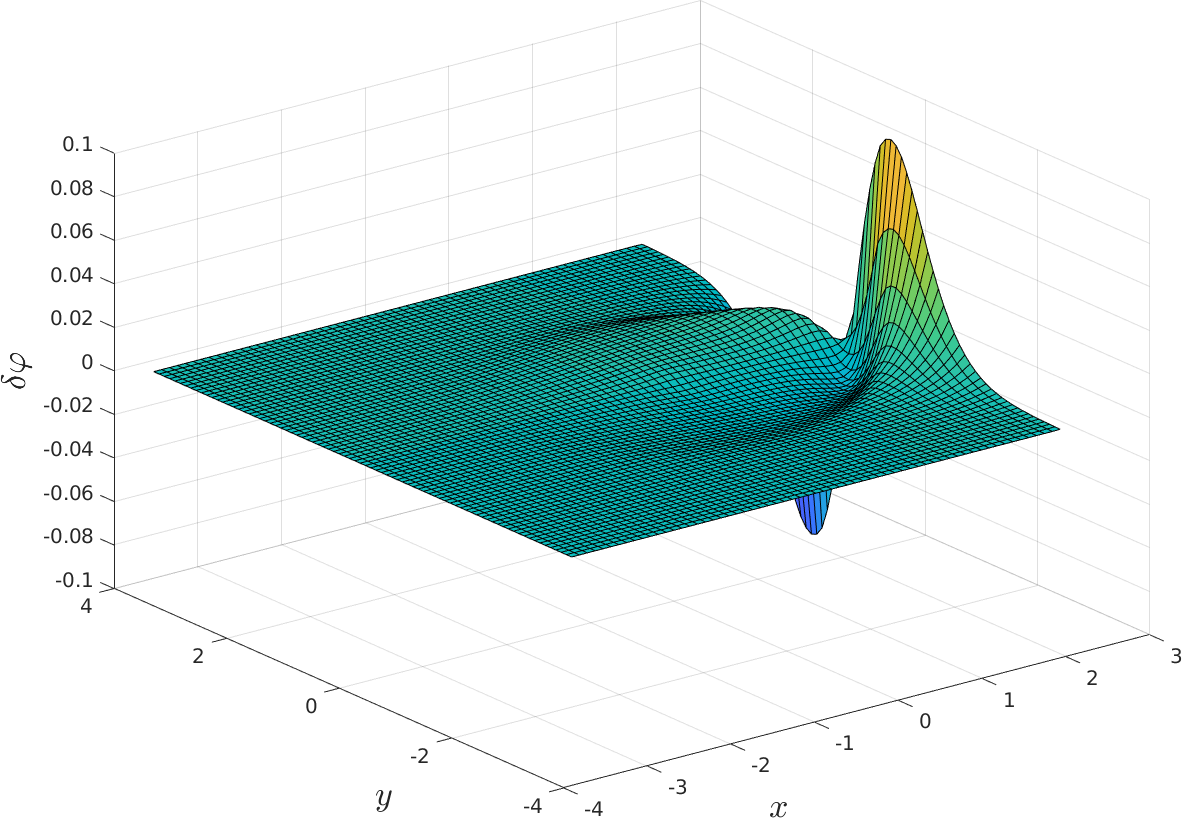}}
\caption{Numerical solution for the fluctuation fields $\df^{\rm glue}$
  $\dtheta$ and $\dvarphi$ for the left half of a bound state of two
  $B=1$ Skyrmions for $\epsilon=0.01$ and vanishing separation
  distance $2a\simeq0$ in the $(x,y)$-plane at $z=0$.
}
\label{fig:dfields}
\end{figure}

We will now solve the coupled PDEs \eqref{eq:eom_general_fluc} with
the cusp condition taken into account by means of splitting the field 
(valid for the linearized equation) \eqref{eq:split}, vanishing
boundary conditions for the fluctuations at infinity and finally the
gluing conditions \eqref{eq:glue1}-\eqref{eq:glue3} at $x=0$
(i.e.~midway between the two spherical $B=1$ compactons), see
fig.~\ref{fig:BC}.
The results are shown in figs.~\ref{fig:binding001} and
\ref{fig:binding00428} for $\epsilon=0.01$ and $\epsilon=0.0428$,
respectively.
The top row of each figure shows the isosurfaces of the fluctuation
fields $\df^{\rm glue}$, $\dtheta$ and $\dvarphi$ at positive
(negative) quarter-maximum (-minimum) levelsets with red, yellow and
blue (green, magenta and orange), respectively.
For large separation distances (right-most panels) the fluctuations
are localized at the gluing boundary ($x=0$), whereas for small or
vanishing separation distances (left-most panels) the fluctuation
fields are turned on throughout the compacton volume.
In particular, $\df$ makes a shallow but negative shell near the
compacton border, whereas $\dtheta$ becomes a dipole with a positive
and negative blob induced in the compacton volume for small separation
distances ($2a\lesssim1$).
The bottom row of the figures shows the isosurfaces of negative energy
density at a quarter of the minimum value for the NLO and N$^2$LO
contributions to the energy from the fluctuation fields $\df^{\rm glue}$, 
$\dtheta$, and $\dvarphi$ with red and blue colors, respectively.
The NLO energy contribution is dominant and is responsible for the
binding of the two $B=1$ Skyrmions and is seen to be localized near
the gluing boundary ($x=0$).
Fig.~\ref{fig:dfields} shows the fluctuation fields in the
$(x,y)$-plane at $z=0$ for the case of vanishing separation distance
$2a\simeq0$.
The nontrivial behavior of the fluctuation fields is only induced by
the gluing conditions at $x=0$ (right-most part of each panel), but
spreads a small perturbation throughout the compacton volume.

\begin{figure}[!htp]
\centering
\mbox{\includegraphics[width=0.49\linewidth]{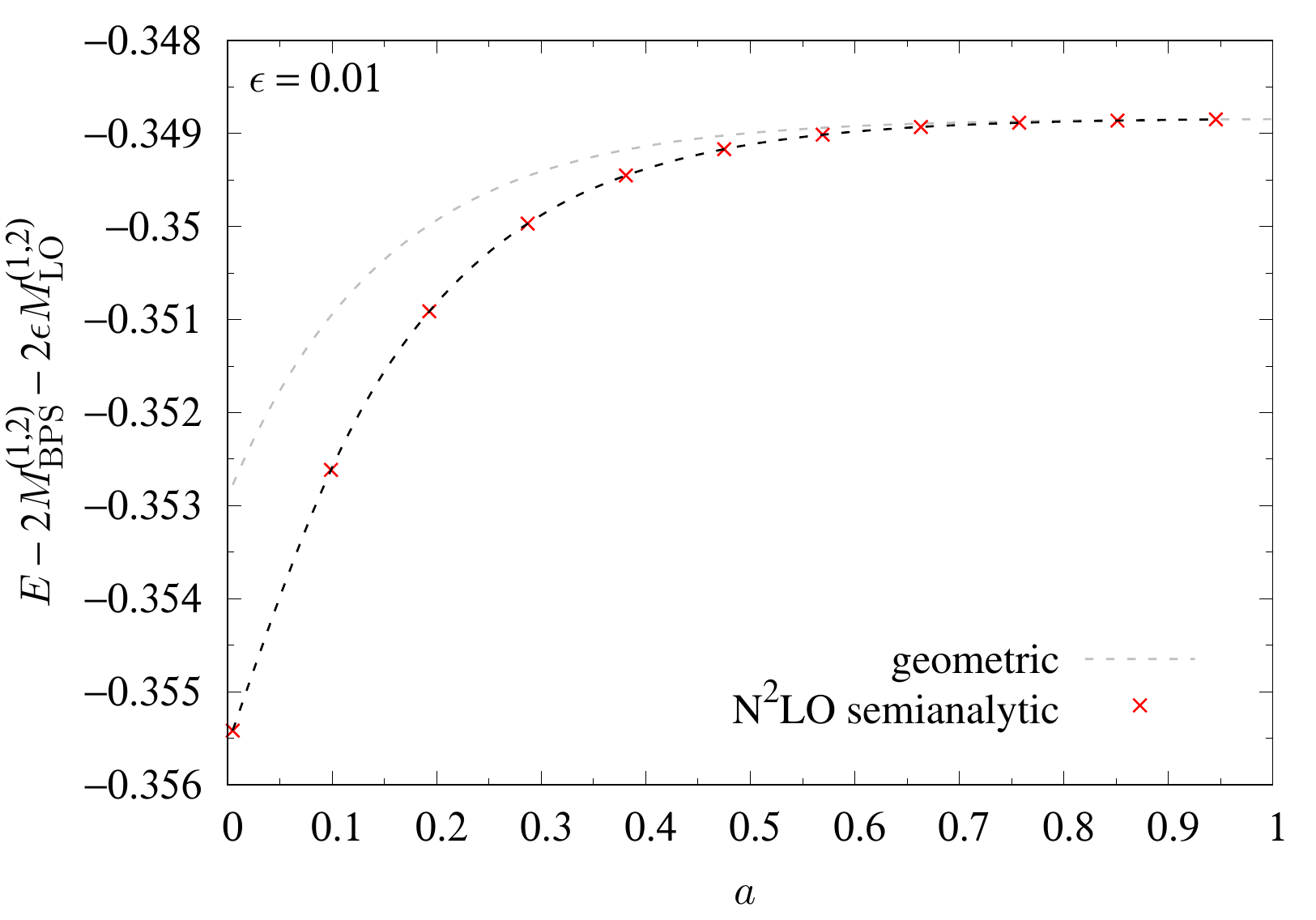}
  \includegraphics[width=0.49\linewidth]{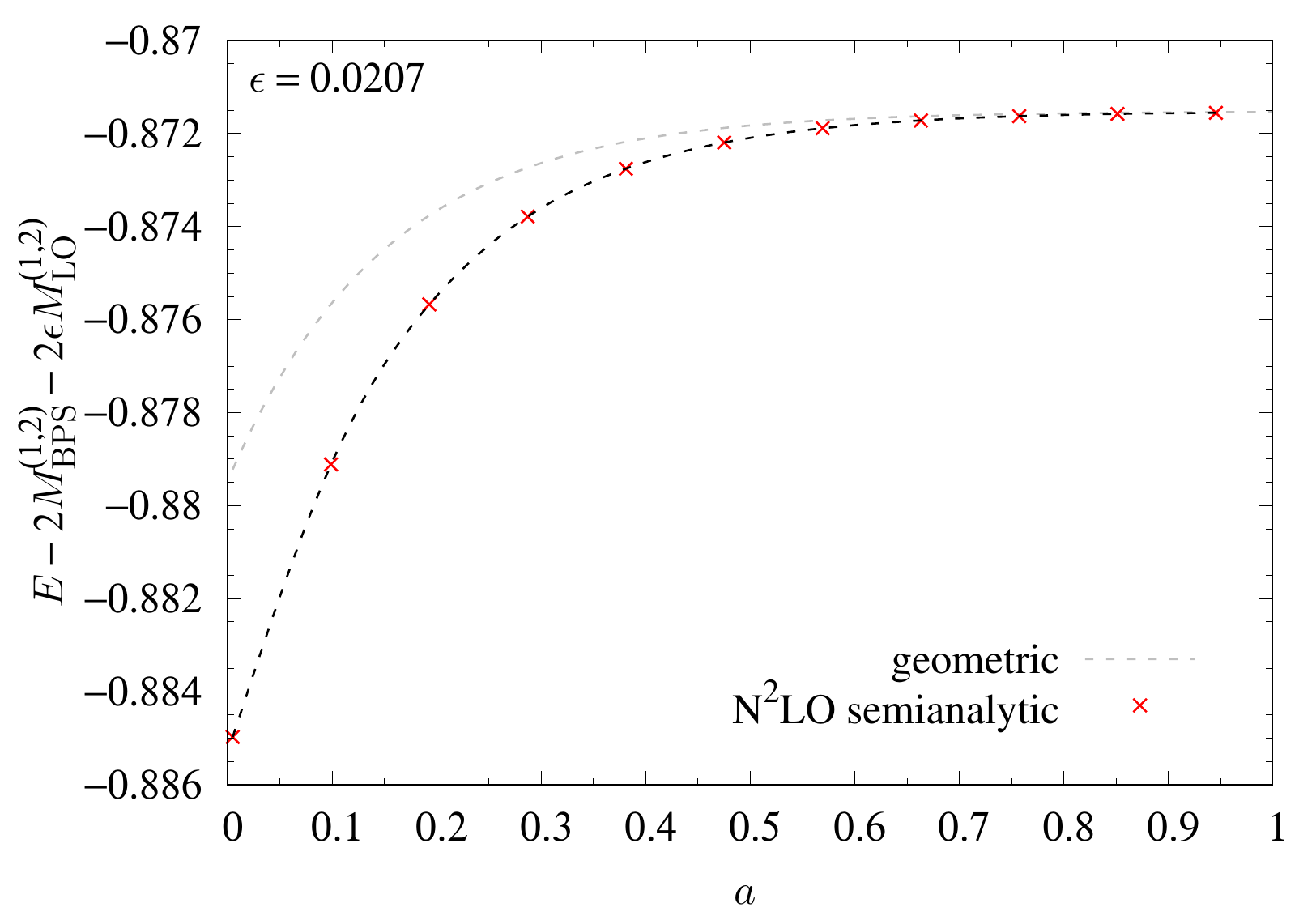}}
\mbox{\includegraphics[width=0.49\linewidth]{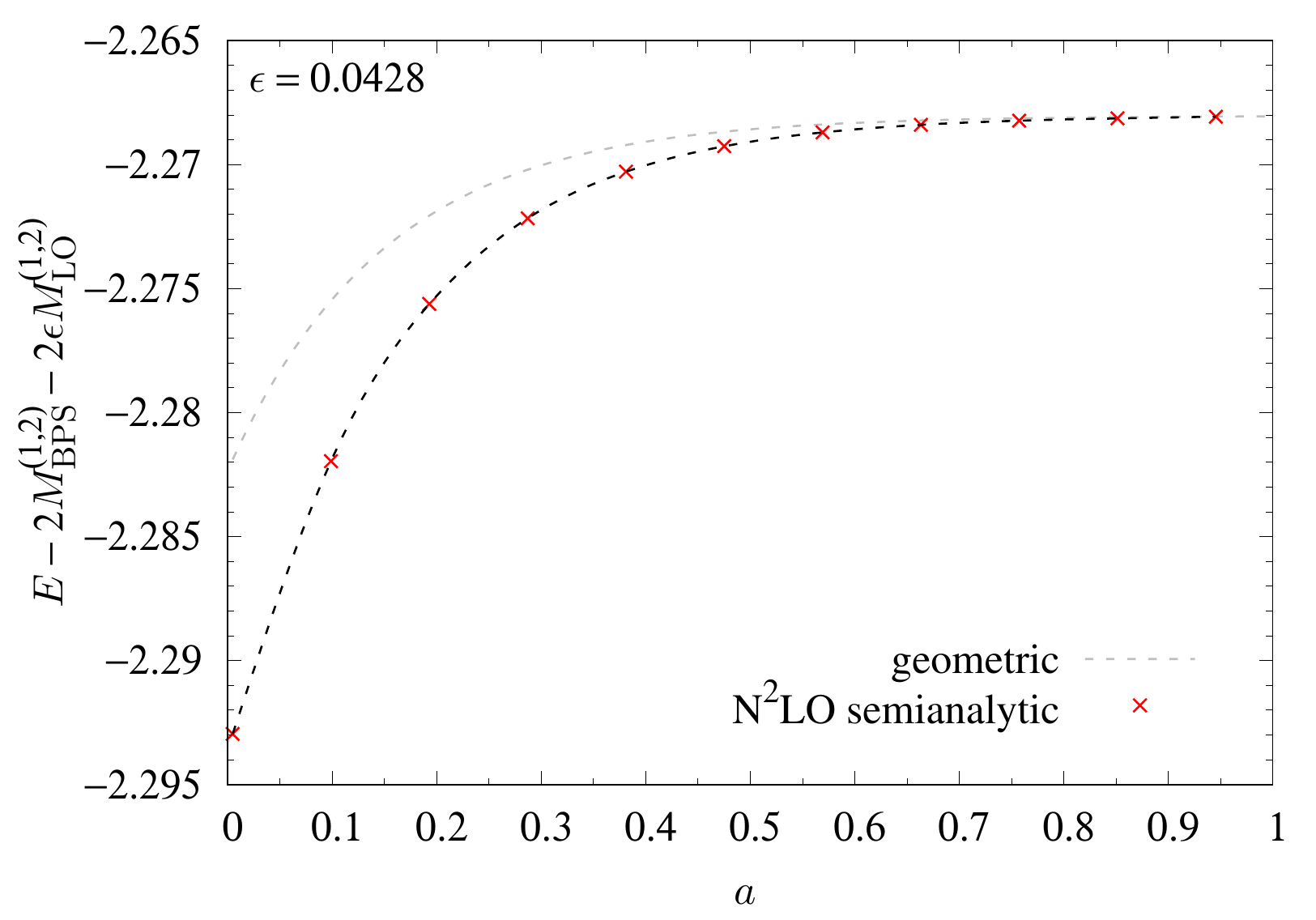}
  \includegraphics[width=0.49\linewidth]{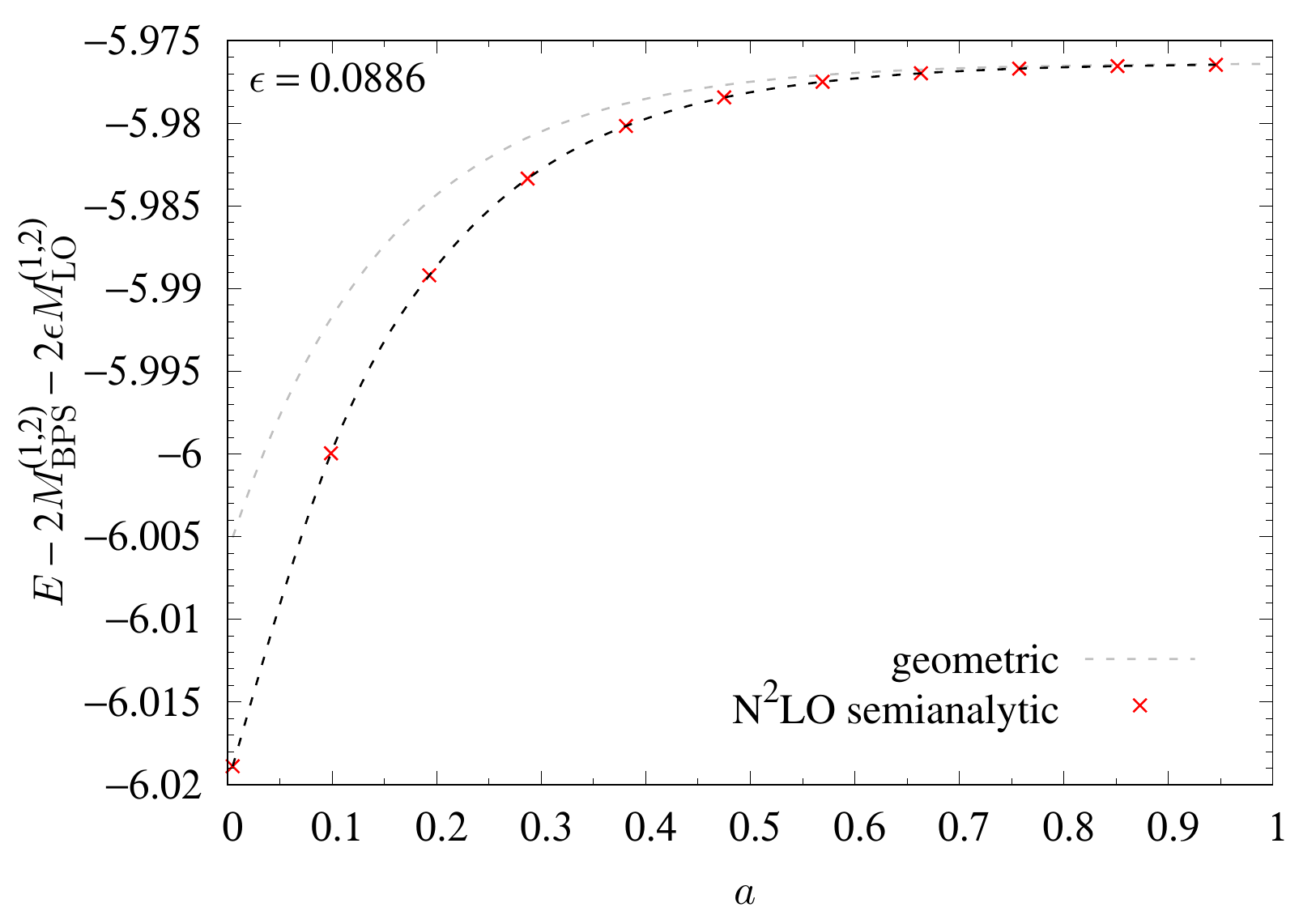}}
\caption{The N$^2$LO corrections to the energy for two $B=1$ Skyrmions
  with separation distance $2a$ and
  $\epsilon=0.01,0.0207,0.0428,0.0886$.
  Asymptotically, i.e.~for large values of $a$, the N$^2$LO
  corrections to the energy become exactly two times those of
  sec.~\ref{sec:axial_perturb}.
  When $a$ tends to zero, the binding energy increases monotonically
  with a maximum at $a=0$, for all values of the perturbation
  parameter $\epsilon$.
  A geometric effect of the binding energy lies simply in part of the
  tail energy of the $B=1$ solutions are cut at the mirror surface
  $x=0$ so as not to over count the $B=1$ energies.
  This reduction of the energy is displayed as geometric with a gray
  dashed line.
  The numerical computations are always slightly below the geometric
  corrections.
  In this figure $c_2=1$, $c_4=8$, $c_6=\tfrac12$, $\mu=1$,
  $m_\pi=3$, $R=(3\pi)^{\frac13}$ and $(s,p)=(1,2)$.
  }
\label{fig:binding_a}
\end{figure}

We are now ready to compute the N$^2$LO energies for the bound state
of two $B=1$ compactons using the energy \eqref{eq:E2general} with the 
fluctuation fields $\df^{\rm glue}$, $\dtheta$ and $\dvarphi$ defined
in eq.~\eqref{eq:Delta_general}.
We have already placed the two $B=1$ Skyrmions in the attractive
channel by means of the Dirichlet boundary condition on $\Phi^2$ in
eq.~\eqref{eq:gluing_condition_Phi}.
We furthermore know that at large separation distances, the attractive
force between the two $B=1$ Skyrmions is exponentially suppressed (due
to the pion mass term), so the minimum of the energy must be at a
finite separation.
In the small $\epsilon\ll1$ or near-BPS limit, we expect that the
minimum of the energy of the bound state of two compactons occurs at
zero separation, as was confirmed both semi-analytically within the
$\epsilon$-expansion scheme as well as numerically for baby-Skyrmions
(compactons) in ref.~\cite{Gudnason:2020tps}.
In order to confirm that this is also the case for $B=1$ compacton
(Skyrmions) in the case of our model choice, we compute the N$^2$LO
energies at different separations $2a$.
The result is shown in fig.~\ref{fig:binding_a} and establishes
numerically that the minimum of the N$^2$LO energy occurs at vanishing
separation of the compactons (i.e.~$2a=0$) for all presented values of
$\epsilon$, i.e.~$\epsilon=0.01,0.0207,0.0428,0.0886$.
The figure also shows the geometric binding energy, which is simply
computed by cutting off the tail contribution to the N$^2$LO energy at
$x=0$ (so as not to over count the two tails of the two compactons).
It is observed that about half of the binding energy is actually
geometric for the presented range of $\epsilon$.

\begin{figure}[!htp]
\centering
\mbox{\subfloat[]{\includegraphics[width=0.49\linewidth]{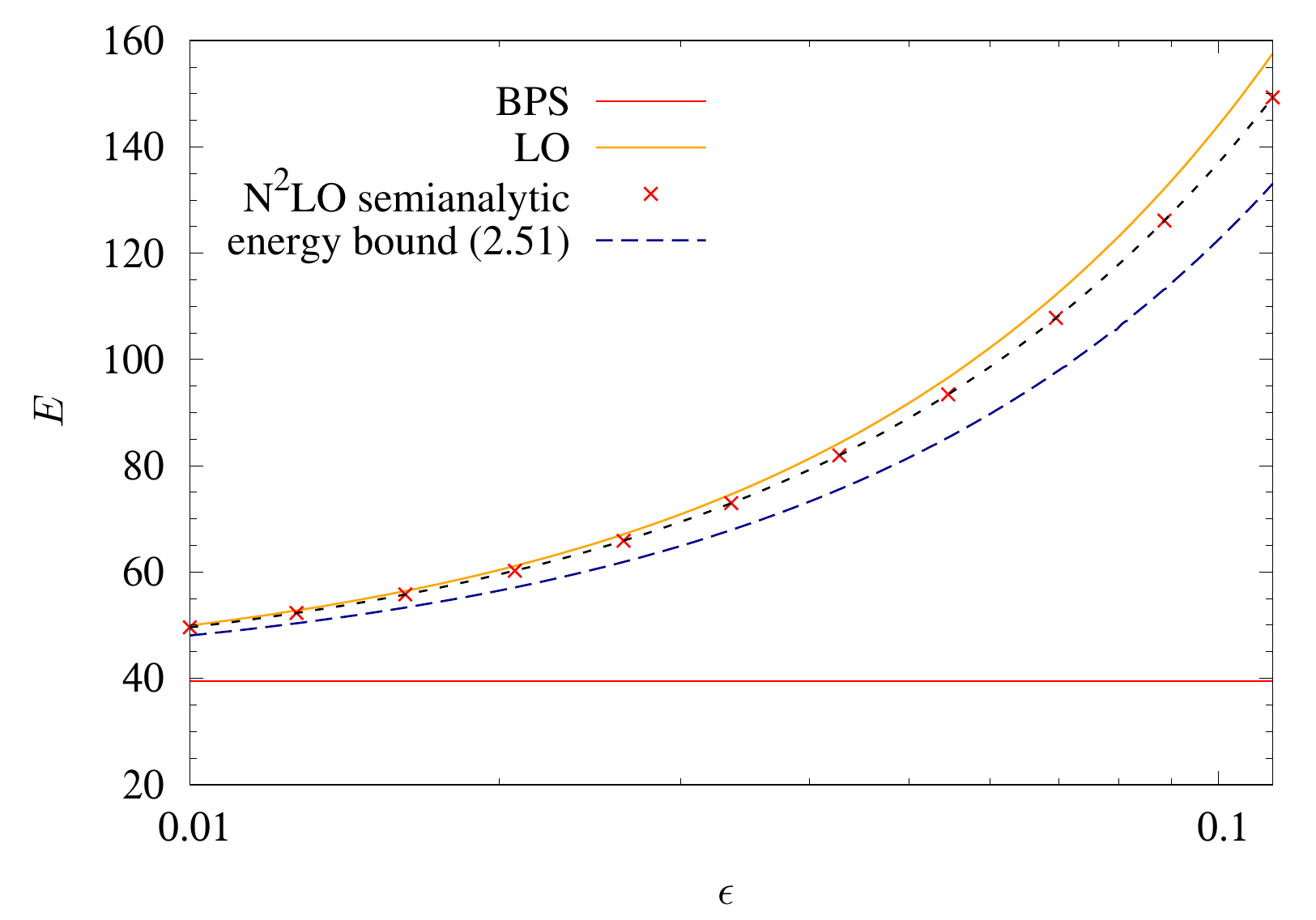}}
  \subfloat[]{\includegraphics[width=0.49\linewidth]{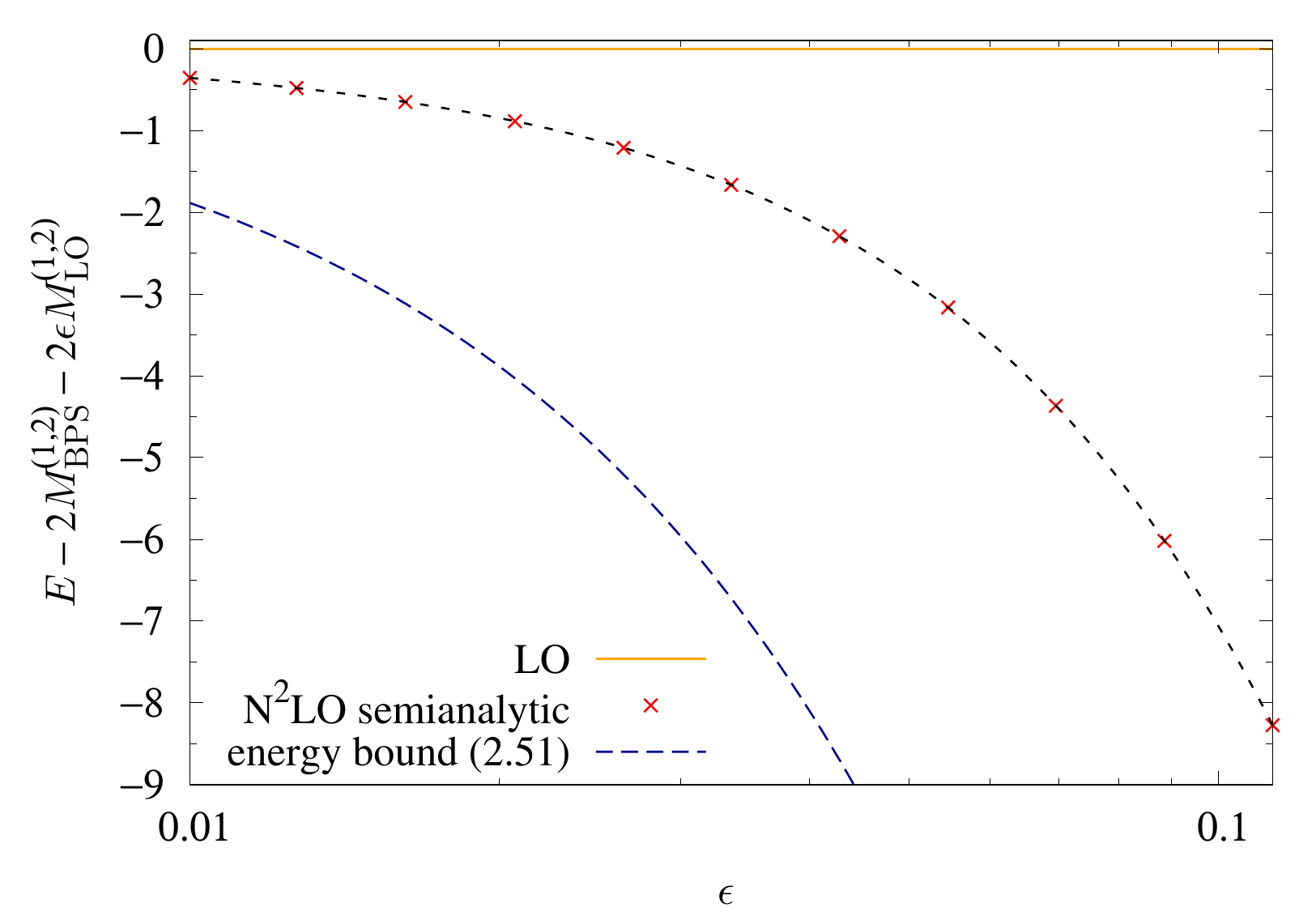}}}
\caption{The (a) energy and (b) N$^2$LO correction to the energy for
  two $B=1$ Skyrmions with separation distance $2a=0$ (they touch in
  one point) as a function of $\epsilon$.
  The BPS energy is shown with a red solid line, the LO correction is
  shown with an orange solid line and the numerically computed N$^2$LO
  corrections are shown with red crosses (linked with black dashed
  lines for ease of reading the figure).
  The energy bound \eqref{eq:bound} is shown with a blue-dashed line.
  In this figure $c_2=1$, $c_4=8$, $c_6=\tfrac12$, $\mu=1$,
  $m_\pi=3$, $R=(3\pi)^{\frac13}$ and $(s,p)=(1,2)$.
}
\label{fig:binding_epsilon}
\end{figure}

The N$^2$LO energy is shown in fig.~\ref{fig:binding_epsilon} as a
function of $\epsilon$ for vanishing separation distance ($2a=0$).
The N$^2$LO energy is below the LO energy as expected and is above the
energy bound \eqref{eq:bound} as it must be.

\begin{figure}[!htp]
\centering
\mbox{\subfloat[]{\includegraphics[width=0.49\linewidth]{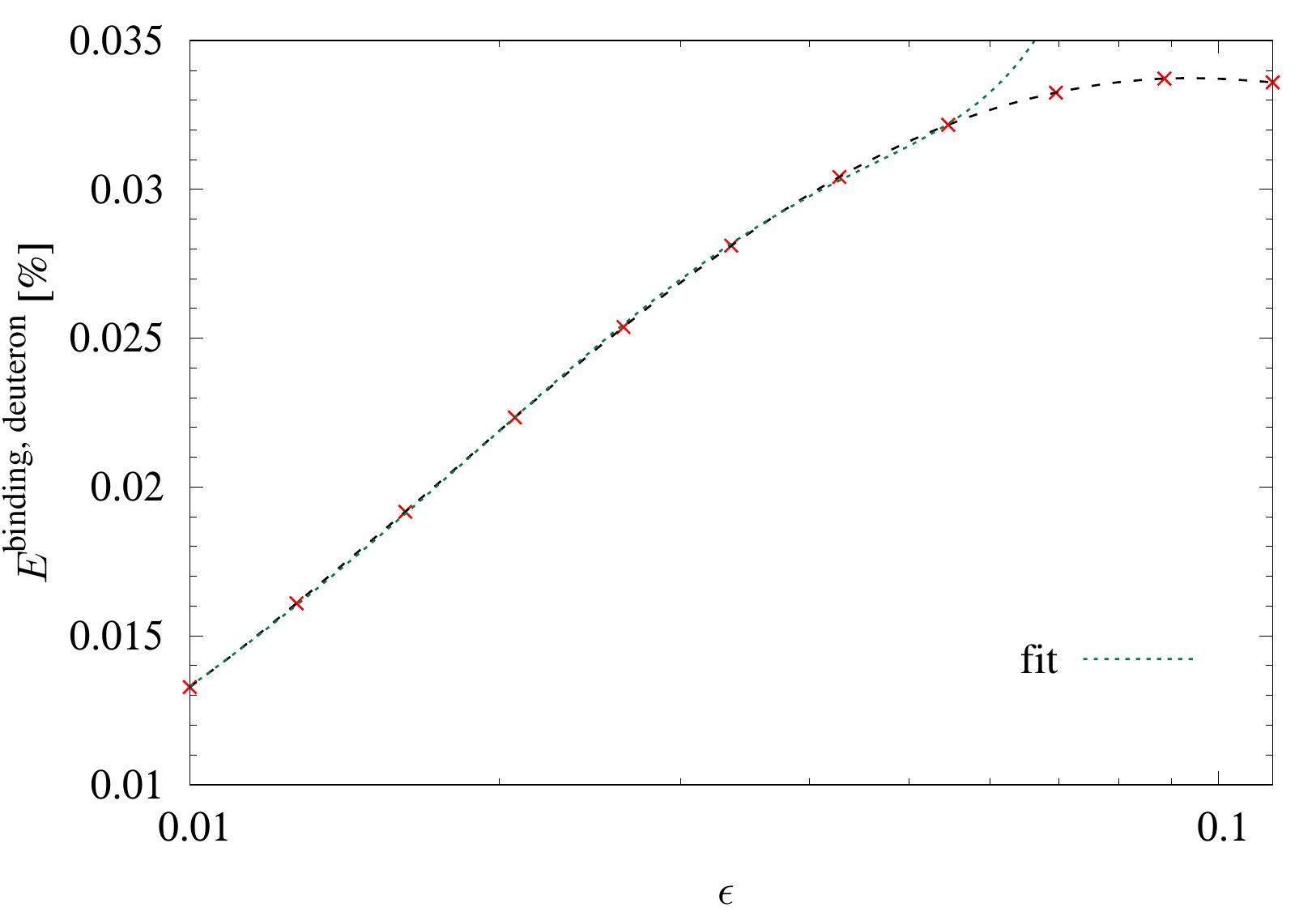}}}
\caption{The binding energy per nucleon in percent for two $B=1$
  Skyrmions, i.e.~the N$^2$LO energy at $a=\infty$ minus that at
  $a=0$, as a function of $\epsilon$.
  The physical binding energy of deuteron per nucleon for comparison 
  is $0.118\%$ \cite{Greene:1986vb}.
  In this figure $c_2=1$, $c_4=8$, $c_6=\tfrac12$, $\mu=1$,
  $m_\pi=3$, $R=(3\pi)^{\frac13}$ and $(s,p)=(1,2)$.
}
\label{fig:binding_deuteron}
\end{figure}

We can now extract the binding energy for the two $B=1$ Skyrmions
(compactons) as a function of $\epsilon$ by comparing the N$^2$LO
energies with those for infinite separations.
The result is shown in fig.~\ref{fig:binding_deuteron}.
A polynomial fit to the classical binding energies in percent yields
\beq
E^{\text{binding, deuteron}}
\simeq 1.605\epsilon -29.56\epsilon^2 + 200.8\epsilon^3.
\eeq
Notice that this fit contains a linear term in $\epsilon$, which is
expected to come from nonanalytic behavior of the solution to the
fluctuation fields.
Unfortunately, the model as calibrated and chosen in order to make the
cusp condition work, the spherical compactons being stable and the
tails to be rapidly decaying, does not provide quite the
phenomenological binding energy of the deuteron of $0.118\%$
\cite{Greene:1986vb} for the range of $\epsilon$ explored.
Clearly this is a crude comparison, we are not considering here
quantum corrections due to the spin, iso-spin rotation, iso-spin
breaking, together with the addition of the electric Coulomb
interaction, which should be included in the phenomenological nuclear 
energies.
It can furthermore be seen from the figure, that the binding energy
tends to a plateau instead continuing its increase, which we
interpret as loss of precision (validity) of the $\epsilon$-expansion
scheme.
This is most likely because $c_4=8$ and hence $\epsilon\simeq0.1129$
yields a Skyrme term coefficient of $\epsilon c_4\sim0.9$ which is no
longer perturbative.
Recall that $c_4\gg c_2R^2$ is needed for the spherical compacton to
be a stable minimum of the energy functional.
If this condition is not satisfied, the two compactons will merge and
form a torus, as is well known in the standard Skyrme model
\cite{Kopeliovich:1987bt,Manton:1987xf,Verbaarschot:1987au}.


\section{Physical units}\label{sec:physunits}

\begin{figure}[!htp]
\centering
\mbox{\subfloat[]{\includegraphics[width=0.49\linewidth]{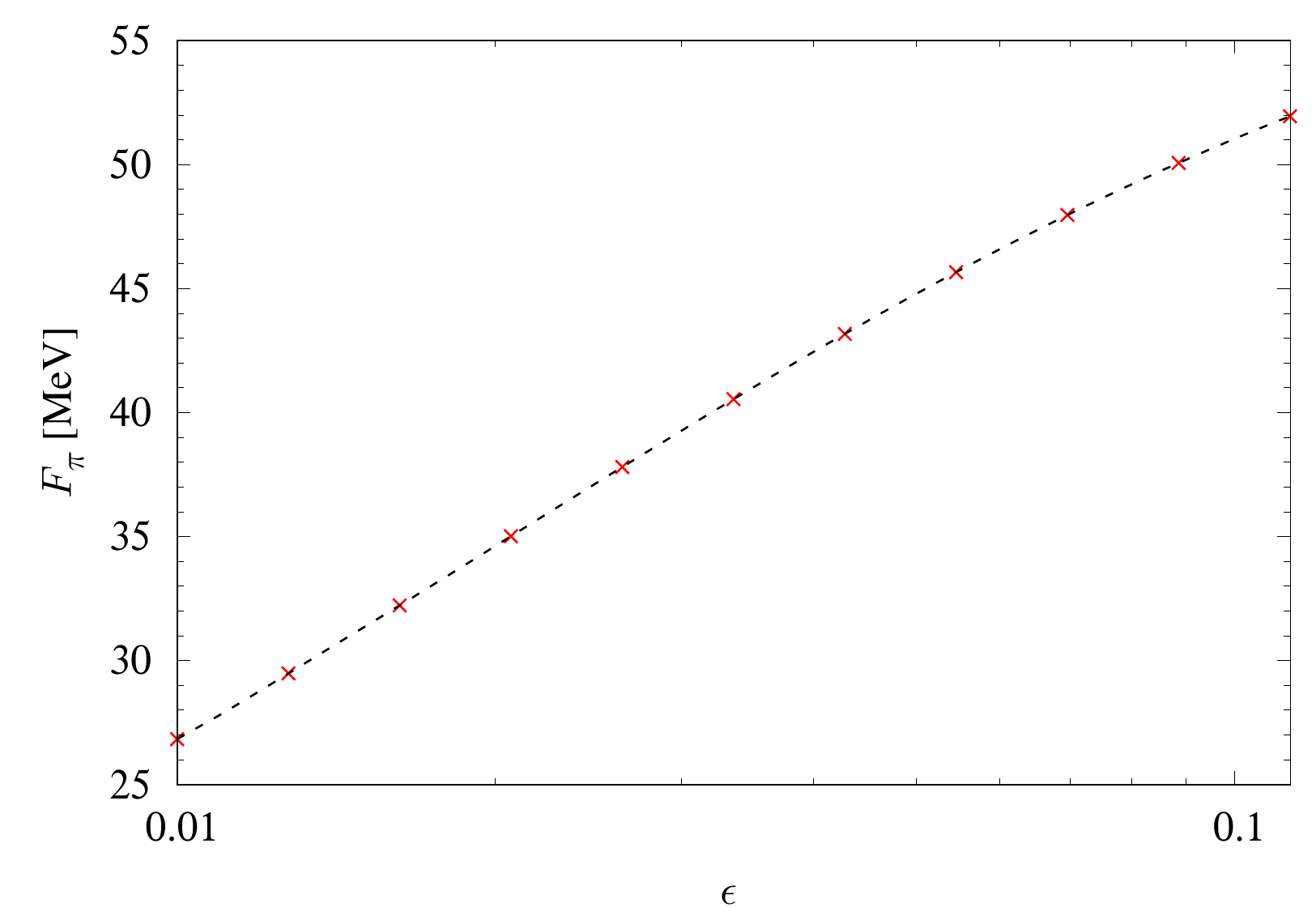}}
  \subfloat[]{\includegraphics[width=0.49\linewidth]{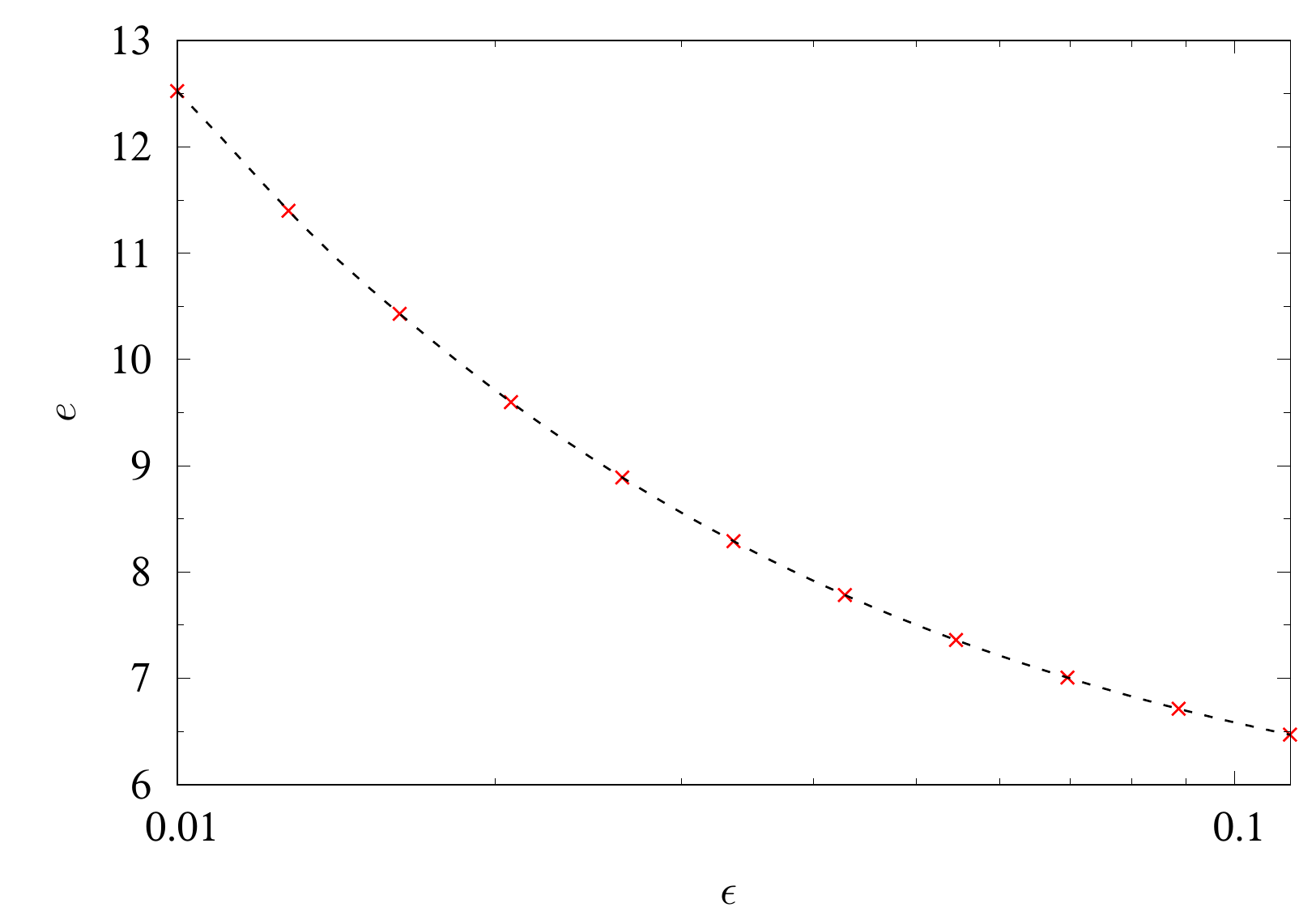}}}
\mbox{\subfloat[]{\includegraphics[width=0.49\linewidth]{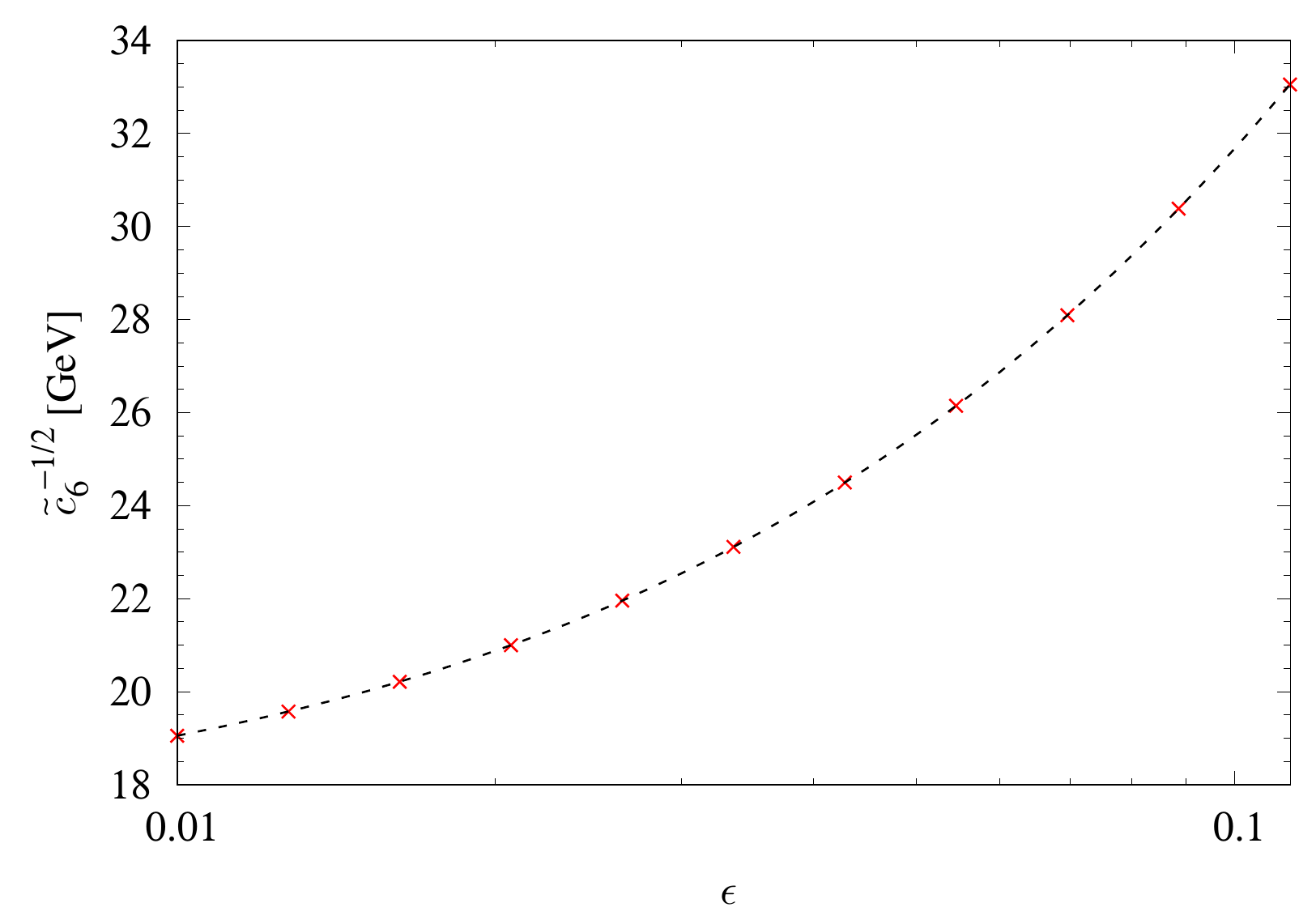}}
  \subfloat[]{\includegraphics[width=0.49\linewidth]{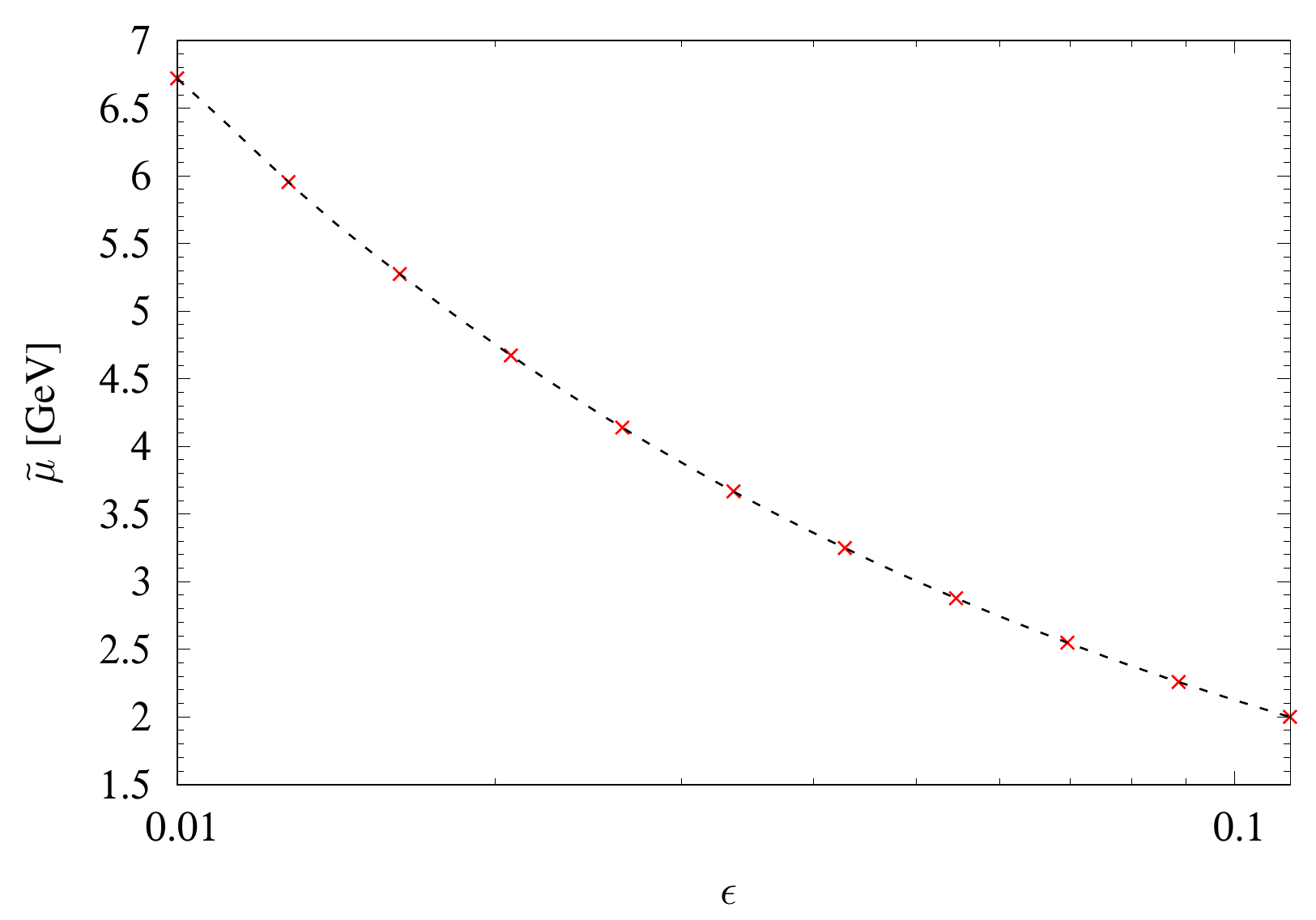}}}
\caption{The Skyrme model parameters (a) $F_\pi$ and (b) $e$ in
  physical units (MeV) as well as the BPS sector's coupling constants
  (c) $\tilde{c}_6^{-\frac12}$ and (d) $\tilde\mu$ in GeV.
  In this figure $c_2=1$, $c_4=8$, $c_6=\tfrac12$, $\mu=1$,
  $m_\pi=3$, $R=(3\pi)^{\frac13}$ and $(s,p)=(1,2)$.
}
\label{fig:physunits}
\end{figure}

It is instructive and straightforward to reinstate physical units in
the model.
Energies and lengths are measured in units
of \cite{Gudnason:2016cdo,Gudnason:2016tiz,Gudnason:2018jia}
\beq
[{\rm mass}] = \frac{F_\pi}{2\epsilon\sqrt{c_2c_4}e},\qquad
[{\rm length}] = \frac{2}{F_\pi e}\sqrt{\frac{c_2}{c_4}},
\eeq
respectively and the calibration of the model is readily performed by
\beq
F_\pi = 2\sqrt{\frac{\epsilon c_2 M_N R}{M R_N}},\qquad
e = \sqrt{\frac{M R}{\epsilon c_4 M_N R_N}},
\eeq
where $M$ and $R$ are the N$^2$LO mass and radius $R$ of the compactons in 
dimensionless units, whereas $M_N$ and $R_N$ are the mass and radius
of the nucleon in MeV.
$e$ is known as the Skyrme coupling constant and should not be
confused with the charge of the electron.
The physical pion mass in MeV is then given by
\beq
\tilde{m}_\pi = \frac{m_\pi F_\pi e}{c_2}\sqrt{\frac{c_4}{2}},
\eeq
the BPS potential mass in MeV is
\beq
\tilde{\mu} = \frac{\mu F_\pi e}{c_2}\sqrt{\frac{c_4}{2\epsilon}},
\eeq
and finally the coefficient of the sextic term in ${\rm MeV}^{-2}$ is
\beq
\tilde{c}_6 = \frac{4c_2c_6}{\epsilon c_4^4e^4F_\pi^2}.
\eeq
The calibration and coupling constants are shown in
fig.~\ref{fig:physunits} and $\tilde{m}_\pi\simeq 2.02$ GeV
independent of $\epsilon$. 

A comment in store is about the large pion mass in physical units.
We recall from sec.~\ref{sec:axial_perturb} that the dimensionless
pion mass parameter, $m_\pi$, was chosen to be abnormally large in
order for the perturbative method to capture the correct asymptotic
behavior of the solution, by imposing only the cusp condition at the
compacton boundary ($r=R$).
Choosing the dimensionless pion mass parameter about $3-6$ times
larger than a usual order-one choice, obviously has an impact on the
mass in physical units (linear relation).
The choice of calibrating the model using the original
Skyrme model (i.e.~$\Lag_2+\Lag_4$), which in our context is a small
perturbation to the BPS sector, gives a large uncertainty in the
physical quantities in physical units and of course many other ways to
calibrate the model could be contemplated.
Nevertheless, the perturbative $\epsilon$-expansion scheme is prone to
require large pion masses to be accurate; something also often seen in
lattice QCD \cite{Aoki:2021kgd}.

\section{Conclusion and discussion}\label{sec:conclusion}

In this paper, we have considered the  Skyrme model in the near-BPS
limit using the perturbative $\epsilon$-expansion scheme developed in
refs.~\cite{Gudnason:2020tps,Gudnason:2021gwc}. The near-BPS systems
we considered consist of a BPS sector, containing a sixth-order derivative
term plus a potential, and a BPS-deformation that is the original
Skyrme model with massive pions. The BPS sector was chosen to give
compacton-type solutions. To this end, parametrizing the BPS
potential as $V_{s,p}(U)\propto(1-(\tr U/2)^s)^p$, we select the
combinations $(s,p)=(1,2)$, $(2,1)$, $(2,2)$, whereas we discarded the
pion mass potential $(s,p)=(1,1)$ since we include it as a
BPS-deformation.

In the $\epsilon$-expansion scheme, the mass of the Skyrmion in the
near-BPS limit is the BPS mass with corrections in powers of $\epsilon$.
The leading-order correction comes from inserting the BPS solution
into the perturbation, i.e.~the kinetic, the Skyrme and the pion mass
terms.
Before explicitly performing the calculation, we checked if
all the BPS solutions lead to a finite LO energy contribution. To test
this, we found a general criterion based on the behavior of the
potential around the vacuum value.
We have shown that, besides the pion mass potential, also the potential
$(s,p)=(2,1)$ generates a divergent LO energy and for this reason we
discarded it from our study.

After that preliminary analysis, we explicitly calculated the LO
energy. As known from ref.~\cite{Speight:2014fqa}, the BPS
configurations that can be correctly used for that purpose must
respect the (generalized) restricted harmonic condition. As shown in
ref.~\cite{Speight:2014fqa}, we verified again that the spherically
symmetric solution of topological charge $B=1$ respects the GRH
criterion.
Moreover, we also checked that the addition of the pion-mass potential
to the BPS-deformation terms does not change the previous result.
Apart from the topological sector $B=1$, we have not been able to
analytically find any other GRH configuration of charge $B>1$,
although we proved that in some cases their existence is necessary.
Given this limitation, the only restricted-harmonic map we could build
for a multi-soliton case was the one made by non-overlapping
$B=1+1+1+\cdots$ spherical compactons.

The risk of using only the $B=1+1+1+\cdots$ configuration as the
background field is the possibility of obtaining meta-stable nuclear
solutions.
Other clusterization, in fact, could be energetically preferred for
the nuclei built with the various near-BPS models considered here.
To avoid such possibility, we analyzed the clusterization problem (at
the leading-order in $\epsilon$) by studying the ratio $E/N$ (energy
per nucleon) for the various topological sectors.
Here, we denoted by $N_{\star}$ the most energetically favored
configuration, analogously to the analysis in
refs.~\cite{Gudnason:2020tps,Gudnason:2021gwc}.
We found that a proper choice of the coefficients of the kinetic and
Skyrme term ($c_4\gg c_2R^2$) leading to $N_{\star}\sim 1$, so that the
single $B=1$ Skyrmion represents the energetically favored fundamental
unit of nuclei, as desired.
In order to obtain physically stable nuclei given the mathematical
results derived by the restricted harmonic analysis, we worked
coherently in that limit.

At the leading-order in the $\epsilon$-expansion there is no binding
energy, since the BPS solution only enjoys compact support (i.e.~it is
a compacton). A further step in the perturbative approximation was
therefore needed. 

The higher-order-in-$\epsilon$, i.e.~the NLO and N$^2$LO corrections
to the mass are computed in the $\epsilon$-expansion scheme by using a
linearized fluctuation field possessing three components, denoted $\df$,
$\dtheta$ and $\dvarphi$.
For a single $B=1$ Skyrmion, only spherically symmetric fluctuations
are turned on and only the $\df$ field, since it is the only sourced
fluctuation.
In order to capture the correct behavior of the fluctuations, a
special cusp condition on the boundary of the compacton must be
imposed making the total field smooth at said boundary.
For a single $B=1$ Skyrmion we were able to test the predictions of
the $\epsilon$-expansion with the full numerical computation.
We finally computed the binding energy of the two $B=1$ Skyrmions
bound state, corresponding to the classical version of the deuteron in
the near-BPS limit in our specific model.
The binding energy is maximal when the two compactons are touching
each other at one point and nonspherical behavior of $\df$ near the
gluing boundary turns on the fluctuation fields $\dtheta$ and
$\dvarphi$.
Although we have not been able to test the accuracy of the
binding energy of the bound state by also performing full
brute-force numerical computations, we rely on the fact that the
$\epsilon$-expansion scheme is accurate for the $B=1$ spherically
symmetric soliton and that the analogous 2-dimensional analysis for
the baby Skyrme model compares rather successful to full numerical
computations \cite{Gudnason:2020tps}.

The classical binding energy of the deuteron bound state comes out about a
factor of 3 too small, but the model is quite constrained by the
necessary conditions making the $\epsilon$-expansion reliable.
Moreover, the various choices that finally select the specific
near-BPS model are not only made for phenomenological reasons
but also for having the possibility of obtaining a
mathematically consistent perturbative expansion in $\epsilon$.
In fact, to that end, we have firstly chosen BPS compacton-type
solutions to simplify the restricted harmonic problem.
Then, we selected among the remaining near-BPS models the ones that
admit finite-energy contribution at every order in the
$\epsilon$-expansion.
In the end, we dealt with the generalized restricted harmonic problem
that pushed us to constrain the BPS-deformation's coefficients to
obtain stable nuclei.
It is therefore clear that there is no reason \textit{a priori} to
think that those constraints get the model close to the one that nature
has chosen.
An important question is, for example, whether the most
phenomenologically viable near-BPS Skyrme model contains compactons or
solitons with tails in the BPS limit that nature has chosen to be
close to.

Nevertheless, this work has shown that the near-BPS model is able to
reproduce the small binding energy for the deuteron (and in principle
for larger nuclei) of the order of the experimental values.
The near-BPS model can therefore be confirmed to be a reasonable
candidate to fix the binding energy problem of the original Skyrme
model and thus to be a reliable nuclear model.
Moreover, the exploration of the near-BPS limit made in this work
clarifies the difficulties, and thus the solutions, for a more
extensive analysis of this and related models.

In light of our new understanding of the near-BPS Skyrme problem, we
can reconsider the study in ref.~\cite{Gillard:2015eia}.
In that work, a BPS model with the potential $(s,p)=(1,2)$ slightly
deformed by just the two-derivative kinetic term was considered.
Such a deformation was coupled to the usual small parameter $\epsilon\ll1$.
Using numerical methods, the full equations of motions of the system
were solved for the cases $1\leq B\leq 8$ for the range $\epsilon\in
[0.2,1]$.
On the contrary, for smaller values of $\epsilon$ ($\epsilon< 0.2$),
all the results were numerically inaccessible (except for the $B=2$
case, where axial symmetry was assumed by Ansatz).
In that range, indeed, the numerical solutions develop spike-like
singularities, indicating that the lattice cannot resolve the field
gradients.
Despite these difficulties, the work gave interesting results. 
First of all, the numerical simulations showed that the near-BPS
solutions for a small value of $\epsilon \sim 0.2$ have different
geometric symmetries (see fig.~2 of ref.~\cite{Gillard:2015eia}).

This numerical outcome gives therefore (partial) confirmation
of the fact that a spherical configuration is far from being a good
approximation to a near-BPS solution at small $\epsilon$ for $B>1$.
This is not in contradiction with our claim.
We worked in fact with specific values of potentials and couplings
such that $N_\star\simeq 1$ so that a $B=1+1+1+\cdots$ configuration
as a background field is the most reasonable candidate BPS
background.
For the specific case considered in ref.~\cite{Gillard:2015eia}
$N_\star=2.197$.
It so happens that by the parameter choices made in our work, the
values of $\epsilon$ needed are about an order of magnitude larger
than those needed for the simplistic model of
ref.~\cite{Gillard:2015eia}, which is not inconsistent because the two
models are fundamentally different.

In particular, we underline the crucial role of the
generalized-restricted-harmonic study to extend the perturbative
method explored here to a larger set of near-BPS models.
An interesting future direction would be to consider non-spherically
symmetric restricted harmonic solutions as the background BPS
solutions for the near-BPS physics.
So far none are known, but our results suggest that there are
undiscovered solutions.

\subsection*{Acknowledgments}

S.~B.~G.~thanks the Outstanding Talent Program of Henan University and
the Ministry of Education of Henan Province for partial support.
The work of S.~B.~G.~is supported by the National Natural Science
Foundation of China (Grants No.~11675223 and No.~12071111) and by the
Ministry of Science and Technology of China (Grant No.~G2022026021L).
The work of M.~B.~and S.~B.~is supported by the INFN special project 
grant ``GAST (Gauge and String Theories)''.

\bibliographystyle{utphys}
\bibliography{refs}

\end{document}